\documentclass[11pt]{article}
\usepackage{amsfonts}
\usepackage{amsmath}

\newtheorem{theorem}{Theorem}

\newtheorem{definition}[theorem]{Definition}

\newtheorem{lemma}[theorem]{Lemma}

\newtheorem{remark}[theorem]{Remark}

\newenvironment{proof}[1][Proof]{\noindent\textbf{#1.} }{\ \rule{0.5em}{0.5em}}
\input{tcilatex}
\begin{document}

\title{Separable bi-Hamiltonian systems with quadratic in momenta first
integrals}
\author{Maciej B\l aszak \\
Institute of Physics, A.Mickiewicz University,\\
Umultowska 85, 61-614 Pozna\'{n}, Poland}
\maketitle

\begin{abstract}
Geometric separability theory of Gel'fand-Zakharevich bi-Hamilto- nian
systems on Riemannian manifolds is reviewed and developed. Particular
attention is paid to the separability of systems generated by the so-called
special conformal Killing tensors, i.e. Benenti systems. Then, infinitely
many new classes of separable systems are constructed by appropriate
deformations of Benenti class systems. All such systems can be lifted to the
Gel'fand-Zakharevich bi-Hamiltonian form.\newline
\end{abstract}

\tableofcontents

\section{Introduction}

The separation of variables for solving by quadratures the Hamilton-Jacobi
(HJ) equations of related Liouville integrable dynamic systems with
quadratic in momenta first integrals has a long history as a part of
analytical mechanics. There are some mile stones of that theory. First, in
1891 St\"{a}ckel initiated a program of classification of separable systems
presenting conditions for separability of the HJ equations in orthogonal
coordinates \cite{s1}-\cite{s3}. Then, in 1904 Levi-Civita found a test for
the separability of a Hamiltonian dynamics in a given system of canonical
coordinates \cite{l}. The next was Eisenhart \cite{Ei}-\cite{Ei1}, who in
1934 inserted a separability theory in the context of Riemannian geometry,
making it coordinate free and introducing the crucial objects of the theory,
i.e. Killing tensors. This approach was then developed by Woodhouse \cite%
{woo}, Klanins \cite{kl}-\cite{KL1} and others. Finally, in 1992, Benenti
\cite{be} -\cite{be1} constructed a particular but very important subclass
of separable systems, based on the so called special conformal Killing
tensors.

The first constructive theory of separated coordinates for dynamical systems
was made by Sklyanin \cite{sk1}. He adopted the method of soliton theory,
i.e. the Lax representation, to systematic derivation of separated
coordinates. In that approach involutive functions appear as coefficients of
characteristic equation (spectral curve) of Lax matrix. The method was
successfully aplied to separation of variables for many integrable systems
\cite{sk1}-\cite{ma1}.

Recently, a modern geometric theory of separability on bi-Poisson manifolds
was constructed \cite{1}-\cite{m3}, related to the so-called
Gel'fand-Zakharevich (GZ) bi-Hamiltonian systems \cite{GZ},\cite{GZ1}. \
Obviously, it contains as a special case Liouville integrable systems with
all constants of motion being quadratic in momenta functions. Indeed, Ibort
et.al. \cite{ib} proved that the Benenti class of systems can be lifted to
the GZ bi-Hamiltonian form.

In the following paper we construct in a systematic way a separability
theory of all Liouville integrable systems on Riemannian manifolds, which
are of the GZ type, including as a special case the Benenti class of
systems. Actually, first we construct a quasi-bi-Hamiltonian theory on the
so called $\omega N$ manifolds and then we lift it to the related GZ
bi-Hamiltonian systems on bi-Poisson manifolds. What is important,
infinitely many classes of separable systems are constructed from
appropriate deformations of the Benenti class of systems. In that sense we
demonstrate the crucial role of this particular class of systems in the
separability theory of dynamic systems on Riemannian manifolds.

The organization of the paper is as follows. In Section 2 we sketch the
geometric separability theory of GZ bi--Hamiltonian systems, which was
recently constructed and is a main tool used in the paper. \ In Section 3 we
recall the basic facts about separable dynamics on Riemannian manifolds.
Section 4 deals with a special case of separable systems, i.e. the so-called
Benenti systems. We review this class of systems systematically as it plays
a crucial role in a separability theory of Gel'fand-Zakharevich systems on
Riemannian manifolds and is of a special importance for the theory developed
in this paper. In Section 5, we construct the simplest new classes of
separable systems being the so-called 1-hole deformations of the Benenti
class. For this example we explain the main ideas of our approach as well as
the methods of systematic construction of separable potentials and
(quasi)-bi-Hamiltonian representations. In Section 6, we develop the
approach to the case of arbitrary k-hole deformations of the Benenti class
of systems, constructing a complete theory of separable GZ systems on
Riemannian manifolds. Finally, in Section 7, we illustrate our theory by a
few simple examples.

\section{Gel'fand-Zakharevich bi-Hamiltonian systems and their separability
on $\protect\omega N$ manifolds}

\subsection{Basic definitions}

In this Section we sketch the basic concepts of the modern geometric
separability theory to make other parts of the paper more clear to the
reader. Let us first remind few basic facts from the Poisson geometry. Given
a manifold $\mathcal{M}$, a \emph{Poisson operator} $\pi $ on $\mathcal{M}$
is a mapping $\pi :T^{\ast }\mathcal{M}\rightarrow T\mathcal{M}$ that is
fibre-preserving (i.e. $\pi |_{T_{x}^{\ast }\mathcal{M}}:T_{x}^{\ast }%
\mathcal{M}\rightarrow T_{x}\mathcal{M}$ for any $x\in \mathcal{M}$) and
such that the induced bracket on the space $C^{\infty }(\mathcal{M})$ of all
smooth real-valued functions on $\mathcal{M}$
\begin{equation}
\left\{ .,.\text{ }\right\} _{\pi }:C^{\infty }(\mathcal{M})\times C^{\infty
}(\mathcal{M})\rightarrow C^{\infty }(\mathcal{M})\text{ \ , \ }\left\{
F,G\right\} _{\pi }\overset{\mathrm{def}}{=}\left\langle dF,\pi
\,dG\right\rangle  \tag{2.1}  \label{0.1}
\end{equation}%
(where $\left\langle .,.\right\rangle $ is the dual map between $T\mathcal{M}
$ and $T^{\ast }\mathcal{M}$). It is skew-symmetric, satisfies Jacobi
identityand the Leibniz rule and the symbol $d$ denotes the operator of
exterior derivative. The second order contravariant tensor field $\pi $ can
always be interpreted as a bivector, $\pi \in \Lambda ^{2}(\mathcal{M})$ and
in a given coordinate system $(x^{1},\ldots ,x^{m})$ on $\mathcal{M}$ we
have
\begin{equation}
\pi =\sum\limits_{i<j}^{m}\pi ^{ij}\frac{\partial }{\partial x_{i}}\wedge
\frac{\partial }{\partial x_{j}}.  \tag{2.2}  \label{0.2}
\end{equation}%
The function $C:\mathcal{M}\rightarrow \mathbb{R}$ is called the \emph{%
Casimir function} of the Poisson operator $\pi $ if for an arbitrary
function $F:\mathcal{M}\rightarrow \mathbb{R}$ we have $\left\{ F,C\right\}
_{\pi }=0$ (or, equivalently, if $\pi dC=0$). A linear combination $\pi
_{\xi }=\pi _{1}-\xi \pi _{0}$ ($\xi \in \mathbb{R}$) of two Poisson
operators $\pi _{0}$ and $\pi _{1}$ is called a \emph{Poisson pencil} if the
operator $\pi _{\xi }$ is Poissonian for any value of the parameter $\xi $.
In this case we say that $\pi _{0}$ and $\pi _{1}$ are \emph{compatible}$.$
Given a Poisson pencil $\pi _{\xi }=\pi _{1}-\xi \pi _{0}$ we can often
construct a sequence of \ vector fields $X_{i}$ on $\mathcal{M}$ that have a
twofold Hamiltonian form (the so-called $\emph{bi-Hamiltonian}$ $\emph{chain}
$)
\begin{equation}
X_{i}=\pi _{1}dh_{i}=\pi _{0}dh_{i+1}  \tag{2.3}  \label{0.3}
\end{equation}%
where $h_{i}:\mathcal{M}\rightarrow \mathbb{R}$ are called the Hamiltonians
of the chain (\ref{0.3}) and where $i\ $\ is some discrete index. This
sequence of vector fields may or may not truncate (depending on the
existence of Casimir functions).

Let us consider a bi-Poisson manifold $(M,\pi _{0},\pi _{1})$ of $\dim
M=2n+m $ where $\pi _{0},\pi _{1}$ is a pair of compatible Poisson tensors
of rank $2n.$ Moreover we assume that the Poisson pencil $\pi _{\xi }$
admits $m,$ polynomial with respect to the pencil parameter $\xi ,$ Casimir
functions of the form
\begin{equation}
h^{(j)}(\xi )=\sum_{i=0}^{n_{j}}h_{i}^{(j)}\xi ^{n_{j}-i},\ \ \ \ \ \ \ \ \
\ \ j=1,...,m,  \tag{2.4}  \label{0.4}
\end{equation}%
such that $n_{1}+...+n_{m}=n$ and $h_{i}^{(j)}$ are functionally
independent. The collection of $n$ bi-Hamiltonian vector fields
\begin{equation}
\pi _{\xi }dh^{(j)}(\xi )=0\Longleftrightarrow X_{i}^{(j)}=\pi
_{1}dh_{i}^{(j)}=\pi _{0}dh_{i+1}^{(j)},\ \ \ \ i=1,...,n_{j},\ \ \
j=1,...,m,  \tag{2.5}  \label{0.5}
\end{equation}%
is called the Gel'fand-Zakharevich system of the bi-Poisson manifold $%
\mathcal{M}$. Notice that each chain starts from a Casimir of $\pi _{0}$ and
terminates with a Casimir of $\pi _{1}.$ Moreover all $h_{i}^{(j)}$ pairwise
commute with respect to both Poisson structures
\begin{equation}
X_{i}^{(j)}(h_{l}^{(k)})=\langle dh_{l}^{(k)},\pi _{0}dh_{i+1}^{(j)}\rangle
=\langle dh_{l}^{(k)},\pi _{1}dh_{i}^{(j)}\rangle =0.  \tag{2.6}
\label{0.6a}
\end{equation}

\subsection{Geometric separability theory}

The necessary condition for separability of Hamiltonian functions $%
h_{i}^{(j)}$ is a projectibility of the Poisson pencil $\pi _{\xi }$ onto a
symplectic leaf $\mathcal{N}$ of $\pi _{0},$ that is a $2n$-dimesional
submanifold defined by fixed values of Casimir functions of $\pi _{0}:\ $ $%
h_{0}^{(1)}=c_{1},...,h_{0}^{(m)}=c_{m}.$ Thus, $\mathcal{N}$ is a
submanifold of a codimension $m$ in $\mathcal{M}.$ The projection is done
through the particular realization of the Marsden-Ratiu scheme \cite{mar},
i.e. along the appropriate transversal distribution. Let $Z_{i},\,i=1,...,m$
be some vector fields transversal to $\mathcal{N}$, spanning an involutive
(integrable) distribution $\mathcal{Z}$ in $\mathcal{M}$ of a constant
dimension $m$ (that is a smooth collection of $m$-dimensional subspaces $%
\mathcal{Z}_{x}\subset $ $T_{x}\mathcal{M}$ at every point $x$ in $\mathcal{M%
}$). The word 'transversal' means here that no vector field $Z_{i}$ is at
any point tangent to the submanifold $\mathcal{N}$ passing through this
point. Hence, the tangent bundle $T\mathcal{M}$ splits into a direct sum
\begin{equation}
T\mathcal{M}=T\mathcal{N}\oplus \mathcal{Z}  \tag{2.7}  \label{0.6}
\end{equation}%
(which means that at any point $x$ in $\mathcal{M}$ we have $T_{x}\mathcal{M}%
=T_{x}\mathcal{N}\oplus \mathcal{Z}_{x}$ with $s$ such that $x\in \mathcal{N}
$ ) and so does its dual
\begin{equation}
T^{\ast }\mathcal{M}=T^{\ast }\mathcal{N}\oplus \mathcal{Z}^{\ast },
\tag{2.8}  \label{0.7}
\end{equation}%
where $T^{\ast }\mathcal{N}$ is the annihilator of $\mathcal{Z}$ and $%
\mathcal{Z}^{\ast }$ is the annihilator of $T\mathcal{N}$. It means that if $%
\alpha $ is a one form in $T^{\ast }\mathcal{N}$ then $\alpha (Z_{i})=0$ for
all $i=1,\ldots ,m$ and if $\beta $ is a one-form in $\mathcal{Z}^{\ast }$
then $\beta $ vanishes on all vector fields tangent to $\mathcal{N}$.
Moreover, we assume that the vector fields $Z_{i}$ which span $\mathcal{Z}$
are chosen in such a way that $dh_{0}^{(j)},\;j=1,...,m$ \ is a basis in $%
\mathcal{Z}^{\ast }$ that is dual to the basis $Z_{i}$ of the distribution $%
\mathcal{Z}$,
\begin{equation}
\left\langle dh_{0}^{(j)},Z_{i}\right\rangle =Z_{i}(h_{0}^{(j)})=\delta
_{ij},  \tag{2.9}  \label{0.8}
\end{equation}%
(it is no restriction since for any distribution $\mathcal{Z}$ transversal
to $\mathcal{N}$ we can choose its basis so that (\ref{0.8}) is satisfied).
The 'orthogonality' condition (\ref{0.8}) together with the involutivity of
the distribution indicates the relation
\begin{equation}
\lbrack Z_{i},Z_{j}]=0,  \tag{2.10}  \label{0.9}
\end{equation}%
where $[.,.]$ stands for the commutator of vector fields.

\begin{theorem}
\label{t1}Poisson tensor $\pi _{1}$ is projectible onto a symplectic leaf $%
\mathcal{N}$ of $\pi _{0}$ along a transversal integrable distribution $%
\mathcal{Z}$ if the normalized vector fields $Z_{i}$ locally generating $%
\mathcal{Z}$ are symmetries of $\pi _{0}$ $(L_{Z_{i}}\pi _{0}=0)$ and
satisfy
\begin{equation}
L_{Z_{i}}\pi _{1}=\sum_{j}Y_{i}^{(j)}\wedge Z_{j},  \tag{2.11a}  \label{0.10}
\end{equation}%
where
\begin{equation}
Y_{i}^{(j)}=\pi _{0}~d(Z_{i}(h_{1}^{(j)}))=[Z_{i},\pi
_{1}dh_{0}^{(j)}]=[Z_{i},X_{1}^{(j)}]  \tag{2.11b}  \label{0.100}
\end{equation}%
and $L_{Z}$ means a Lie derivative in the direction of $Z$. The projection
of $\pi _{1}$ onto $\mathcal{N}$ along $\mathcal{Z}$ is equivalent to the
projection of the deformed Poisson tensor
\begin{equation}
\pi _{1D}=\pi _{1}-\sum_{j}X_{1}^{(j)}\wedge Z_{j},\ \ \ \ \ X_{1}^{(j)}=\pi
_{1}dh_{0}^{(j)}=\pi _{0}dh_{1}^{(j)}  \tag{2.12}  \label{0.11}
\end{equation}%
onto its symplectic leaf $\mathcal{N}.$ Moreover, $\pi _{1}$ and $\pi _{1D}$
are compatible.
\end{theorem}

The proof as well as the details of the whole construction the reader can
find in \cite{m3}, \cite{m4} and \cite{mb}. Notice that
\begin{equation*}
L_{Z_{i}}\pi _{1D}=0,
\end{equation*}%
so $Z_{i}$ are also symmetries of $\pi _{1D}.$

Let us denote the projections of $\pi _{0},\pi _{1}$ by $\theta _{0},\theta
_{1}$ and restrictions of $(h_{1}^{(1)},...,$ $h_{n_{m}}^{(m)})|_{\mathcal{N}%
}$ to $\mathcal{N}$ by $%
(H_{1}^{(1)},...,H_{n_{m}}^{(m)}):=(H_{1},...,H_{n}). $ Notice that the
projection of the Poisson pencil $\pi _{\xi }$ is a Poisson pencil $\theta
_{\xi }$ on $\mathcal{N}$ and in a generic case both Poisson tensors $\theta
_{0},\theta _{1}$ are nondegenerate. Hence, $\mathcal{N}$ is a bi-symplectic
manifold as is endowed with two symplectic forms $\omega _{0},\omega _{1}$
defined by
\begin{equation}
\{F,G\}_{\theta _{i}}=\omega _{i}(X_{F},X_{G}),\ \ \ \ X_{F}=\theta _{0}dF,\
\ \ i=0,1.  \tag{2.13}  \label{0.13}
\end{equation}%
This simply means that $\omega _{0}=\theta _{0}^{-1}$ and $\omega
_{1}=\omega _{0}\theta _{1}\omega _{0}.$ Indeed
\begin{eqnarray}
\omega _{0}(X_{F},X_{G}) &=&<\omega _{0}X_{F},X_{G}>=<\omega _{0}\theta
_{0}dF,\theta _{0}dG>  \TCItag{2.14a}  \label{0.14a} \\
&=&<dF,\theta _{0}dG>=\{F,G\}_{\theta _{0}},  \notag
\end{eqnarray}%
\begin{eqnarray}
\omega _{1}(X_{F},X_{G}) &=&<\omega _{0}\theta _{1}\omega
_{0}X_{F},X_{G}>=<\omega _{0}X_{F},\theta _{1}\omega _{0}X_{G}>
\TCItag{2.14b}  \label{0.14b} \\
&=&<dF,\theta _{1}dG>=\{F,G\}_{\theta _{1}}.  \notag
\end{eqnarray}%
Moreover, one can construct the tensor field $N:=\theta _{1}\theta
_{0}^{-1}=\theta _{1}\omega _{0},$ of type $(1,1),$ called a \emph{recursion
operator }of $\mathcal{N}$ and its dual $N^{\ast }=\omega _{0}\theta _{1}.$%
Notice that%
\begin{equation}
N\theta _{0}=\theta _{1},~\ \ \ \ \ N^{\ast }\omega _{0}=\omega _{1}.
\tag{2.15}  \label{0.15a}
\end{equation}%
The important property of $N$ is that its Nijenhuis torsion
\begin{equation}
T(N)(X,Y):=[NX,NY]-N([NX,Y]+[X,NY]-N[X,Y])  \tag{2.16}  \label{0.14}
\end{equation}%
vanishes as a consequence of the compatibility between $\theta _{0}$ and $%
\theta _{1}$ and hence implies that $\omega _{1}$ is closed \cite{mag}. Such
manifolds are known as the so-called $\omega N$ manifolds. The generic case
means that $2n$-dimensional $\omega N$ manifold is endowed with a recursion
operator $N$ which has at every point $n$ distinct double eigenvalues $%
\lambda _{1},...,\lambda _{n},$ which are functionally independent on $%
\mathcal{N}$. Choosing $\lambda _{i}$ as the canonical position coordinates,
we can always supplement a set of local coordinates $(\lambda _{i},\mu _{i})$
on $\mathcal{N}$ by the canonically conjugate momenta $\mu _{i}$.

\begin{definition}
\label{d1}A set of local coordinates $(\lambda _{i},\mu _{i})$ on $\omega N$
manifold $\mathcal{N}$ is called a set of Darboux-Nijenhuis (DN) coordinates
if they are canonical with respect to $\theta _{0}$ and diagonalize the
recursion operator with diagonal elements beeing its eigenvalues.
\end{definition}

It means that in the $(\lambda ,\mu )$ coordinates
\begin{equation}
\theta _{0}=\left(
\begin{array}{cc}
0 & I_{n} \\
-I_{n} & 0%
\end{array}%
\right) ,\ \ \ \theta _{1}=\left(
\begin{array}{cc}
0 & \Lambda _{n} \\
-\Lambda _{n} & 0%
\end{array}%
\right) ,\ \ \ N=\left(
\begin{array}{cc}
\Lambda _{n} & 0 \\
0 & \Lambda _{n}%
\end{array}%
\right) ,\ \   \tag{2.17}  \label{0.15}
\end{equation}%
where $\ \Lambda _{n}=\mathrm{diag}(\lambda _{1},...,\lambda _{n}),$ and
their differentials span the $T^{\ast }\mathcal{N}$ which is an eigenspace
of $N^{\ast }$ (the adjoint of $N$), as
\begin{equation}
N^{\ast }d\lambda _{i}=\lambda _{i}d\lambda _{i},\ \ \ \ N^{\ast }d\mu
_{i}=\lambda _{i}d\mu _{i},\ \ \ i=1,...,n.  \tag{2.18}  \label{0.16}
\end{equation}

As well known, $n$ functionally independent Hamiltonian functions $%
H_{i},i=1,...,n$ are said to be separable in the canonical coordinates $%
(\lambda ,\mu )$ if there are $n$ relations, called the separation
conditions (Sklyanin \cite{sk1}), of the form
\begin{equation}
\varphi _{i}(\lambda ^{i},\mu _{i};H_{1},...,H_{n})=0,\;\;\;i=1,...,n,\ \
\det \left[ \frac{\partial \varphi _{i}}{\partial H_{j}}\right] \neq 0,
\tag{2.19}  \label{0.17}
\end{equation}%
which guarantee the solvability of the appropriate Hamilton-Jacobi equations
and involutivity of $H_{i}$. A special case, when all separation relations (%
\ref{0.17}) are affine in $H_{i}$ is given by the set of equations
\begin{equation}
\sum_{k=1}^{n}\phi _{i}^{k}(\lambda _{i},\mu _{i})H_{k}=\psi _{i}(\lambda
_{i},\mu _{i}),\;\;\;i=1,...,n,  \tag{2.20}  \label{0.18}
\end{equation}%
where $\phi $ and $\psi $ are arbitrary smooth functions of their arguments,
are called the \emph{St\"{a}ckel separation conditions} and the related
dynamic systems are called the St\"{a}ckel separable ones.

\begin{theorem}
\label{t2}\cite{m3} Let $\mathcal{N}$ be a generic $\omega N$ manifold and
let $(H_{1},...,H_{n})$ be a set of $n$ functionally independent
Hamiltonians on $\mathcal{N}$. Then, the foliation defined by $%
(H_{1},...,H_{n})$ is separable in DN coordinates and the subspace
spanned by $(dH_{1},...,dH_{n})$ is invariant with respect to
$N^{\ast }$ if there exist some functions $\alpha _{ij}$ such that
\begin{equation}
N^{\ast }dH_{i}=\sum_{j=1}^{n}\alpha _{ij}dH_{j},\ \ \ i=1,...,n.  \tag{2.21}
\label{0.19}
\end{equation}%
Hence, the distribution defined by $(H_{1},...,H_{n}),$ spanned by the
Hamiltonian vector fields $X_{H_{i}},$ is invariant with respect to $N.$
\end{theorem}

The formula (\ref{0.19}) can be written in the equivalent form
\begin{equation}
\theta _{1}dH_{i}=\sum_{j=1}^{n}\alpha _{ij}\theta _{0}dH_{j},\ \ \
i=1,...,n,  \tag{2.22}  \label{0.20}
\end{equation}%
which will be called a \emph{quasi-bi-Hamiltonian representation} of
separable dynamics. Notice that the projection of bi-Hamiltonian chains (\ref%
{0.5}) onto $\mathcal{N}$ takes the form of (\ref{0.20}). Indeed from (\ref%
{0.11}) we get
\begin{eqnarray}
\theta _{1}dH_{i}^{(j)} &=&\pi _{1D}|_{\mathcal{N}}~dh_{i}^{(j)}|_{\mathcal{N%
}}=(\pi _{1D}~dh_{i}^{(j)})|_{\mathcal{N}}  \notag \\
&=&(\pi _{1}dh_{i}^{(j)})|_{\mathcal{N}}-\sum_{k=1}^{m}\left(
Z_{k}(h_{i}^{(j)})X_{1}^{(k)}\right) |_{\mathcal{N}}  \notag \\
&=&(\pi _{0}dh_{i+1}^{(j)})|_{\mathcal{N}}-\sum_{k=1}^{m}\left(
Z_{k}(h_{i}^{(j)})\pi _{0}dh_{1}^{(k)}\right) |_{\mathcal{N}}  \TCItag{2.23}
\label{0.21a} \\
&=&\pi _{0}|_{\mathcal{N}}\ dh_{i+1}^{(j)}|_{\mathcal{N}}-%
\sum_{k=1}^{m}Z_{k}(h_{i}^{(j)})|_{\mathcal{N}}\ \pi _{0}|_{\mathcal{N}}\
dh_{1}^{(k)}|_{\mathcal{N}}  \notag \\
&=&\theta _{0}dH_{i+1}^{(j)}-\sum_{k=1}^{m}Z_{k}(h_{i}^{(j)})|_{\mathcal{N}%
}\theta _{0}dH_{1}^{(k)},  \notag
\end{eqnarray}%
where $\pi |_{\mathcal{N}}$ means a plain restriction of $\pi $ to $\mathcal{%
N}$.

\begin{theorem}
\label{t3}\cite{m3} A $n$-tuple $(H_{1},...,H_{n})$ of separable functions
on $\omega N$ manifold $\mathcal{N}$ is St\"{a}ckel separable \emph{iff }%
additionally to the condition (\ref{0.19}) we have
\begin{equation}
N^{\ast }d\alpha _{ij}=\sum_{k=1}^{n}\alpha _{ik}d\alpha _{kj},\ \ \
i,j=1,...,n.  \tag{2.24}  \label{0.22}
\end{equation}%
Equivalently, a $n$-tuple $(h_{1}^{(1)},...,h_{n_{m}}^{(m)})$ of functions
from Gel'fand-Zakharevich bi-Hamiltonian chains on bi-Poisson manifold $%
\mathcal{M}$ is St\"{a}ckel separable $\emph{iff}$ a Poisson pencil $\pi
_{\xi }$ is projectible onto a symplectic leaf $\mathcal{N}$ of $\pi _{0}$
and $h_{i}^{(j)}$ functions are affine in Casimir coordinates $%
c_{i},i=1,...,m,$ i.e.
\begin{equation}
Z_{k}(Z_{l}(h_{i}^{(j)})=0  \tag{2.25}  \label{0.23}
\end{equation}%
for all $k,l,j=1,...,m$ and $i=1,...,n_{m}.$
\end{theorem}

Hence, in (2.21)
\begin{equation}
Z_{k}(h_{i}^{(j)})=\alpha _{i,k}^{(j)}  \tag{2.26}  \label{0.24}
\end{equation}%
are $c$-independent, so
\begin{equation}
N^{\ast }dH_{i}^{(j)}=dH_{i+1}^{(j)}-\sum_{k=1}^{m}\alpha
_{i,k}^{(j)}dH_{1}^{(k)}.  \tag{2.27}  \label{0.25}
\end{equation}%
In such a case the St\"{a}ckel separation conditions (\ref{0.18}) take the
form
\begin{equation}
\sum_{j=1}^{m}\sigma _{j}^{i}(\lambda _{i},\mu _{i})h^{(j)}(\lambda
_{i})=\psi _{i}(\lambda _{i},\mu _{i}),\ \ \ \ \ \ \ \ \ i=1,...,n,
\tag{2.28}  \label{0.26}
\end{equation}%
where $h^{(j)}(\lambda _{i})$ is a Casimir $h^{(j)}(\xi )$ evaluated at $\xi
=\lambda _{i}$ and $\sigma _{j}^{i}$ is a Casimir multiplier. For the
majority of known integrable systems $\psi _{i}(\lambda _{i},\mu _{i})=\psi
(\lambda _{i},\mu _{i})$ and $\sigma _{j}^{i}(\lambda _{i},\mu _{i})=\sigma
_{j}(\lambda _{i},\mu _{i}),$ and then, the separation conditions (\ref{0.26}%
) can be presented in a compact form as $n$ copies of the so-called \emph{%
separation curve }%
\begin{equation}
\sum_{j=1}^{m}\sigma _{j}(\xi ,\mu )h^{(j)}(\xi )=\psi (\xi ,\mu ),\ \ \ \ \
\ \ (\xi ,\mu )=(\lambda _{i},\mu _{i}),\ \ \ \ i=1,...,n.  \tag{2.29}
\label{0.27}
\end{equation}%
For the uniqness of the representation (\ref{0.27}) we chose the
normalization condition $\sigma _{m}=1.$

Let us define the following matrix
\begin{equation}
\alpha (\xi )=\left(
\begin{array}{ccc}
Z_{1}(h^{(1)}(\xi ) & \cdots & Z_{1}(h^{(m)}(\xi ) \\
\cdots & \cdots & \cdots \\
Z_{m}(h^{(1)}(\xi ) & \cdots & Z_{m}(h^{(m)}(\xi )%
\end{array}%
\right) .  \tag{2.30}  \label{0.28}
\end{equation}

\begin{lemma}\cite{m3}
\label{l1}:
\end{lemma}

\begin{description}
\item[(i)] \emph{Determinant of $\alpha (\xi )$ coincides with the minimal
polynomial of $N$
\begin{equation}
\det \alpha (\xi )=\sqrt{\det (\xi I-N)}.  \tag{2.31}  \label{0.29}
\end{equation}%
}

\item[(ii)] \emph{\ Casimir multipliers $\sigma _{j}^{i}(\lambda _{i},\mu
_{i})$ are given by the relation
\begin{equation}
\alpha (\lambda _{i})\sigma ^{i}(\lambda _{i},\mu _{i})=0,  \tag{2.32}
\label{0.30}
\end{equation}%
where $\sigma ^{i}=(\sigma _{1}^{i},...,\sigma _{m-1}^{i},1)^{T}$ and $%
\alpha (\lambda _{i})$ is the matrix $\alpha (\xi )$ evaluated at $\xi
=\lambda _{i}.$}
\end{description}

Up to now we sketch the passage from degenerated Poisson pencil $\pi
_{\lambda }$ to nondegenerated one $\theta _{\lambda }:~\ \pi _{\lambda
}\rightarrow \theta _{\lambda }.$ According to the further applications, we
are interested in the inverse passage: $\theta _{\lambda }\rightarrow \pi
_{\lambda },$ as we would like to construct GZ bi-Hamiltonian chains from a
given quasi-bi-Hamiltonian ones.

\begin{lemma}
\label{l2}Let $\theta $ be a Poisson tensor on manifold $\mathcal{N}$ of $%
\dim \mathcal{N}=m,$ and let $\pi $ be a second order contravariant tensor
field on $\mathcal{M}=\mathcal{N}\otimes \underset{k}{\underbrace{\mathbb{R}%
\otimes \ldots \otimes \mathbb{R}}}$ of the same rank as $\theta ,$ i.e.
with $k$ additional Casimir functions $\varphi _{1},...,\varphi _{k}$.
Moreover, let the restricton of $\pi $ to $\mathcal{N}$ (the level set of $%
\varphi _{1}=...=\varphi _{k}=0$) coincides with $\theta .$ Then, the
operator $\pi $ is also Poisson.
\end{lemma}

\begin{proof}
Let us complete the set of functions $\varphi _{i}$ with some functions $%
x_{j}$ which parametrized $\mathcal{N}$ submanifold, to a coordinate system $%
(x,\varphi )$ on \emph{M}$.$ Then, the matrix of the operator $\pi $ has
last $k$ rows and columns equal to zero, while the $m$ dimensional upper
left block coincides with $\theta ,$ which is Poisson.
\end{proof}

Let $\theta _{0}$ and $\theta _{1}$ be compatible Poisson tensors on $%
\mathcal{N}$. Then, let $\pi _{0}$ and $\pi _{1D}$ be Poisson extensions of $%
\theta _{0}$ and $\theta _{1}$ from $\mathcal{N}$ to $\mathcal{M}$, such as
in Lemma \ref{l2}, with a common set of Casimirs $\varphi _{1},...,\varphi
_{k}.$ Of course, $\pi _{0}$ and $\pi _{1D}$ are still compatible. Now, let $%
\pi _{1}$ be another Poisson extension of $\theta _{1}$ from $\mathcal{N}$
to $\mathcal{M}$ with a new set of Casimirs $\psi _{1},...,\psi _{k}.$ The
question is: when $\pi _{1}$ is compatible with $\pi _{0}$?

Assume that there are vector fields $Z_{i},i=1,...,k$ transversal to the
submanifold $\mathcal{N}$, such that $L_{Z_{i}}\pi _{0}=0,i=1,...,k,$ and $%
\pi _{1}$ has the form
\begin{equation}
\pi _{1}=\pi _{1D}+\sum_{i=1}^{k}X_{i}\wedge Z_{i},  \tag{2.33}
\label{0.33a}
\end{equation}%
where $X_{i}$ are some vector fields.

\begin{lemma}
\label{l3}A sufficient condition for compatibility of $\pi _{0}$ and $\pi
_{1}$ is that $X_{i}$ are Hamiltonian vector fields with respect to $\pi
_{0}.$
\end{lemma}

The Jacobi identity for a tensor $\pi $ is equivalent to the relation $[\pi
,\pi ]_{S}=0,$ where $[.,.]_{S}$ is a Schouten bracket. Hence, the
compatibility of two Poisson tensors $\pi _{0}$ and $\pi _{1}$ is equivalent
to the relation $[\pi _{0},\pi _{1}]_{S}=0.$ Before we start the proof we
will therefore remind some basic facts about Schouten bracket.

The Schouten bracket \cite{S1}, \cite{S2} in general is a bilinear mapping $%
\Lambda ^{q}(\mathcal{M})\times \Lambda ^{r}(\mathcal{M})\rightarrow \Lambda
^{q+r-1}(\mathcal{M})$ that with every $q$-vector and $r$-vector associates
a $(q+r-1)$-vector $[Q,R]_{S}$ (Schouten bracket of $Q$ and $R$) that is
skewsymmetric in the sense that
\begin{equation}
\lbrack R,Q]_{S}=(-1)^{qr}[Q,R]_{S}  \tag{2.34}  \label{0.34a}
\end{equation}%
and that satisfies the Leibniz identity
\begin{equation}
\lbrack Q,R\wedge P]_{S}=[Q,R]_{S}\wedge P+(-1)^{(q-1)r}R\wedge \lbrack
Q,P]_{S},  \tag{2.35}  \label{0.35a}
\end{equation}%
where $P$ is any $p$-vector $\in \Lambda ^{p}(\mathcal{M}).$ Using the above
properties of the Schouten bracket one can show that if $X,Y$ are some
vector fields, then $[X,Y]_{S}=[X,Y]=L_{X}Y$ is a usual Lie bracket
(commutator) of vector fields and if moreover $P$ is a bivector then
\begin{equation}
\lbrack X\wedge Y,P]_{S}=Y\wedge \lbrack X,P]_{S}-X\wedge \lbrack Y,P]_{S},\
\ \ \ [X,P]_{S}=L_{X}P.  \tag{2.36}  \label{0.36a}
\end{equation}

\begin{proof}
From properties of the Schouten bracket
\begin{eqnarray}
\lbrack \pi _{0},\pi _{1}]_{S} &=&[\pi _{0},\pi
_{1D}+\sum_{i=1}^{k}X_{i}\wedge Z_{i}]_{S}=\sum_{i=1}^{k}[\pi
_{0},X_{i}\wedge Z_{i}]_{S}  \notag \\
&=&\sum_{i=1}^{k}(Z_{i}\wedge L_{X_{i}}\pi _{0}-X_{i}\wedge L_{Z_{i}}\pi
_{0})=\sum_{i=1}^{k}Z_{i}\wedge L_{X_{i}}\pi _{0}.  \TCItag{2.27}
\label{0.37a}
\end{eqnarray}%
Hence, the most general condition for the compatibility is
\begin{equation}
\sum_{i=1}^{k}Z_{i}\wedge L_{X_{i}}\pi _{0}=0,  \tag{2.38}  \label{0.38a}
\end{equation}%
with the strong solution $L_{X_{i}}\pi _{0}=0,i=1,...,k,$ which takes the
place when $X_{i}$ are Hamiltonian vector fields with respect to $\pi _{0}$,
i.e. there are some functions $f_{1},...,f_{k}$ such that
\begin{equation}
X_{i}=\pi _{0}df_{i}.  \tag{2.39}  \label{0.39a}
\end{equation}
\end{proof}

Notice that when in addition we impose the normalization conditions $%
Z_{i}(\varphi _{j})=\delta _{ij},i,j=1,...,k,$ and commutativity of $X_{i}$
then $\pi _{1}d\varphi _{j}=X_{j}$ and $X_{j}$ are bi-Hamiltonian vector
fields
\begin{equation}
\pi _{0}df_{i}=\pi _{1}d\varphi _{i}=X_{i},  \tag{2.40}  \label{0.40a}
\end{equation}%
which is just our case (\ref{0.11}) with $\varphi
_{i}=h_{0}^{(i)},f_{i}=h_{1}^{(i)}.$

Finally, let us mention the problem of explicit construction of DN
coordinates $(\lambda ,\mu ).$ Assume that $\mathcal{N}$ is parametrized by
a set of coordinates $\{v_{i}\}_{i=1}^{2n},$ not necessary canonical. The
first part of the transformation $(\lambda ,\mu )\rightarrow (v),$ i.e. $%
\lambda \rightarrow v$ is reconstructed from the characteristic equation of $%
N$
\begin{equation}
\sqrt{\det (\xi I-N)}=\det \alpha (\xi )=0\ \ \ \Longrightarrow \ \ \
\lambda _{i}=\lambda _{i}(v),\ \ \ \ \ i=1,...,n.  \tag{2.41}  \label{0.31}
\end{equation}%
The reconstruction of missing part of the transformation: $\mu \rightarrow
v, $ in general case, is far from being trivial and is discussed in \cite{7}-%
\cite{m3}. In the following we restrict ourselves to the canonical point
transformations $(\lambda ,\mu )\rightarrow (q,p),$ where reconstruction of
the remaining $n$ relations $p_{i}=p_{i}(\lambda ,\mu )$ from the first $n$
ones $q_{i}=q_{i}(\lambda )$ is standard.

In the next Sections we apply this theory to dynamic systems on Riemannian
manifolds, constructing all those which can be lifted to the
Gel'fand-Zakharevich bi-Hamiltonian chains.

\section{Introduction to separable Riemannian dynamics}

\subsection{Hamiltonian representation}

Let $(Q,g)$ be a Riemann (pseudo-Riemann) manifold with covariant metric
tensor $g$ and local coordinates $q^{1},...,q^{n}.$ Moreover, let $G:=g^{-1}$
be a contravariant metric tensor satisfying $\sum_{j=1}^{n}g_{ij}G^{jk}=%
\delta _{i}^{k}.$ The Levi-Civita connection components are defined by
\begin{equation}
\Gamma _{jk}^{i}=\frac{1}{2}\sum_{l=1}^{n}G^{il}(\partial
_{k}g_{lj}+\partial _{j}g_{kl}-\partial _{l}g_{jk}),~\ \ \ \ \ \ \ \ \ \ \ \
\ \partial _{i}\equiv \frac{\partial }{\partial q^{i}}.  \tag{3.1}
\label{1.1}
\end{equation}%
The equations
\begin{equation}
q_{tt}^{i}+\Gamma _{jk}^{i}q_{t}^{j}q_{t}^{k}=G^{ik}\partial _{k}V(q),\ \ \
\ \ \ \ \ i=1,...,n,\ \ \ \ \ \ \ q_{t}\equiv \frac{dq}{dt}  \tag{3.2}
\label{1.2}
\end{equation}%
describe the motion of a particle in the Riemannian space with the metric $%
g. $ Eqs.(\ref{1.2}) can be obtained by varying the Lagrangian
\begin{equation}
\mathcal{L}(q,q_{t})=\frac{1}{2}\sum_{i,j}g_{ij}q_{t}^{i}q_{t}^{j}-V(q)
\tag{3.3}  \label{1.3}
\end{equation}%
and are called Euler-Lagrange equations. Obviously, for $G=I$ equations (\ref%
{1.2}) reduce to Newton equations of motion.

One can pass in a standard way to the Hamiltonian description of dynamics,
where the Hamiltonian function takes the form
\begin{equation}
H(q,p)=\sum_{i=1}^{n}q_{t}^{i}\frac{\partial \mathcal{L}}{\partial q_{t}^{i}}%
-\mathcal{L}=\frac{1}{2}\sum_{i,j=1}^{n}G^{ij}p_{i}p_{j}+V(q),\ \ \ \ \
p_{i}:=\frac{\partial \mathcal{L}}{\partial q_{t}^{i}}%
=\sum_{j}g_{ij}q_{t}^{j}  \tag{3.4}  \label{1.7}
\end{equation}%
and equations of motion are
\begin{equation*}
\left(
\begin{array}{c}
q \\
p%
\end{array}%
\right) _{t}=\theta _{0}dH=\left(
\begin{array}{cc}
0 & I \\
-I & 0%
\end{array}%
\right) \left(
\begin{array}{c}
\frac{\partial H}{\partial q} \\
\frac{\partial H}{\partial \partial p}%
\end{array}%
\right) =X_{H}
\end{equation*}%
\begin{equation}
\Updownarrow  \tag{3.5}  \label{1.8}
\end{equation}%
\begin{equation*}
q_{t}^{i}=\frac{\partial H}{\partial p_{i}},\ \ \ \ p_{it}=-\frac{\partial H%
}{\partial q^{i}}.
\end{equation*}%
$X_{H}$ denotes the Hamiltonian vector field and the whole dynamics takes
place on the phase space $\mathcal{N}=T^{\ast }G$ in local coordinates $%
(q^{1},...,q^{n},p_{1},...,p_{n}).$

Of special importance is the geodesic motion $V(q)\equiv 0,$ with
Euler-Lagrange equations
\begin{equation}
q_{tt}^{i}+\Gamma _{jk}^{i}q_{t}^{j}q_{t}^{k}=0,\ \ \ \ i=1,...,n  \tag{3.6}
\label{1.9}
\end{equation}%
and Hamiltonian representation
\begin{equation}
\left(
\begin{array}{c}
q \\
p%
\end{array}%
\right) _{t}=\theta _{0}dE=X_{E},\ \ \ \ \ \ \ \ E=\frac{1}{2}%
\sum_{i,j=1}^{n}G^{ij}p_{i}p_{j}.  \tag{3.7}  \label{1.10}
\end{equation}

\subsection{Separated coordinates}

We are going to present a complete theory of a subclass of one-particle
dynamics, containing Liouville integrable, separable, (quasi-)bi-Hamiltonian
systems with $n$ quadratic in momenta constants of motion. St\"{a}ckel \cite%
{s1}-\cite{s3} gave the first characterization of the Riemann
(pseudo-Riemann) manifold $(Q,g)$ on which the equations of geodesic motion
can be solved by separation of variables. He proved that if in a system of
orthogonal coordinates $(\lambda ,\mu )$ there exists a non-singular matrix $%
\varphi =(\varphi _{k}^{l}(\lambda _{k}))$, called a \emph{St\"{a}ckel matrix%
} such that the geodesic Hamiltonians $E_{r}$ are of the form
\begin{equation}
E_{r}=\frac{1}{2}\sum_{i=1}^{n}(\varphi ^{-1})_{r}^{i}\mu _{i}^{2},
\tag{3.8}  \label{a}
\end{equation}%
then $E_{r}$ are functionally independent, pairwise commute with respect to
the canonical Poisson bracket and the Hamilton-Jacobi equation associated to
$E_{1}$ is separable.

Then, Eisenhart \cite{Ei}-\cite{Ei1} gave a coordinate-free representation
for St\"{a}ckel geodesic motion introducing a special family of \emph{%
Killing tensors}. As known, a $(1,1)$-type tensor $B=(B_{j}^{i})$ (or a $%
(2,0)$-type tensor $B=(B^{ij}))$ is called a Killing tensor with respect to $%
g$ if
\begin{equation*}
\left\{ \,\tsum (BG)^{ij}p_{i}p_{j}\,,\,E\right\} _{\theta _{0}}=0\ \ \
\text{or}\ \ \ \left\{ \,\tsum (B)^{ij}p_{i}p_{j}\,,E\right\} _{\theta
_{0}}=0.
\end{equation*}
He proved \cite{Ei}-\cite{Ei1} that the geodesic Hamiltonians can be
transformed into the St\"{a}ckel form (\ref{a}) if the contravariant metric
tensor $G=g^{-1}$ has $(n-1)$ commuting independent contravariant Killing
tensors $A_{r}$ of a second order such that
\begin{equation}
E_{r}=\frac{1}{2}\sum_{i,j}A_{r}^{ij}p_{i}p_{j},  \tag{3.9}  \label{b}
\end{equation}%
admitting a common system of closed eigenforms $\alpha _{i}$
\begin{equation}
(A_{r}^{\ast }-v_{r}^{i}G)\alpha _{i}=0,\;\;d\alpha _{i}=0,\;\;i=1,...,n,
\tag{3.10}  \label{c}
\end{equation}%
where $v_{r}^{i}$ are eigenvalues of $(1,1)$ Killing tensor $K_{r}=A_{r}g$ $%
(K_{r}^{\ast }=gA_{r}^{\ast }).$

From now on, separated canonical coordinates will be denoted by $(\lambda
,\mu )$ and natural canonical coordinates by $(q,p).$ For $n$ degrees of
freedom, let us consider $n$ St\"{a}ckel Hamiltonian functions in separated
coordinates in the following form
\begin{equation}
H_{r}=\frac{1}{2}\sum_{i=1}^{n}v_{r}^{i}G^{ii}\mu _{i}^{2}+V_{r}(\lambda )=%
\frac{1}{2}\mu ^{T}K_{r}G\mu +V_{r}(\lambda ),\;\;\;r=1,...,n,  \tag{3.11}
\label{6}
\end{equation}%
where $\mu =(\mu _{1},...,\mu _{n})^{T},$and $V_{r}(\lambda )$ are
appropriate potentials separable in $(\lambda ,\mu )$ coordinates. For the
integrable system (\ref{6}) the St\"{a}ckel separation conditions (\ref{0.18}%
) take the general form
\begin{equation}
\sum_{k=1}^{n}\phi _{i}^{k}(\lambda ^{i})H_{k}=\frac{1}{2}f_{i}(\lambda
^{i})\mu _{i}^{2}+\gamma _{i}(\lambda ^{i})\;\;\;i=1,...,n,  \tag{3.12}
\label{13}
\end{equation}%
where $f_{i},\gamma _{i},\phi _{i}^{k}$ are arbitrary smooth functions of
its argument and the normalization $\phi _{i}^{n}=1,\;i=1,...,n$ is assumed.
To get the explicit form of $H_{k}=H_{k}(\lambda ,\mu )$ one has to solve
the system of linear equations (\ref{13}). The results are the following
\begin{equation}
G^{ii}=(-1)^{i+1}\frac{f_{i}(\lambda ^{i})\det W^{i1}}{\det W}%
,\;\;\;v_{r}^{i}=(-1)^{r+1}\frac{\det W^{ir}}{\det W^{i1}},  \tag{3.13}
\label{14}
\end{equation}%
\begin{equation}
V_{r}=\sum_{i=1}^{n}(-1)^{i+r}\gamma _{i}(\lambda ^{i})\frac{\det W^{ir}}{%
\det W},  \tag{3.14}  \label{rr}
\end{equation}%
where
\begin{equation}
W=\left(
\begin{array}{ccccc}
\phi _{1}^{1}(\lambda ^{1}) & \phi _{1}^{2}(\lambda ^{1}) & \cdots & \phi
_{1}^{n-1}(\lambda ^{1}) & 1 \\
\vdots & \vdots & \cdots & \vdots & \vdots \\
\phi _{n}^{1}(\lambda ^{n}) & \phi _{n}^{2}(\lambda ^{n}) & \cdots & \phi
_{n}^{n-1}(\lambda ^{n}) & 1%
\end{array}%
\right)  \tag{3.15}  \label{u}
\end{equation}%
and $W^{ik}$ is the $(n-1)\times (n-1)$ matrix obtained from $W$ after we
cancel its $i$th row and $k$th column. Then the St\"{a}ckel matrix $\varphi $
is given by
\begin{equation}
\varphi =\left(
\begin{array}{ccccc}
\frac{\phi _{1}^{1}(\lambda ^{1})}{f_{1}(\lambda ^{1})} & \frac{\phi
_{1}^{2}(\lambda ^{1})}{f_{1}(\lambda ^{1})} & \cdots & \frac{\phi
_{1}^{n-1}(\lambda ^{1})}{f_{1}(\lambda ^{1})} & \frac{1}{f_{1}(\lambda ^{1})%
} \\
\vdots & \vdots & \cdots & \vdots & \vdots \\
\frac{\phi _{n}^{1}(\lambda ^{n})}{f_{n}(\lambda ^{n})} & \frac{\phi
_{n}^{2}(\lambda ^{n})}{f_{n}(\lambda ^{n})} & \cdots & \frac{\phi
_{n}^{n-1}(\lambda ^{n})}{f_{n}(\lambda ^{n})} & \frac{1}{f_{n}(\lambda ^{n})%
}%
\end{array}%
\right) .  \tag{3.16}  \label{d}
\end{equation}

In our further considerations we restrict to the case when $\phi
_{i}^{k}(\lambda _{i})$ are monomials. Actually, we assume that
\begin{equation}
\phi _{i}^{k}(\lambda _{i})=\phi ^{k}(\lambda _{i})=(\lambda ^{i})^{m_{k}},\
\ \ \ m_{n}=0<m_{n-1}<...<m_{1}\in \mathbb{N},  \tag{3.17}  \label{1.19}
\end{equation}%
which is a sufficient condition for the existence of bi-Hamiltonian
representation of Gel'fand-Zakharevich type with separation conditions and
the separation curve in the form
\begin{equation}
\sum_{j=1}^{m}\sigma _{j}(\lambda ^{i})h^{(j)}(\lambda ^{i})=\frac{1}{2}%
f(\lambda ^{i})\mu _{i}^{2}+\gamma (\lambda ^{i}),\ \ \ \ \ i=1,...,n,
\tag{3.18a}  \label{3.18aa}
\end{equation}%
\begin{equation}
\sum_{j=1}^{m}\sigma _{j}(\xi )h^{(j)}(\xi )=\frac{1}{2}f(\xi )\mu
^{2}+\gamma (\xi ).\ \ \ \ \ \ \ \ \ \ \ \ \ \ \ \ \ \ \ \ \ \ \ \
\tag{3.18b}  \label{3.18bb}
\end{equation}

\subsection{Integration by quadratures}

Let us consider a set of coordinates $\{\lambda ^{i},\mu _{i}\}_{i=1}^{n}$
on $T^{\ast }Q$ canonical with respect to $\theta _{0}$. One can try to
linearize the system (\ref{6}) through a canonical transformation $(\lambda
,\mu )\rightarrow (b,a)$ in the form $b^{j}=\frac{\partial W}{\partial a_{i}}%
,\mu _{i}=\frac{\partial W}{\partial \lambda ^{i}},$ where $W(\lambda ,a)$
is a generating function that solves the related Hamilton-Jacobi (HJ)
equations
\begin{equation}
E_{r}(\lambda ,\frac{\partial W}{\partial \lambda })=a_{r},\;\;\;\;%
\;r=0,...,n.  \tag{3.19}  \label{5}
\end{equation}%
The HJ equations (\ref{5}) are nonlinear partial differential equations that
are very difficult to solve. In general it is a hopeless task. However,
there are rare cases when one can find \ a solution of (\ref{5}) \ in the
separable form
\begin{equation}
W(\lambda ,a)=\sum_{i=1}^{n}W_{i}(\lambda _{i},a)  \tag{3.20}  \label{20}
\end{equation}%
that turns the HJ equations into a set of decupled ordinary differential
equations that can be solved by quadratures. Such $(\lambda ,\mu )$
coordinates are called \emph{separated coordinates. }In the $(b,a)$
coordinates the flow $d/dt_{j}$ associated with every Hamiltonian $H_{j}$ is
trivial
\begin{equation}
\frac{da_{i}}{dt_{j}}=0,\text{ \ }\frac{db^{i}}{dt_{j}}=\delta _{ij}\text{ ,}%
\ \ \ \ \ \text{\ \ }i,j=1,\ldots ,n  \tag{3.21}  \label{8}
\end{equation}%
and the implicit form of the trajectories $\lambda (t_{j})$ is given by
\begin{equation}
b^{i}(\lambda ,a)=\frac{\partial W}{\partial a_{i}}=\delta
_{ij}t_{j}+const,\,\,\,\,i,j=1,...,n.  \tag{3.22}  \label{9}
\end{equation}%
As explained in the previous section, separated coordinates are defined by $%
n $ relations of the form
\begin{equation}
\varphi _{i}(\lambda ^{i},\mu _{i},H_{1},...,H_{n})=0,\ \ \ \ \ \ \ i=1,...,n
\tag{3.23}  \label{2.20}
\end{equation}%
joining each pair $(\lambda ^{i},\mu _{i})$ of conjugate coordinates and all
Hamiltonians $H_{i},\ i=1,...,n.$ Fixing the values of the Hamiltonians $%
H_{j}=const=a_{j}$ one obtains an explicit factorization of the Liouville
tori given by the equations

\bigskip\
\begin{equation*}
\varphi _{i}(\lambda ^{i},\mu _{i},a_{1},...,a_{n})=0,\ \ \ \ \ \ \ i=1,...,n
\end{equation*}%
\begin{equation}
\Updownarrow  \tag{3.24}  \label{2.21}
\end{equation}%
\begin{equation*}
\varphi _{i}(\lambda ^{i},\frac{dW_{i}}{d\lambda ^{i}},a_{1},...,a_{n})=0,\
\ \ \ \ \ \ i=1,...,n
\end{equation*}%
i.e. a decupled system of ordinary differential equations.

Let us now solve explicitly the Hamilton-Jacobi equations (\ref{5}) and
related dynamic equations, with respect to the evolution parameter $t_{r}$,
written in $(\lambda ,\mu )$ coordinates when separation conditions are
given in the form (\ref{13}), (\ref{1.19}):%
\begin{equation*}
\frac{1}{2}f(\lambda ^{i})\left( \frac{\partial W}{\partial \lambda ^{i}}%
\right) ^{2}+\gamma (\lambda ^{i})=a_{1}(\lambda ^{i})^{m_{1}}+a_{2}(\lambda
^{i})^{m_{2}}+...+a_{n}\equiv a(\lambda ^{i})
\end{equation*}%
\begin{equation*}
\Downarrow \ \ W=\sum_{i=1}^{n}W_{i}(\lambda _{i},a)
\end{equation*}%
\begin{equation*}
\frac{1}{2}f(\lambda ^{i})\left( \frac{dW_{i}}{d\lambda ^{i}}\right)
^{2}+\gamma (\lambda ^{i})=a_{1}(\lambda ^{i})^{m_{1}}+a_{2}(\lambda
^{i})^{m_{2}}+...+a_{n}\equiv a(\lambda ^{i})
\end{equation*}%
\begin{equation*}
\Downarrow
\end{equation*}%
\begin{equation*}
W_{i}(\lambda _{i},a)=\int^{\lambda ^{i}}\sqrt{\frac{a(\xi )-\gamma (\xi )}{%
\frac{1}{2}f(\xi )}}d\xi
\end{equation*}%
\begin{equation*}
\Downarrow
\end{equation*}%
\begin{equation*}
W(\lambda ,a)=\sum_{k=1}^{n}\int^{\lambda ^{k}}\sqrt{\frac{a(\xi )-\gamma
(\xi )}{\frac{1}{2}f(\xi )}}d\xi
\end{equation*}%
\begin{equation*}
\Downarrow
\end{equation*}%
\begin{equation*}
b^{i}=\frac{\partial W}{\partial a_{i}}=\sum_{k=1}^{n}\int^{\lambda ^{k}}%
\frac{\xi ^{m_{i}}}{\sqrt{2f(\xi )[a(\xi )-\gamma (\xi )]}}d\xi
\end{equation*}%
\begin{equation*}
\Downarrow \ \ 2f(\xi )[a(\xi )-\gamma (\xi )]\equiv \psi (\xi )
\end{equation*}%
\begin{equation}
\sum_{k=1}^{n}\int^{\lambda ^{k}}\frac{\xi ^{m_{i}}}{\sqrt{\psi (\xi )}}d\xi
=t_{r}\delta _{ri}+const_{i},\ \ \ \ \ \ i=1,...,n,  \tag{3.25}  \label{2.26}
\end{equation}%
where eqs.(\ref{2.26}) are implicit solutions called the\emph{\ inverse
Jacobi problem.}

\section{Systems of Benenti type}

\subsection{Separable geodesics}

Among all St\"{a}ckel systems a particularly important subclass consists of
these considered by Benenti \cite{be}, \cite{ben}, \cite{be1} and
constructed with the help of the so-called\emph{\ conformal Killing tensor. }%
Let $L=(L_{j}^{i})$ be a second order mixed type tensor on $Q$ and let $%
\overline{L}:M\rightarrow \mathbb{R}$ be a function on $M$ defined as $%
\overline{L}=\frac{1}{2}\dsum\limits_{i,j=1}^{n}(LG)^{ij}p_{i}p_{j}$. If
\begin{equation}
\{\overline{L},E\}_{\theta _{0}}=\kappa E,\ \ \ \ \mathrm{\ where}\ \ \ \ \
\ \kappa =\{\varepsilon ,E\}_{\theta _{0}},\ \ \ \ \varepsilon =Tr(L),
\tag{4.1}  \label{2.5}
\end{equation}%
then $L$ is called a conformal Killing tensor with the associated potential $%
\varepsilon =Tr(L)$. If we assume additionally that $L$ has simple
eigenvalues and its Nijenhuis torsion vanishes, then $L$ is called a \emph{%
special conformal Killing tensor }\cite{c}.\emph{\ }

For the Riemannian manifold $(Q,g,L),$ geodesic flow has $n$ constants of
motion of the form
\begin{equation}
E_{r}=\frac{1}{2}\sum_{i,j=1}^{n}A_{r}^{ij}p_{i}p_{j}=\frac{1}{2}%
\sum_{i,j=1}^{n}(K_{r}G)^{ij}p_{i}p_{j},\ \ \ \ \ \ \ \ \ \ r=1,...,n,
\tag{4.2}  \label{3.4}
\end{equation}%
where $A_{r}$ and $K_{r}$ are Killing tensors of type $(2,0)$ and $(1,1)$,
respectively. Moreover, as was shown by Benenti \cite{be},\cite{ben}, all
the Killing tensors $K_{r}$ with a common set of eigenvectors, are
constructed from $L$ by
\begin{equation}
K_{r+1}=\sum_{k=0}^{r}\rho _{k}L^{r-k},  \tag{4.3}  \label{2.7a}
\end{equation}%
where $\rho _{r}$ are coefficients of the characteristic polynomial of $L$%
\begin{equation}
\det (\xi I-L)=\xi ^{n}+\rho _{1}\xi ^{n-1}+...+\rho _{n},\ \ \ \rho _{0}=1.
\tag{4.4}  \label{2.8b}
\end{equation}%
From (\ref{2.7a}) we immediately find that \
\begin{equation}
K_{r+1}-LK_{r}=\rho _{r}I,\ \ \ \ -LK_{n}=\rho _{n}I,\ \ \ \ K_{1}=I.
\tag{4.5}  \label{2.7b}
\end{equation}

\begin{lemma}
From (\ref{2.7a}) and (\ref{2.7b}) it follows that appropriate Killing
tensors $K_{r}$ are given by the following 'cofactor' formula
\begin{equation}
cof(\xi I-L)=\sum_{i=0}^{n-1}K_{n-i}\xi ^{i},  \tag{4.6}  \label{2.7}
\end{equation}%
where $cof(A)$ stands for the matrix of cofactors, so that $cof(A)A=(\det
A)I.$ Notice that $K_{1}=I,$ hence $A_{1}=G$ and $E_{1}\equiv E.$
\end{lemma}

\begin{proof}
\begin{eqnarray*}
(\xi I-L)(\sum_{i=0}^{n-1}K_{n-i}\xi ^{i}) &=&\sum_{i=0}^{n-1}(K_{n-i}\xi
^{i+1}-LK_{n-i}\xi ^{i}) \\
&=&-LK_{n}+(K_{n}-LK_{n-1})\xi +... \\
&&+(K_{2}-LK_{1})\xi ^{n-1}+K_{1}\xi ^{n} \\
&=&I(\rho _{n}+\rho _{n-1}\xi +...+\rho _{1}\xi ^{n-1}+\xi ^{n}) \\
&=&I\det (\xi I-L).
\end{eqnarray*}
\end{proof}

According to the above results, the functions $E_{r},$ satisfy
\begin{equation}
\{E_{s},E_{r}\}_{\theta _{0}}=0,  \tag{4.7}  \label{2.8}
\end{equation}%
and thus constitute a system of $n$ constants of motion in involution with
respect to the Poisson structure $\theta _{0}$. So, for a given metric
tensor $g$, the existence of a special conformal Killing tensor $L$ is a
sufficient condition for the geodesic flow on $Q$ to be a Liouville
integrable Hamiltonian system with all constants of motion quadratic in
momenta.

It turns out that with the tensor $L$ we can (generically) associate a
coordinate system on $Q$ in which the geodesic flows associated with all the
functions $E_{r}$ separate. Namely, let $(\lambda ^{1}(q),...,\lambda
^{n}(q))$ be $n$ distinct, functionally independent eigenvalues of $L$, i.e.
solutions of the characteristic equation \ $\det (\xi I-L)=0$. Solving these
relations with respect to $q$ we get the transformation $\lambda \rightarrow
q$
\begin{equation}
q^{i}=\alpha _{i}(\lambda ),\;\;\;\;\;i=1,...,n.  \tag{4.8}  \label{2.9}
\end{equation}%
The remaining part of the transformation to the separation coordinates can
be obtained as a canonical transformation reconstructed from the generating
function $W(p,\lambda )=\sum_{i}p_{i}\alpha _{i}(\lambda )$ in the following
way
\begin{equation}
\mu _{i}=\frac{\partial W(p,\lambda )}{\partial \lambda ^{i}}\Longrightarrow
p_{i}=\beta _{i}(\lambda ,\mu )\;\;\;\;\;i=1,...,n.  \tag{4.9}  \label{3.8}
\end{equation}%
In the $(\lambda ,\mu )$ coordinates the tensor $L$ is diagonal
\begin{equation}
L=\mathrm{diag}(\lambda ^{1},...,\lambda ^{n})\equiv \Lambda _{n},
\tag{4.10}  \label{2.11}
\end{equation}%
while the geodesic Hamiltonians have the following form \cite{1}
\begin{equation}
E_{r}=-\frac{1}{2}\sum_{i=1}^{n}\frac{\partial \rho _{r}}{\partial \lambda
^{i}}\frac{f_{i}(\lambda ^{i})}{\Delta _{i}}\mu
_{i}^{2},\;\;\;\;\;\;r=1,...,n,  \tag{4.11}  \label{3.9}
\end{equation}%
where
\begin{equation}
\Delta _{i}=\prod\limits_{k=1,...n,\,k\neq i}(\lambda ^{i}-\lambda ^{k}),
\tag{4.12}  \label{2.13}
\end{equation}%
$\rho _{r}(\lambda )$ are symmetric polynomials (Vi\'{e}te polynomials)
defined by the relation
\begin{equation}
\det (\xi I-\Lambda )=(\xi -\lambda ^{1})(\xi -\lambda
^{2})...(\xi -\lambda ^{n})=\sum_{r=0}^{n}\rho _{r}\xi ^{n-r}
\tag{4.13}  \label{3.10}
\end{equation}%
and $f_{i}$ are arbitrary smooth functions of one real argument. From (\ref%
{3.9}) it immediately follows that in $(\lambda ,\mu )$ variables the
contravariant metric tensor $G$ and all the Killing tensors $K_{r}$ are
diagonal
\begin{equation}
G^{ij}=\frac{f_{i}(\lambda ^{i})}{\Delta _{i}}\delta
^{ij},\;\;\;(K_{r})_{j}^{i}=-\frac{\partial \rho _{r}}{\partial \lambda ^{i}}%
\delta _{j}^{i}.  \tag{4.14}  \label{3.11}
\end{equation}

\begin{remark}
When $f_{i}(\lambda ^{i})$ is a polynomial of order $\leq n$ the metric is
flat, if the order of $f$ is equal $n+1$ the metric is of constant curvature.
\end{remark}

\bigskip Moreover, as
\begin{equation}
\Lambda ^{\ast }d\lambda ^{i}=\lambda ^{i}d\lambda ^{i},\ \ \ K_{r}^{\ast
}d\lambda ^{i}=-\frac{\partial \rho _{r}}{\partial \lambda ^{i}}d\lambda
^{i}=v_{r+1}^{i}d\lambda ^{i},  \tag{4.15}  \label{4.15a}
\end{equation}%
then multiplying both sides of eq.(\ref{4.15a}) by $G$ we get
\begin{equation}
(GK_{r}^{\ast }-v_{r}^{i}G)d\lambda _{i}=0\Leftrightarrow (A_{r}^{\ast
}-v_{r}^{i}G)d\lambda _{i}=0,\;\;\;i=1,...,n,  \tag{4.16}  \label{ee}
\end{equation}%
i.e. the tensorial Eisenhart realization (\ref{c}) of the St\"{a}ckel
results.

To ensure that in the $(\lambda ,\mu )$ coordinates the geodesic
Hamiltonians (\ref{3.9}) are separable it is sufficient to observe that in
these coordinates they actually have the St\"{a}ckel form (\ref{a}) with the
related St\"{a}ckel matrix
\begin{equation}
\varphi =\left(
\begin{array}{ccccc}
\frac{(\lambda ^{1})^{n-1}}{f_{1}(\lambda ^{1})} & \frac{(\lambda ^{1})^{n-2}%
}{f_{1}(\lambda ^{1})} & \cdots & \frac{\lambda ^{1}}{f_{1}(\lambda ^{1})} &
\frac{1}{f_{1}(\lambda ^{1})} \\
\vdots & \vdots & \cdots & \vdots & \vdots \\
\frac{(\lambda ^{n})^{n-1}}{f_{n}(\lambda ^{n})} & \frac{(\lambda ^{n})^{n-2}%
}{f_{n}(\lambda ^{n})} & \cdots & \frac{\lambda ^{n}}{f_{n}(\lambda ^{n})} &
\frac{1}{f_{n}(\lambda ^{n})}%
\end{array}%
\right) .  \tag{4.17}  \label{2.23}
\end{equation}%
It means that St\"{a}ckel separation conditions (\ref{13}) take the
particular form
\begin{equation}
E_{1}(\lambda ^{i})^{n-1}+E_{2}(\lambda ^{i})^{n-2}+...+E_{n}=\frac{1}{2}%
f_{i}(\lambda ^{i})\mu _{i}^{2},\ \ \ \ \ \ \ i=1,...,n.  \tag{4.18}
\label{2.24}
\end{equation}%
For $f_{i}(\lambda ^{i})=f(\lambda ^{i})$ eqs. (\ref{2.24}) can be
represented by $n$ different copies $(\xi ,\mu )=(\lambda ^{i},\mu
_{i}),i=1,...,n$ of some curve
\begin{equation}
E_{1}\xi ^{n-1}+E_{2}\xi ^{n-2}+...+E_{n}=\frac{1}{2}f(\xi )\mu ^{2}
\tag{4.19}  \label{2.25}
\end{equation}%
called the separation curve of geodesic motion.

\subsection{(Quasi-)bi-Hamiltonian chains}

The special conformal Killing tensor $L$ can be lifted from $Q$ to a $(1,1)$%
-type tensor on $\mathcal{N}=T^{\ast }Q$ where it takes the form
\begin{equation}
N=\left(
\begin{array}{cc}
L & 0 \\
F & L^{\ast }%
\end{array}%
\right) ,\;\;\;\;F_{j}^{i}=\frac{\partial }{\partial q^{j}}(p^{T}L)_{i}-%
\frac{\partial }{\partial q^{i}}(p^{T}L)_{j},\ \ \ L^{\ast }=L^{T},
\tag{4.20}  \label{2.27}
\end{equation}%
The lifted $(1,1)$ tensor $N$ is Nijehuis torsion free, like the $L$ one,
hence it defines a recursion operator on $\mathcal{N}$.\emph{\ }An important
property of $N$ is that when it acts on the canonical Poisson tensor $\theta
_{0}$ it produces another Poisson tensor
\begin{equation}
\theta _{1}=N\theta _{0}=\left(
\begin{array}{cc}
0 & L \\
-L^{\ast } & F%
\end{array}%
\right) ,  \tag{4.21}  \label{2.28}
\end{equation}%
compatible with the canonical one (actually $\theta _{0}$ is compatible with
$N^{k}\theta _{0}$ for any integer $k$) making $\mathcal{N}$ the $\omega N$
manifold. It is now possible to show that the geodesic Hamiltonians $E_{r}$
satisfy on $\mathcal{N}=T^{\ast }Q$ the set of relations \cite{ib} \emph{\ }
\begin{equation*}
\theta _{1}dE_{r}=\theta _{0}dE_{r+1}-\rho _{r}\theta
_{0}dE_{1},\;\;E_{n+1}=0,\;\;\;r=1,...,n.
\end{equation*}%
\begin{equation}
\Updownarrow  \tag{4.22}  \label{2.30a}
\end{equation}%
\begin{equation*}
N^{\ast }dE_{r}=dE_{r+1}-\rho _{r}dE_{1},\ \ \ \ N^{\ast }=\theta
_{0}^{-1}\theta _{1}=\left(
\begin{array}{cc}
L^{\ast } & -F \\
0 & L%
\end{array}%
\right) ,
\end{equation*}%
which are a particular case of the quasi-bi-Hamiltonian chain (\ref{0.20})
\cite{mor}. Obviously $\{E_{s},E_{r}\}_{\theta _{1}}=0,$ as follows from (%
\ref{2.8}) and (\ref{2.30a}). Notice that relation (\ref{2.7b}) is
immediately reconstructed from (\ref{2.30a}). Indeed, according to
\begin{equation}
\left( \frac{\partial E}{\partial p_{1}},...,\frac{\partial E}{\partial p_{1}%
}\right) ^{T}=KGp,\ \ \ \ \ \ p=(p_{1},...,p_{n})^{T},  \tag{4.23}
\label{2.30b}
\end{equation}%
$d_{p}$ components of (\ref{2.30a}) are
\begin{equation*}
LK_{r}Gp=K_{r+1}Gp-\rho _{r}Gp
\end{equation*}%
\begin{equation}
\Downarrow  \tag{4.24}  \label{2.30c}
\end{equation}%
\begin{equation*}
LK_{r}=K_{r+1}-\rho _{r}I.
\end{equation*}%
On the other hand%
\begin{equation*}
LK_{r}Gp=K_{r+1}Gp-\rho _{r}Gp
\end{equation*}%
\begin{equation}
\ \Downarrow  \tag{4.25}  \label{2.30d}
\end{equation}%
\begin{equation*}
\overline{L}_{r}\equiv \frac{1}{2}p^{T}LK_{r}Gp=\frac{1}{2}p^{T}K_{r+1}Gp-%
\frac{1}{2}\rho _{r}p^{T}Gp\ =E_{r+1}-\rho _{r}E_{1},
\end{equation*}%
so, for $r=1$ we get the relation%
\begin{equation}
\overline{L}\equiv \overline{L}_{1}=E_{2}-\rho _{1}E_{1}  \tag{4.26}
\label{2.30e}
\end{equation}%
and immediate verification that $L$ is a special conformal Killing tensor,
as
\begin{equation}
\{\overline{L},E_{1}\}_{\theta _{0}}=\{E_{1},\rho _{1}\}_{\theta
_{0}}E_{1}=\kappa E_{1},\ \ \ \ \rho _{1}=-TrL.  \tag{4.27}  \label{2.30f}
\end{equation}

As was presented in this section, tensors $L^{\ast }$ and $A_{r}^{\ast }$
have a common set of closed eigenforms (\ref{4.15a}), (\ref{ee}), hence
obviously
\begin{equation}
(G^{(s)}K_{r}^{\ast }-v_{r}^{i}G^{(s)})d\lambda _{i}=0\Leftrightarrow
(A_{r}^{(s)\ast }-v_{r}^{i}G^{(s)})d\lambda _{i}=0,\;\;\;i=1,...,n,
\tag{4.28}  \label{28}
\end{equation}%
where
\begin{equation}
A_{r}^{(s)}=L^{s}A_{r}=K_{r}L^{s}G=K_{r}G^{(s)},\ \ \ \ s=\pm 1,\pm 2,...\ .
\tag{4.29}  \label{29}
\end{equation}%
It means that $A_{r}^{(s)}$ are contravariant Killing tensors of the metric $%
G^{(s)}$ and functions
\begin{equation}
E_{r}^{(s)}=\frac{1}{2}p^{T}A_{r}^{(s)}p,\ \ \ \ r=1,...,n  \tag{4.30}
\label{30}
\end{equation}%
are in involution. Moreover, $L$ is the special conformal Killing tensor for
each $G^{(s)}.$ Indeed, from (\ref{2.30c}) we have
\begin{equation*}
L^{s+1}K_{r}G~=L^{s}K_{r+1}G~-\rho _{r}L^{s}G~
\end{equation*}%
\begin{equation*}
\Downarrow
\end{equation*}%
\begin{equation*}
LK_{r}L^{s}G~=K_{r+1}L^{s}G~-\rho _{r}L^{s}G~
\end{equation*}%
\begin{equation*}
\Downarrow
\end{equation*}%
\begin{equation*}
LK_{r}G^{(s)}=K_{r+1}G^{(s)}-\rho _{r}G^{(s)}
\end{equation*}%
\begin{equation*}
\Downarrow
\end{equation*}%
\begin{equation}
\overline{L}_{r}^{(s)}=E_{r+1}^{(s)}-\rho _{r}E_{1}^{(s)},  \tag{4.31}
\label{31}
\end{equation}%
where
\begin{equation}
\overline{L}_{r}^{(s)}=\frac{1}{2}p^{T}LA_{r}^{(s)}p.  \tag{4.31a}
\label{31a}
\end{equation}%
Hence, for $\overline{L}^{(s)}\equiv \overline{L}_{1}^{(s)},$ we find the
condition (\ref{2.30f})%
\begin{equation}
\{\overline{L}^{(s)},E_{1}^{(s)}\}_{\theta _{0}}=\{E_{1}^{(s)},\rho
_{1}\}_{\theta _{0}}E_{1}^{(s)}=\kappa E_{1}^{(s)}.  \tag{4.32}  \label{32}
\end{equation}

Let us denote by $G^{(0)}$ a basic flat contravariant metric, which in $%
\lambda $ coordinates takes the form
\begin{equation}
\left( G^{(0)}\right) ^{ij}=\frac{1}{\Delta _{i}}\delta ^{ij},  \tag{4.33}
\label{33}
\end{equation}%
i.e. $f_{i}(\lambda ^{i})=1,\ i=1,...,n$ (\ref{3.11}). It means that in the
appropriate separation curve for geodesic motion (\ref{2.25}) $f(\xi )=1.$
Moreover, the metric tensor $G$ for which the separation curve (\ref{2.25})
contains $f(\xi )$ in the form of Laurent polynomial
\begin{equation}
f(\xi )=\sum_{i=m_{1}}^{i=m_{2}}a_{i}\xi ^{i},\ \ \ m_{1},m_{2}\in \mathbb{N}%
,\ \ \ a_{i}=const,  \tag{4.34}  \label{34}
\end{equation}%
is constructed from the basic metric $G^{(0)}$ in the following way
\begin{equation}
G=f(L)G^{(0)}.  \tag{4.35}  \label{35}
\end{equation}

Now we pass to the bi-Hamiltonian representation. On the extended phase
space $\mathcal{M}=T^{\ast }Q\times \mathbb{R}$, the extended geodesic
Hamiltonians
\begin{equation}
e_{r}=E_{r}+c\rho _{r},\;\;\;\;r=1,...,n,\ \ \ \ e_{0}=c,  \tag{4.36}
\label{2.31}
\end{equation}%
where $c$ is an additional coordinate, satisfy the following bi-Hamiltonian
chain\emph{\ }\cite{ib}\emph{\ }
\begin{equation}
\begin{array}{l}
\pi _{0}de_{0}=0 \\
\pi _{0}de_{1}=X_{1}=\pi _{1}de_{0} \\
\pi _{0}de_{2}=X_{2}=\pi _{1}de_{1} \\
\,\,\,\,\,\,\,\,\,\,\,\,\,\,\,\,\,\,\,\,\,\,\,\,\,\,\,\,\,\,\,\,\vdots \\
\pi _{0}de_{n}=X_{n}=\pi _{1}de_{n-1} \\
\qquad \qquad \,\,\,\,0=\pi _{1}de_{n}\,\,%
\end{array}
\tag{4.37}  \label{2.32}
\end{equation}%
with the Poisson operators $\pi _{0}$ and $\pi _{1}$

\begin{equation}
\pi _{0}=\left(
\begin{array}{c|c}
\theta _{0} & 0 \\ \hline
0 & 0%
\end{array}%
\right) \text{ \ , \ }\pi _{1}=\left(
\begin{array}{c|c}
\theta _{1} & \theta _{0}de_{1} \\ \hline
-(\theta _{0}de_{1})^{T} & 0%
\end{array}%
\right) .  \tag{4.38}  \label{2.33}
\end{equation}%
Both Poisson tensors $\pi _{0}$ and $\pi _{1}$ are compatible and
degenerated. The Casimir of $\pi _{0}$ is $e_{0}$ and the Casimir of $\pi
_{1}$ is $e_{n}$ and they start and terminate the bi-Hamiltonian chain (\ref%
{2.32}). Of course $\{e_{s},e_{r}\}_{\pi _{0}}=\{e_{s},e_{r}\}_{\pi _{1}}=0.$
The projections of $\pi _{0},\pi _{1}$ onto a symplectic leaf of $\pi _{0}$ (%
$c=const$) along $Z=\frac{\partial }{\partial c}$ reconstructs our
nondegenerate Poisson tensors $\theta _{0},\theta _{1}.$ If we introduce the
Poisson pencil $\pi _{\xi }=\pi _{1}-\xi \pi _{0},$ the chain (\ref{2.32})
can be written in a compact form
\begin{equation}
\pi _{\xi }de(\xi )=0,\;\;\;e(\xi )=\sum_{r=0}^{n}e_{n-r}\xi ^{r},
\tag{4.39}  \label{2.34}
\end{equation}%
where $e_{\xi }$ is a Casimir of the Poisson pencil $\pi _{\xi },$ depending
polynomial on $\xi ,$ i.e. we deal with a Gel'fand-Zakharevich type system.
The projection of (\ref{2.32}) onto symplectic leaf of $\pi _{0}:$ $c=0$
reconstructs the quasi-bi-Hamiltonian chain (\ref{2.30a}) on $\mathcal{N}$
in a compact form
\begin{equation*}
\theta _{\xi }d(E(\xi ))+\rho (\xi )\theta _{0}d(E_{1})=0\;
\end{equation*}%
\begin{equation}
\Updownarrow  \tag{4.40}  \label{2.35a}
\end{equation}%
\begin{equation*}
(N^{\ast }-\xi I)d(E(\xi ))+\rho (\xi )d(E_{1})=0,
\end{equation*}%
where
\begin{equation}
E(\xi )=\sum_{r=0}^{n-1}E_{n-r}\xi ^{r},\;\rho (\xi )=\det (\xi
I-L)=\sum_{r=0}^{n}\rho _{n-r}(q)\xi ^{r}.  \tag{4.40a}  \label{40a}
\end{equation}%
If we start with a bi-Hamiltonian chain (\ref{2.32}) or a
quasi-bi-Hamiltonian chain (\ref{2.30a}) written in the 'physical'
(original) coordinates $(q,p)$ then we can usually find the functions $%
q(\lambda )$ that constitute the first half of the transformation $(\lambda
,\mu )\rightarrow (q,p)$ simply by taking the functions $\rho _{r}(q)$ from
the Hamiltonians (\ref{2.31}) or directly from (\ref{2.8b}) and solve the
system of equations $\rho _{r}(q)=\rho _{r}(\lambda )$ , $r=1,\ldots ,n$
with respect to $q$. The remaining functions $p(\lambda ,\mu )$ are
calculated according to (\ref{3.8}).

\subsection{Separable potentials}

What potentials can be added to geodesic Hamiltonians $E_{r}$ without
destroying their separability within the above schema? It turns out that
there exists a sequence of basic separable potentials $V_{r}^{(k)}$,\ $k=\pm
1,\pm 2,...,$ which can be added to geodesic Hamiltonians $E_{r}$ such that
the new Hamiltonians
\begin{equation}
H_{r}(q,p)=E_{r}(q,p)+V_{r}^{(k)}(q),\;\;\;\;r=1,...,n,  \tag{4.41}
\label{2.36}
\end{equation}%
are in involution with respect to $\theta _{0}$ and $\theta _{1}$ and are
still separable in the same coordinates $(\lambda ,\mu )$. It means that $%
H_{r}$ follow the quasi-bi-Hamiltonian chain (\ref{2.30a})
\begin{equation}
N^{\ast }dH_{r}=dH_{r+1}-\rho _{r}dH_{1},\ \ \ \ H_{n+1}=0,\ \ \ \ r=1,...,n,
\tag{4.42}  \label{2.36a}
\end{equation}%
while for potentials we have
\begin{equation}
L^{\ast }dV_{r}=dV_{r+1}-\rho _{r}dV_{1},\ \ \ \ r=1,...,n,  \tag{4.43}
\label{2.36b}
\end{equation}%
\begin{equation*}
\Updownarrow
\end{equation*}%
\begin{equation*}
dV_{r+1}=\sum_{k=0}^{r}\rho _{k}(L^{\ast })^{r-k}dV_{1}
\end{equation*}%
\begin{equation*}
\Updownarrow (\ref{2.7a})
\end{equation*}%
\begin{equation}
dV_{r+1}=K_{r+1}^{\ast }dV_{1}\Longrightarrow d(K_{r+1}^{\ast }dV_{1})=0.
\tag{4.44}  \label{3.26c}
\end{equation}%
The relations (\ref{3.26c}) were derived for the first time by Benenti \cite%
{be}, \cite{ben}.

\begin{theorem}
The basic separable potentials $V_{r}^{(m)}$ are given by the following
recursion relation \cite{5}, \cite{8}%
\begin{equation}
V_{r}^{(m+1)}=V_{r+1}^{(m)}-\rho _{r}V_{1}^{(m)},  \tag{4.45}  \label{2.37}
\end{equation}%
and its inverse
\begin{equation}
V_{r}^{(-m-1)}=V_{r-1}^{(-m)}-\frac{\rho _{r-1}}{\rho _{n}}V_{n}^{(-m)},\ \
\ \ \ V_{r}^{(0)}=-\delta _{r,n}.  \tag{4.46}  \label{2.38}
\end{equation}
\end{theorem}

\begin{proof}
The proof is inductive. We show it for positive potentials. Assuming that
potentials $V_{r}^{(m)}$ fulfil condition (\ref{2.36b}), we prove that
potentials $V_{r}^{(m+1)}$ fulfil the same condition. The condition (\ref%
{2.36b}) is true for the first nontrivial potentials $V_{r}^{(n)}=\rho _{r}$%
, which are coefficients of characteristic polynomials of a special
conformal Killing tensor $L$ \cite{ib}%
\begin{equation}
d\rho _{r+1}=L^{\ast }d\rho _{r}+\rho _{r}d\rho _{1},\ \ \ \ r=1,...,n.
\tag{4.47}  \label{2.38a}
\end{equation}%
Notice that (\ref{2.38a}) is a special case of the condition (\ref{0.22})
and follows from the St\"{a}ckel separability. Then we have
\begin{eqnarray*}
&&L^{\ast }dV_{r}^{(m+1)}+\rho _{r}dV_{1}^{(m+1)} \\
&=&L^{\ast }d(V_{r+1}^{(m)}-\rho _{r}V_{1}^{(m)})-\rho
_{r}d(V_{2}^{(m)}-\rho _{1}V_{1}^{(m)}) \\
&=&L^{\ast }dV_{r+1}^{(m)}-\rho _{r}L^{\ast
}dV_{1}^{(m)}-V_{1}^{(m)}(L^{\ast }d\rho _{r}+\rho _{r}d\rho _{1}) \\
&&+\rho _{r}(dV_{2}^{(m)}-\rho _{1}dV_{1}^{(m)}) \\
&=&L^{\ast }dV_{r+1}^{(m)}-\rho _{r}L^{\ast }dV_{1}^{(m)}-V_{1}^{(m)}d\rho
_{r+1}+\rho _{r}L^{\ast }dV_{1}^{(m)} \\
&=&L^{\ast }dV_{r+1}^{(m)}-V_{1}^{(m)}d\rho _{r+1}=dV_{r+2}^{(m)}-\rho
_{r+1}dV_{1}^{(m)}-V_{1}^{(m)}d\rho _{r+1} \\
&=&d(V_{r+2}^{(m)}-\rho _{r+1}V_{1}^{(m)})=dV_{r+1}^{(m+1)}.
\end{eqnarray*}
\end{proof}

Notice that $V_{r}^{(m)}=-\delta _{r,n-m},m=0,...,n-1,$ $V_{r}^{(n)}=\rho
_{r},V_{r}^{(-1)}=\frac{\rho _{r-1}}{\rho _{n}}$. Such a notation will be
useful in the case of deformed Benenti systems.

Now, again the extended Hamiltonians $h_{r}:\mathcal{N}\times \mathbb{R}%
\rightarrow \mathbb{R}$
\begin{equation}
h_{r}=H_{r}+c\rho _{r}  \tag{4.48}  \label{2.40}
\end{equation}%
satisfy the bi-Hamiltonian chain (compare with (\ref{2.32})) \
\begin{equation}
\begin{array}{l}
\pi _{0}dh_{0}=0 \\
\pi _{0}dh_{1}=X_{1}=\pi _{1}dh_{0} \\
\pi _{0}dh_{2}=X_{2}=\pi _{1}dh_{1} \\
\,\,\,\,\,\,\,\,\,\,\,\,\,\,\,\,\,\,\,\,\,\,\,\,\,\,\,\,\,\,\,\,\vdots \\
\pi _{0}dh_{n}=X_{n}=\pi _{1}dh_{n-1} \\
\qquad \qquad \,\,\,\,\,0=\pi _{1}dh_{n}\,\,%
\end{array}
\tag{4.49}  \label{2.41}
\end{equation}%
with $\pi _{0}$ as in (\ref{2.33}),
\begin{equation}
\pi _{1}=\left(
\begin{tabular}{cc}
$\theta _{1}$ & $\theta _{0}dh_{1}$ \\
$-(\theta _{0}dh_{1})^{T}$ & $0$%
\end{tabular}%
\right)  \tag{4.50}  \label{2.42}
\end{equation}%
and $\{h_{s},h_{r}\}_{\pi _{0}}=\{h_{s},h_{r}\}_{\pi _{1}}=0.$ If we use the
following notation
\begin{equation*}
H(\xi )=E(\xi )+V(\xi ),\;\;\;\;\;\;V(\xi )=\sum_{j=0}^{n-1}V_{n-j}(q)\xi
^{j},
\end{equation*}%
then the recursion formulas (\ref{2.37}) and (\ref{2.38}) can be written in
the compact form \cite{lu}
\begin{equation}
V(\xi )^{(k+1)}=\lambda V(\xi )^{(k)}-\det (\xi I-L)V_{1}^{(k)}  \tag{4.51}
\label{2.43}
\end{equation}%
and
\begin{equation}
V(\xi )^{(-k-1)}=\frac{1}{\xi }\left( V(\xi )^{(-k)}-\frac{\det (\xi I-L)}{%
\det L}V_{n}^{(-k)}\right)  \tag{4.52}  \label{2.44}
\end{equation}%
and our bi-Hamiltonian chain (\ref{2.41}) is given by
\begin{equation}
\pi _{\xi }dh(\xi )=0,\;\;\;h(\xi )=H(\xi )+c\rho (\xi ),  \tag{4.53}
\label{2.45}
\end{equation}%
while the corresponding quasi-bi-Hamiltonian (\ref{2.36a}) chain takes the
form
\begin{equation*}
\theta _{\xi }dH(\xi )+\rho (\xi )\theta _{0}dH_{1}(q,p)=0,\ \ \rho (\xi )=%
\frac{\partial h(\xi )}{\partial c}
\end{equation*}%
\begin{equation}
\Updownarrow  \tag{4.54}  \label{2.46a}
\end{equation}%
\begin{equation*}
(N^{\ast }-\xi I)dH(\xi )+\rho (\xi )dH_{1}(q,p)=0.
\end{equation*}%
In $(\lambda ,\mu )$ coordinates the full (i.e. with a non-zero potential
part) Hamiltonians (\ref{2.40}) of bi-Hamiltonian chain (\ref{2.41}) attain
the form
\begin{equation}
h_{r}(\lambda ,\mu ,c)=-\sum_{i=1}^{n}\frac{\partial \rho _{r}}{\partial
\lambda ^{i}}\frac{\frac{1}{2}f_{i}(\lambda ^{i})\mu _{i}^{2}+\gamma
_{i}(\lambda ^{i})}{\Delta _{i}}+c\rho _{r}(\lambda ),\;\;r=1,...,n.
\tag{4.55}  \label{2.47}
\end{equation}

\begin{lemma}
Nontrivial potentials $V_{r}^{(n-1+k)}$ and $V_{r}^{(-k)}$ $k=1,2,...$ enter
the separation curve
\begin{equation}
H_{1}\xi ^{n-1}+H_{2}\xi ^{n-2}+...+H_{n}=\frac{1}{2}f(\xi )\mu ^{2}+\gamma
(\xi )  \tag{4.56}  \label{2.50b}
\end{equation}%
as $\gamma (\xi )=-\xi ^{n-1+k},-\xi ^{-k}$, hence $\gamma _{i}(\lambda
^{i})=-(\lambda ^{i})^{n-1+k}$ and $\gamma _{i}(\lambda ^{i})=-(\lambda
^{i})^{-k},$ respectively.
\end{lemma}

\begin{proof}
Potentials $V_{r}^{(n)}=\rho _{r}$ are coefficients of characteristic
equation of the special conformal Killing tensor $L$%
\begin{equation}
\xi ^{n}+\rho _{1}\xi ^{n-1}+...+\rho _{n}=0.  \tag{4.57}  \label{2.50}
\end{equation}%
Then, we define $V^{(n+k)}$ potentials by a generating equation
\begin{equation}
\xi ^{n+k}+V_{1}^{(n+k)}\xi ^{n-1}+...+V_{n}^{(n+k)}=0.  \tag{4.58}
\label{2.61}
\end{equation}%
The recursion formula (\ref{2.37}) is reconstructed as follows. From (\ref%
{2.61}) we have
\begin{equation}
\xi ^{n+k+1}+V_{1}^{(n+k)}\xi ^{n}+...+V_{n}^{(n+k)}\xi =0.  \tag{4.59}
\label{2.62}
\end{equation}%
Elimination of $\xi ^{n}$ via (\ref{2.50}) leads to the form
\begin{equation*}
\xi ^{n+k+1}+(V_{2}^{(n+k)}-\rho _{1}V_{1}^{(n+k)})\xi
^{n-1}+...+(V_{n}^{(n+k)}-\rho _{n-1}V_{1}^{(n+k)})\xi -\rho
_{n}V_{1}^{(n+k)}=0
\end{equation*}%
\begin{equation}
\Downarrow  \tag{4.60}  \label{2.63}
\end{equation}%
\begin{equation*}
V_{r}^{(n+k+1)}=V_{r+1}^{(n+k)}-\rho _{r}V_{1}^{(n+k)}.
\end{equation*}%
For the inverse potentials we have from (\ref{2.50})
\begin{equation}
\frac{1}{\rho _{n}}\xi ^{n-1}+\frac{\rho _{1}}{\rho _{n}}\xi ^{n-2}+...+%
\frac{\rho _{n-1}}{\rho _{n}}+\xi ^{-1}=0,  \tag{4.61}  \label{2.65}
\end{equation}%
so $V_{r}^{(-1)}=\frac{\rho _{r-1}}{\rho _{n}}.$ Then, we define $V^{(-k)}$
potentials by a generating equation
\begin{equation}
V_{1}^{(-k)}\xi ^{n-1}+...+V_{n}^{(-k)}+\xi ^{-k}=0.  \tag{4.62}
\label{2.66}
\end{equation}%
The recursion formula (\ref{2.38}) is reconstructed as follows. From (\ref%
{2.66}) we have%
\begin{equation}
V_{1}^{(-k)}\xi ^{n-2}+...+V_{n}^{(-k)}\xi ^{-1}+\xi ^{-k-1}=0.  \tag{4.63}
\label{2.67}
\end{equation}%
Substituting (\ref{2.65}) to eliminate $\xi ^{-1}$ we get
\begin{equation*}
V_{1}^{(-k)}\xi ^{n-2}+...+V_{n}^{(-k)}(-\frac{1}{\rho _{n}}\xi ^{n-1}-...-%
\frac{\rho _{n-1}}{\rho _{n}})+\xi ^{-k-1}=0
\end{equation*}%
\begin{equation}
\Downarrow  \tag{4.64}  \label{2.68}
\end{equation}%
\begin{equation*}
V_{r}^{(-k-1)}=V_{r-1}^{(-k)}-\frac{\rho _{r-1}}{\rho _{n}}V_{n}^{(-k)}.
\end{equation*}
\end{proof}

\section{One-hole deformation of Benenti systems}

Separable systems on Riemannian manifolds considered by Benenti belong to
important but very particular subclass of such systems. In this context, a
question about classification of all separable systems on Riemannian
manifolds, with $n$ quadratic in momenta constants of motion, arises. The
classification can be made with respect to the admissible forms of St\"{a}%
ckel separability conditions. The right hand side of the conditions (\ref{13}%
) is always the same for the class of systems considered
\begin{equation}
r.h.s.=\frac{1}{2}f_{i}(\lambda ^{i})\mu _{i}^{2}+\gamma _{i}(\lambda
^{i})=\psi (\lambda ^{i},\mu _{i}),  \tag{5.1}  \label{3.1}
\end{equation}%
so different classes of separable systems are described by different forms
of the l.h.s. of St\"{a}ckel conditions. In the simplest Benenti case, it is
given by the following polynomial form
\begin{equation}
l.h.s.=H_{1}\xi ^{n-1}+H_{2}\xi ^{n-2}+...+H_{n},\ \ \ \ \xi =\lambda
^{1},...,\lambda ^{n}.  \tag{5.2}  \label{3.2}
\end{equation}%
We will show, that all other classes, given by some polynomial in $\lambda ,$
are appropriate deformations of the Benenti class.

First, let us define a 1-hole deformation of the Benenti class. Consider the
following separability condition
\begin{equation}
\widetilde{H}_{1}\xi ^{(n+1)-1}+\widetilde{H}_{2}\xi ^{(n+1)-2}+...+%
\widetilde{H}_{n+1}=\psi (\xi ,\mu ),\ \ \ \widetilde{H}_{n_{1}}=0  \tag{5.3}
\label{3.3}
\end{equation}%
and the Benenti separability condition with the same $\psi $ representation

\begin{equation}
H_{1}\xi ^{n-1}+H_{2}\xi ^{n-2}+...+H_{n}=\psi (\xi ,\mu ).  \tag{5.4}
\label{3.4a}
\end{equation}%
Notice that all Benenti systems are classified by different forms of $\psi ,$
i.e. by $f_{i}(\lambda ^{i})$ and $\gamma _{i}(\lambda ^{i}).$ A missing
monomial (a hole) in (\ref{3.3}) is $\widetilde{H}_{n_{1}}\xi
^{(n+1)-n_{1}}. $ Using the characteristic equation of a conformal Killing
tensor $L$ of the Benenti system (\ref{3.4a})%
\begin{equation*}
\xi ^{n}+\rho _{1}\xi ^{n-1}+...+\rho _{n}=0,
\end{equation*}%
for the elimination of $\xi ^{n},$ equation (\ref{3.3}) can be transformed
to the form
\begin{equation}
(\widetilde{H}_{2}-\rho _{1}\widetilde{H}_{1})\xi ^{n-1}+...+(\widetilde{H}%
_{n+1}-\rho _{n}\widetilde{H}_{1})=\psi (\xi ,\mu )  \tag{5.5}  \label{3.5}
\end{equation}%
and hence, comparing it with (\ref{3.4a}) we find
\begin{equation}
H_{r}=\widetilde{H}_{r+1}-\rho _{r}\widetilde{H}_{1},\ \ \ \ r=1,...,n,
\tag{5.6}  \label{3.6}
\end{equation}%
with the inverse
\begin{equation}
\widetilde{H}_{r+1}=H_{r}-\frac{\rho _{r}}{\rho _{n_{1}-1}}H_{n_{1}-1},\ \ \
r=0,...,n,  \tag{5.7}  \label{3.7}
\end{equation}%
where $H_{0}=0$, $\rho _{0}=1.$

Notice, that formula (\ref{3.7}) applies separately to the geodesic and the
potential parts, i.e.%
\begin{equation}
\widetilde{E}_{r+1}=E_{r}-\frac{\rho _{r}}{\rho _{n_{1}-1}}E_{n_{1}-1},
\tag{5.8a}  \label{3.8a}
\end{equation}%
\begin{equation}
\widetilde{V}_{r+1}=V_{r}-\frac{\rho _{r}}{\rho _{n_{1}-1}}V_{n_{1}-1},\ \ \
r=0,...,n.  \tag{5.8b}  \label{3.8bb}
\end{equation}

\subsection{Geodesic part}

Let us first look onto geodesic Hamiltonians
\begin{equation}
\widetilde{E}_{r}=\frac{1}{2}p^{T}(\rho _{n_{1}-1}K_{r-1}-\rho
_{r-1}K_{n_{1}-1})\frac{1}{\rho _{n_{1}-1}}Gp,\ r=1,...,n+1.  \tag{5.9}
\label{3.9a}
\end{equation}%
Using the known relation for the Benenti chain
\begin{equation*}
\rho _{r}I=K_{r+1}-LK_{r}
\end{equation*}%
we get~
\begin{equation}
\begin{array}{c}
\widetilde{E}_{1}=-\frac{1}{\rho _{n_{1}-1}}E_{n_{1}-1}=-\frac{1}{\rho
_{n_{1}-1}}\frac{1}{2}p^{T}K_{n_{1}-1}G\ p \\
\\
=\frac{1}{2}p^{T}\widetilde{G}p\Longrightarrow \widetilde{G}=-\frac{1}{\rho
_{n_{1}-1}}K_{n_{1}-1}G%
\end{array}
\tag{5.10}  \label{3.10b}
\end{equation}%
and%
\begin{equation}
\begin{array}{c}
\widetilde{E}_{r}=\frac{1}{2}p^{T}[\frac{1}{\rho _{n_{1}-1}}%
(K_{n_{1}}K_{r-1}-K_{n_{1}-1}K_{r})G]p \\
\\
\ \ \ \ \ \ \ \ \ \ \ \ \ \ \ \ \ \ \ =\frac{1}{2}p^{T}\widetilde{K}_{r}%
\widetilde{G}\ p\Longrightarrow \widetilde{K}%
_{r}=K_{r}-K_{r-1}K_{n_{1}}(K_{n_{1}-1})^{-1}.%
\end{array}
\tag{5.11}  \label{3.11b}
\end{equation}%
Although we know from the construction that $\widetilde{E}_{r}$ are in
involution, as they are St\"{a}ckel geodesics, but here we prove it in a
coordinate free way.

\begin{lemma}
$\widetilde{K}_{r}$ are Killing tensors of the metric $\widetilde{G}$ which
share the common set of eigenfunctions, i.e. $\widetilde{E}_{r}$ are in
involution.
\end{lemma}

\begin{proof}
From involutivity of Benenti Hamiltonians we have
\begin{eqnarray}
0 &=&\{H_{r},H_{s}\}_{\theta _{0}}=\{E_{r}+V_{r},E_{s}+V_{s}\}_{\theta
_{0}}=\{E_{r},V_{s}\}_{\theta _{0}}+\{V_{r},E_{s}\}_{\theta _{0}}  \notag \\
&&  \TCItag{5.12}  \label{3.12b} \\
&\Longrightarrow &\{E_{r},V_{s}\}_{\theta _{0}}=\{E_{s},V_{r}\}_{\theta
_{0}}\Longrightarrow \{E_{r},\rho _{s}\}_{\theta _{0}}=\{E_{s},\rho
_{r}\}_{\theta _{0}}.  \notag
\end{eqnarray}%
Then,
\begin{eqnarray*}
\{\widetilde{E}_{r+1},\widetilde{E}_{s+1}\}_{\theta _{0}} &=&\{E_{r},-\frac{%
\rho _{s}}{\rho _{n_{1}-1}}E_{n_{1}-1}\}_{\theta _{0}}+\{-\frac{\rho _{r}}{%
\rho _{n_{1}-1}}E_{n_{1}-1},E_{s}\}_{\theta _{0}} \\
&&+\{\frac{\rho _{r}}{\rho _{n_{1}-1}}E_{n_{1}-1},\frac{\rho _{s}}{\rho
_{n_{1}-1}}E_{n_{1}-1}\}_{\theta _{0}} \\
&=&\{E_{r},\rho _{n_{1}-1}\}_{\theta _{0}}\frac{\rho _{s}}{\rho
_{n_{1}-1}^{2}}E_{n_{1}-1}+\{\rho _{n_{1}-1},E_{s}\}_{\theta _{0}}\frac{\rho
_{r}}{\rho _{n_{1}-1}}E_{n_{1}-1} \\
&&+\{\rho _{r},E_{n_{1}-1}\}_{\theta _{0}}\frac{\rho _{s}}{\rho
_{n_{1}-1}^{2}}E_{n_{1}-1}+\{E_{n_{1}-1},\rho _{s}\}_{\theta _{0}}\frac{\rho
_{r}}{\rho _{n_{1}-1}}E_{n_{1}-1} \\
&=&0.
\end{eqnarray*}
\end{proof}

\subsection{Deformed potentials}

Let us analyze the basic deformed potentials. The first potentials are the
following. From (\ref{3.8bb}) and the Benenti potentials we have $\widetilde{%
V}_{r}^{(m)}=-\delta _{r-1,n-m}$ for $m<n+1,\ m\neq (n+1)-n_{1}$ and
\begin{equation}
\widetilde{V}_{r}^{(n+1)-n_{1}}=\frac{\rho _{r-1}}{\rho _{n_{1}-1}},\ \ \
\widetilde{V}_{r}^{(n+1)}=\rho _{r}-\frac{\rho _{r-1}\rho _{n_{1}}}{\rho
_{n_{1}-1}},...\ \ .  \tag{5.13}  \label{3.13a}
\end{equation}%
Notice that $\widetilde{V}_{n_{1}}^{(m)}=0$ for $m\geq n+1$ and $\widetilde{V%
}_{n_{1}}^{(n+1)-n_{1}}=1.$

\begin{lemma}
Nontrivial basic potentials $\widetilde{V}_{r}^{(n+1-n_{1})},\widetilde{V}%
_{r}^{(n+k)}$ and $\widetilde{V}_{r}^{(-k)}$ $k=1,2,...$ enter the
separation curve
\begin{equation}
\widetilde{H}_{1}\xi ^{n}+\widetilde{H}_{2}\xi ^{n-1}+...+\widetilde{H}%
_{n+1}=\frac{1}{2}f(\xi )\mu ^{2}+\gamma (\xi ),\ \ \widetilde{H}_{n_{1}}=0,
\tag{5.14}  \label{3.14c}
\end{equation}%
as $\gamma (\xi )=-\xi ^{(n+1)-n_{1}},-\xi ^{n+k},-\xi ^{-k}.$
\end{lemma}

\begin{proof}
We will show the following generating equations for the potentials
considered
\begin{equation}
\xi ^{n+k}+\widetilde{V}_{1}^{(n+k)}\xi ^{n}+..+0\xi ^{n+1-n_{1}}+...+%
\widetilde{V}_{n+1}^{(n+k)}=0,\ \ \ k=1,2,...,  \tag{5.15a}  \label{3.15c}
\end{equation}%
\begin{equation}
\xi ^{-k}+\widetilde{V}_{1}^{(-k)}\xi ^{n}+...+0\xi ^{n+1-n_{1}}+...+%
\widetilde{V}_{n+1}^{(-k)}=0,\ \ \ k=1,2,...,  \tag{5.15b}  \label{3.15d}
\end{equation}%
\begin{equation}
\widetilde{V}_{1}^{(n+1-n_{1})}\xi ^{n}+...+\xi ^{n+1-n_{1}}+...+\widetilde{V%
}_{n+1}^{(n+1-n_{1})}=0.  \tag{5.15c}  \label{3.15e}
\end{equation}%
For the first two equations we have ($m>n$ or $m<0$)
\begin{equation*}
\xi ^{m}+\widetilde{V}_{1}^{(m)}\xi ^{n}+...+\widetilde{V}_{n+1}^{(m)}=0,\ \
\ \widetilde{V}_{n_{1}}^{(m)}=0
\end{equation*}%
\begin{equation*}
\Updownarrow
\end{equation*}%
\begin{equation*}
\xi ^{m}+\widetilde{V}_{1}^{(m)}(-\rho _{1}\xi ^{n-1}-...-\rho _{n})+...+%
\widetilde{V}_{n+1}^{(m)}=0
\end{equation*}%
\begin{equation*}
\Updownarrow
\end{equation*}%
\begin{equation*}
\xi ^{m}+(\widetilde{V}_{2}^{(m)}-\rho _{1}\widetilde{V}_{1}^{(m)})\xi
^{n-1}+...+(\widetilde{V}_{n+1}^{(m)}-\rho _{n}\widetilde{V}_{1}^{(m)})=0
\end{equation*}%
\begin{equation*}
\Updownarrow
\end{equation*}%
\begin{equation*}
\xi ^{m}+V_{1}^{(m)}\xi ^{n-1}+...+V_{n}^{(m)}=0
\end{equation*}%
which reveal the known deformation relations (\ref{3.8bb})
\begin{equation}
V_{r}^{(m)}=\widetilde{V}_{r+1}^{(m)}-\rho _{r}\widetilde{V}%
_{1}^{(m)}\Longleftrightarrow \widetilde{V}_{r+1}^{(m)}=V_{r}^{(m)}-\frac{%
\rho _{r}}{\rho _{n_{1}-1}}V_{n_{1}-1}^{(m)}  \tag{5.16}  \label{3.16c}
\end{equation}%
between nontrivial basic potentials from Benenti class and respective
deformed potentials. For the last case (\ref{3.15e}) we have
\begin{equation*}
\widetilde{V}_{1}^{(n+1-n_{1})}\xi ^{n}+...+\xi ^{n+1-n_{1}}+...+\widetilde{V%
}_{n+1}^{(n+1-n_{1})}=0,\
\end{equation*}%
\begin{equation*}
\Updownarrow
\end{equation*}%
\begin{equation*}
\xi ^{n}+\frac{\widetilde{V}_{2}^{(n+1-n_{1})}}{\widetilde{V}%
_{1}^{(n+1-n_{1})}}\xi ^{n-1}+...+\frac{1}{\widetilde{V}_{1}^{(n+1-n_{1})}}%
\xi ^{n+1-n_{1}}+...+\frac{\widetilde{V}_{n+1}^{(n+1-n_{1})}}{\widetilde{V}%
_{1}^{(n+1-n_{1})}}=0
\end{equation*}%
\begin{equation*}
\Updownarrow
\end{equation*}%
\begin{equation*}
\xi ^{n}+\rho _{1}\xi ^{n-1}+...+\rho _{n}=0
\end{equation*}%
\begin{equation*}
\Downarrow
\end{equation*}%
\begin{equation*}
\widetilde{V}_{r}^{(n+1)-n_{1}}=\frac{\rho _{r-1}}{\rho _{n_{1}-1}}
\end{equation*}%
which is a special case of deformations (\ref{3.16c}) related to the trivial
Benenti potential $V_{r}^{(n+1-n_{1})}=-\delta _{r,n_{1}-1}.$
\end{proof}

Alternatively, the basic potentials $\widetilde{V}^{(m)},m>n+1$ can be
constructed recursively as in the Benenti case.

\begin{lemma}
The basic separable potentials $\widetilde{V}_{r}^{(m)},m>n+1$ are given by
the following recursion relation
\begin{equation}
\widetilde{V}_{r}^{(m+1)}=\widetilde{V}_{r+1}^{(m)}-\widetilde{V}_{r}^{(n+1)}%
\widetilde{V}_{1}^{(m)}-\widetilde{V}_{r}^{(n+1-n_{1})}\widetilde{V}%
_{n_{1}+1}^{(m)},  \tag{5.21}  \label{3.21cc}
\end{equation}%
where $\widetilde{V}_{r}^{(n+1)-n_{1}}$ and $\widetilde{V}_{r}^{(n+1)}$ are
given by (\ref{3.13a}).
\end{lemma}

\begin{proof}
The potentials $\widetilde{V}_{r}^{(n+1)-n_{1}}$ enter the separation curve
in the form (\ref{3.15e}), while the potentials $\widetilde{V}_{r}^{(n+1)},%
\widetilde{V}_{r}^{(m)},\widetilde{V}_{r}^{(m+1)}$ enter the separation
curve in the form (\ref{3.15c}) with $k=1,m-n,m+1-n.$ The recursion formula (%
\ref{3.21cc}) is reconstructed as follows. Multiplying equation (\ref{3.15c}%
) for $k=m-n$ by $\xi $ we have
\begin{equation*}
\xi ^{m+1}+\widetilde{V}_{1}^{(m)}\xi ^{n+1}+..+\widetilde{V}%
_{n_{1}+1}^{(m)}\xi ^{n+1-n_{1}}+...+\widetilde{V}_{n+1}^{(m)}\xi =0.
\end{equation*}%
Substituting $\xi ^{n+1}$ from (\ref{3.15c}) for $k=1$ and $\xi ^{n+1-n_{1}}$
from (\ref{3.15e}) we get
\begin{equation*}
\xi ^{m+1}+\widetilde{V}_{1}^{(m)}\left( -\widetilde{V}_{1}^{(n+1)}\xi
^{n}-...-\widetilde{V}_{n+1}^{(n+1)}\right) +...\ \ \ \ \ \ \ \ \ \ \ \ \ \
\ \ \ \ \ \ \ \ \ \ \ \ \ \ \ \ \ \ \ \ \ \ \
\end{equation*}%
\begin{equation*}
+\widetilde{V}_{n_{1}+1}^{(m)}\left( -\widetilde{V}_{1}^{(n+1-n_{1})}\xi
^{n}-...-\widetilde{V}_{n+1}^{(n+1-n_{1})}\right) +...+\widetilde{V}%
_{n+1}^{(m)}\xi =0.
\end{equation*}%
A comparison with the separation curve for the potential $\widetilde{V}%
_{r}^{(m+1)}$ (eq. (\ref{3.15c}) with $k=m+1-n$) reveals the formula (\ref%
{3.21cc}).
\end{proof}

Of course, all Hamiltonian functions $\widetilde{H}_{r}$ are in involution
with respect to $\theta _{0}$ as the proof of Lemma 6 is valid when we
change $E_{i}\rightarrow H_{i}$ and $\widetilde{E}_{i}\rightarrow \widetilde{%
H}_{i}.$

\subsection{Quasi-bi-Hamiltonian representation}

The invariance of the subspace spanned by $(d\widetilde{H}_{1},...,d%
\widetilde{H}_{n+1})$ with respect to $N^{\ast }$ (\ref{0.19}) is proved in
the following theorem.

\begin{theorem}
Hamiltonian functions $\widetilde{H}_{r}$ fulfill the following quasi-bi-
Hamiltonian chain
\begin{equation*}
d\widetilde{H}_{r+1}=N^{\ast }d\widetilde{H}_{r}+\alpha _{r}d\widetilde{H}%
_{1}+\beta _{r}d\widetilde{H}_{n_{1}+1}
\end{equation*}%
\begin{equation}
\Updownarrow  \tag{5.17}  \label{3.14a}
\end{equation}%
\begin{equation*}
\theta _{0}d\widetilde{H}_{r+1}=\theta _{1}d\widetilde{H}_{r}+\alpha
_{r}\theta _{0}d\widetilde{H}_{1}+\beta _{r}\theta _{0}d\widetilde{H}%
_{n_{1}+1},
\end{equation*}%
where $\alpha _{r}=\widetilde{V}_{r}^{(n+1)},\beta _{r}=\widetilde{V}%
_{r}^{(n+1)-n_{1}}.$
\end{theorem}

\begin{proof}
We use the property of the Benenti chain
\begin{equation}
dH_{r+1}=N^{\ast }dH_{r}+V_{r}^{(n)}dH_{1},  \tag{5.18a}  \label{3.15a}
\end{equation}%
\begin{equation}
d\rho _{r+1}=L^{\ast }d\rho _{r}+V_{r}^{(n)}d\rho _{1},\ \ \
V_{r}^{(n)}=\rho _{r},  \tag{5.18b}  \label{3.15b}
\end{equation}%
and recursion relations (\ref{3.7}), (\ref{3.8bb}). Hence, we have
\begin{eqnarray*}
r.h.s.(\ref{3.14a}) &=&N^{\ast }d\widetilde{H}_{r}+\alpha _{r}d\widetilde{H}%
_{1}+\beta _{r}d\widetilde{H}_{n_{1}+1} \\
&=&N^{\ast }d(H_{r-1}-\frac{\rho _{r-1}}{\rho _{n_{1}-1}}H_{n_{1}-1})+\left(
\rho _{r}-\frac{\rho _{r-1}\rho _{n_{1}}}{\rho _{n_{1}-1}}\right) d(-\frac{1%
}{\rho _{n_{1}-1}}H_{n_{1-1}}) \\
&&+\frac{\rho _{r-1}}{\rho _{n_{1}-1}}(H_{n_{1}}-\frac{\rho _{n_{1}}}{\rho
_{n_{1}-1}}H_{n_{1}-1}) \\
&=&N^{\ast }dH_{r-1}-\frac{\rho _{r-1}}{\rho _{n_{1}-1}}N^{\ast
}dH_{n_{1}-1}-\rho _{r-1}H_{n_{1}-1}N^{\ast }d(\frac{1}{\rho _{n_{1}-1}}) \\
&&-\frac{1}{\rho _{n_{1}-1}}H_{n_{1}-1}N^{\ast }d\rho _{r-1}+\frac{\rho
_{r-1}}{\rho _{n_{1}-1}}dH_{n_{1}}-\frac{\rho _{r-1}}{\rho _{n_{1}-1}^{2}}%
H_{n_{1}-1}d\rho _{n_{1}} \\
&&-\rho _{r}d\left( \frac{H_{n_{1}-1}}{\rho _{n_{1}-1}}\right) \\
&=&N^{\ast }dH_{r-1}+\rho _{r-1}dH_{1}-\rho _{n_{1}-1}H_{n_{1}-1}N^{\ast }d(%
\frac{1}{\rho _{n_{1}-1}}) \\
&&-\frac{1}{\rho _{n_{1}-1}}H_{n_{1}-1}N^{\ast }d\rho _{r-1}-\frac{\rho
_{r-1}}{\rho _{n_{1}-1}^{2}}H_{n_{1}-1}d\rho _{n_{1}}-\rho _{r}d\left( \frac{%
H_{n_{1}-1}}{\rho _{n_{1}-1}}\right) \\
&=&dH_{r}-\rho _{r}d\left( \frac{H_{n_{1}-1}}{\rho _{n_{1}-1}}\right) +\frac{%
\rho _{r-1}}{\rho _{n_{1}-1}^{2}}H_{n_{1}-1}(N^{\ast }d\rho _{r-1}-d\rho
_{n_{1}}) \\
&&-\frac{1}{\rho _{n_{1}-1}}H_{n_{1}-1}N^{\ast }d\rho _{r-1}
\end{eqnarray*}%
\begin{eqnarray*}
&=&dH_{r}-\rho _{r}d\left( \frac{H_{n_{1}-1}}{\rho _{n_{1}-1}}\right) -\frac{%
\rho _{r-1}}{\rho _{n_{1}-1}^{2}}H_{n_{1}-1}d\rho _{1}-\frac{1}{\rho
_{n_{1}-1}}H_{n_{1}-1}N^{\ast }d\rho _{r-1} \\
&=&dH_{r}-\rho _{r}d\left( \frac{H_{n_{1}-1}}{\rho _{n_{1}-1}}\right) -\frac{%
H_{n_{1}-1}}{\rho _{n_{1}-1}}(N^{\ast }d\rho _{r-1}+\rho _{r-1}d\rho _{1}) \\
&=&dH_{r}-\rho _{r}d\left( \frac{H_{n_{1}-1}}{\rho _{n_{1}-1}}\right) -\frac{%
H_{n_{1}-1}}{\rho _{n_{1}-1}}d\rho _{r} \\
&=&d\left( H_{r}-\frac{\rho _{r}}{\rho _{n_{1}-1}}H_{n_{1}-1}\right) =d%
\widetilde{H}_{r+1}=l.h.s.(\ref{3.14a})\ \ \ \ \
\end{eqnarray*}
\end{proof}

Notice that from (\ref{3.14a}) it follows that also $\{\widetilde{H}_{r},%
\widetilde{H}_{s}\}_{\theta _{1}}=0$ is valid. Of course, formula (\ref%
{3.14a}) works separately for $\widetilde{E}_{r}$ and $\widetilde{V}_{r}$ in
the form
\begin{equation}
d\widetilde{E}_{r+1}=N^{\ast }d\widetilde{E}_{r}+\alpha _{r}d\widetilde{E}%
_{1}+\beta _{r}d\widetilde{E}_{n_{1}+1},  \tag{5.19a}  \label{3.16a}
\end{equation}%
\begin{equation}
d\widetilde{V}_{r+1}=L^{\ast }d\widetilde{V}_{r}+\alpha _{r}d\widetilde{V}%
_{1}+\beta _{r}d\widetilde{V}_{n_{1}+1}.  \tag{5.19b}  \label{3.16b}
\end{equation}%
Repeating recursive transformation (\ref{3.16b}) we end up with the
equivalent form
\begin{equation}
d\widetilde{V}_{r+1}=A_{r}d\widetilde{V}_{1}+B_{r}d\widetilde{V}_{n_{1}+1},
\tag{5.20}  \label{3.17a}
\end{equation}%
where
\begin{equation}
A_{r}=K_{r+1}^{\ast }-\frac{\rho _{n_{1}}}{\rho _{n_{1}-1}}K_{r}^{\ast },\ \
\ B_{r}=\frac{1}{\rho _{n_{1}-1}}K_{r}^{\ast }.  \tag{5.20a}  \label{3.17b}
\end{equation}

In analogy to the Benenti case (\ref{2.30b}), (\ref{2.30c}), the $d_{p}$
part of (\ref{3.16a}) gives us immediately the analog of formula (\ref{2.7b}%
) for the one-hole case
\begin{equation}
\widetilde{K}_{r+1}=L\widetilde{K}_{r}+\alpha _{r}I+\beta _{r}\widetilde{K}%
_{n_{1}+1}.  \tag{5.21}  \label{3.17c}
\end{equation}%
Let us introduce $\widetilde{L}$ function as it was done for the Benenti
case
\begin{equation}
\widetilde{L}=\frac{1}{2}p^{T}L\widetilde{G}~p.  \tag{5.22}  \label{3.17d}
\end{equation}%
Then, from (\ref{3.17c}) we find
\begin{equation}
\widetilde{L}=\widetilde{E}_{2}-\alpha _{1}\widetilde{E}_{1}-\beta _{1}%
\widetilde{E}_{n_{1}+1}  \tag{5.23}  \label{5.23}
\end{equation}%
and
\begin{equation}
\{\widetilde{L},\widetilde{E}_{1}\}_{\theta _{0}}=\{\widetilde{E}_{1},\alpha
_{1}\}_{\theta _{0}}\widetilde{E}_{1}+\{\widetilde{E}_{1},\beta
_{1}\}_{\theta _{0}}\widetilde{E}_{n_{1}+1}=\kappa ^{0}\widetilde{E}%
_{1}+\kappa ^{n_{1}}\widetilde{E}_{n_{1}+1}.  \tag{5.24}  \label{5.24}
\end{equation}%
Thus, for $\widetilde{G}$, $L$ is not a conformal Killing tensor.

\subsection{Bi-Hamiltonian representation}

Adding two Casimir coordinates $c_{1},c_{2}$ (with respect to $\theta _{0}$)
and extending Hamiltonians $\widetilde{H}_{r}$ to the form
\begin{equation}
\widetilde{h}_{r}=\widetilde{H}_{r}+\alpha _{r}c_{1}+\beta _{r}c_{2}
\tag{5.25}  \label{3.18a}
\end{equation}%
one can put the quasi-bi-Hamiltonian chain (\ref{3.14a}) from $T^{\ast }Q$
into a bi-Hamiltonian chain on $T^{\ast }Q\times \mathbb{R}\times \mathbb{R}$%
, being a composition of two bi-Hamiltonian sub-chains
\begin{equation}
\begin{array}{l}
\widetilde{\pi }_{0}d\widetilde{h}_{0}=0 \\
\widetilde{\pi }_{0}d\widetilde{h}_{1}=\widetilde{X}_{1}=\widetilde{\pi }%
_{1}d\widetilde{h}_{0} \\
\,\,\,\,\,\,\,\,\,\,\,\,\,\,\,\,\,\,\,\,\,\,\,\,\,\,\,\,\,\,\,\,\vdots \\
\widetilde{\pi }_{0}d\widetilde{h}_{n_{1}-1}=\widetilde{X}_{n_{1}-1}=%
\widetilde{\pi }_{1}d\widetilde{h}_{n_{1}-2} \\
\qquad \qquad \,\,\,\,\ \ \ \ \ \ \ \ \ \,\,0=\widetilde{\pi }_{1}d%
\widetilde{h}_{n_{1}-1}\,\,%
\end{array}
\tag{5.26a}  \label{3.19a}
\end{equation}%
\begin{equation*}
\end{equation*}%
\begin{equation}
\begin{array}{l}
\widetilde{\pi }_{0}d\widetilde{h}_{n_{1}}=0 \\
\widetilde{\pi }_{0}d\widetilde{h}_{n_{1}+1}=\widetilde{X}_{n_{1}+1}=%
\widetilde{\pi }_{1}d\widetilde{h}_{n_{1}} \\
\,\,\,\,\,\,\,\,\,\,\,\,\,\,\,\,\,\,\,\,\,\,\,\,\,\,\,\,\,\,\,\,\vdots \\
\widetilde{\pi }_{0}d\widetilde{h}_{n+1}=\widetilde{X}_{n+1}=\widetilde{\pi }%
_{1}d\widetilde{h}_{n} \\
\qquad \qquad \ \ \ \ \ \ \ \,\,\,\,0=\widetilde{\pi }_{1}d\widetilde{h}%
_{n+1}\,\,,%
\end{array}
\tag{5.26b}  \label{3.19b}
\end{equation}%
where $\widetilde{h}_{0}=c_{1},\ \widetilde{h}_{n_{1}}=c_{2}$ and
\begin{equation*}
\widetilde{\pi }_{0}=\left(
\begin{array}{c|c}
\theta _{0} & 0\ \ 0 \\ \hline
\begin{array}{c}
0 \\
0%
\end{array}
& 0%
\end{array}%
\right) \text{,\ }\widetilde{\pi }_{1}=\left(
\begin{array}{c|c}
\theta _{1} & \theta _{0}d\widetilde{h}_{1}\ \ \ \theta _{0}d\widetilde{h}%
_{n_{1}+1} \\ \hline
\begin{array}{c}
-(\theta _{0}d\widetilde{h}_{1})^{T} \\
-(\theta _{0}d\widetilde{h}_{n_{1+}1})^{T}%
\end{array}
& 0%
\end{array}%
\right) .
\end{equation*}%
$\widetilde{\pi }_{0}$ and $\widetilde{\pi }_{1}$ are compatible Poisson
tensors on $T^{\ast }Q\times \mathbb{R}\times \mathbb{R}$ as in $(q,p,c)$
coordinates $\widetilde{\pi }_{1}$ tensor has the form (\ref{0.33a})
\begin{equation*}
\widetilde{\pi }_{1}=\widetilde{\pi }_{1D}+\sum_{i=1}^{k}X_{i}\wedge Z_{i},
\end{equation*}%
where
\begin{equation*}
\widetilde{\pi }_{1D}=\left(
\begin{array}{c|c}
\theta _{1} & 0\ \ \ 0 \\ \hline
\begin{array}{c}
0 \\
0%
\end{array}
& 0%
\end{array}%
\right) ,\ \ \ Z_{1}=(0,...,0,1,0)^{T},\ \ \ Z_{2}=(0,...,0,0,1)^{T},
\end{equation*}%
\begin{equation*}
X_{1}=\widetilde{\pi }_{0}d\widetilde{h}_{1}=(\theta _{0}d\widetilde{h}%
_{1},0)^{T},\ \ \ X_{2}=\widetilde{\pi }_{0}d\widetilde{h}_{n_{1}+1}=(\theta
_{0}d\widetilde{h}_{n_{1}+1},0)^{T}
\end{equation*}%
and vector fields $Z_{i}$ and $X_{i}$ fulfil condition from Lemma \ref{l3}.\
The reduction of both chains onto the symplectic leaf $c_{1}=c_{2}=0$ of $%
\widetilde{\pi }_{0}$ along $Z_{1}=\frac{\partial }{\partial c_{1}},Z_{2}=%
\frac{\partial }{\partial c_{2}}$ reconstructs immediately the
quasi-bi-Hamiltonian chain (\ref{3.14a}).

If we introduce the Poisson pencil $\widetilde{\pi }_{\xi }=\widetilde{\pi }%
_{1}-\xi \widetilde{\pi }_{0}$ both chains (\ref{3.19a}), (\ref{3.19b}) can
be written in a compact form
\begin{equation}
\widetilde{\pi }_{\xi }d\widetilde{h}(\xi )=0,\ \ \widetilde{h}(\xi )=(%
\widetilde{h}^{(1)}(\xi ),\widetilde{h}^{(2)}(\xi ))^{T},  \tag{5.27}
\label{3.24}
\end{equation}%
where
\begin{equation}
\widetilde{h}^{(1)}(\xi )=\widetilde{h}_{0}\xi ^{n_{1}-1}+...+\widetilde{h}%
_{n_{1}-1},\ \ \ \widetilde{h}^{(2)}(\xi )=\widetilde{h}_{n_{1}}\xi
^{n+1-n_{1}}+...+\widetilde{h}_{n+1}  \tag{5.27a}  \label{3.24a}
\end{equation}%
and the operation is applied to each component of the vector. The
quasi-bi-Hamiltonian chain (\ref{3.14a}) takes the form
\begin{equation*}
\widetilde{\theta }_{\xi }d\widetilde{H}(\xi )+\widetilde{\alpha }(\xi
)\left( \widetilde{\theta }_{0}d\widetilde{H}_{(1)}\right) =0
\end{equation*}%
\begin{equation}
\Updownarrow  \tag{5.28}  \label{3.24b}
\end{equation}%
\begin{equation*}
(N^{\ast }-\xi I)d\widetilde{H}(\xi )+\widetilde{\alpha }(\xi )d\widetilde{H}%
_{(1)}=0,
\end{equation*}%
where%
\begin{equation*}
\widetilde{\theta }_{\xi }=\widetilde{\theta }_{1}-\xi \widetilde{\theta }%
_{0},\ \ \ \ \widetilde{H}_{(1)}=(\widetilde{H}_{1},\widetilde{H}%
_{n_{1}+1})^{T},
\end{equation*}%
\begin{equation}
\widetilde{H}^{(1)}(\xi )=\widetilde{H}_{1}\xi ^{n_{1}-2}+...+\widetilde{H}%
_{n_{1}-1},\ \ \widetilde{H}_{\xi }^{(2)}=\widetilde{H}_{n_{1}+1}\xi
^{n-n_{1}}+...+\widetilde{H}_{n+1}  \tag{5.28a}  \label{3.24c}
\end{equation}%
and
\begin{equation}
\widetilde{\alpha }(\xi )=\left(
\begin{array}{cc}
\frac{\partial \widetilde{h}^{(1)}(\xi )}{\partial c_{1}} & \frac{\partial
\widetilde{h}^{(2)}(\xi )}{\partial c_{1}} \\
&  \\
\frac{\partial \widetilde{h}^{(1)}(\xi )}{\partial c_{2}} & \frac{\partial
\widetilde{h}^{(2)}(\xi )}{\partial c_{2}}%
\end{array}%
\right) .  \tag{5.28b}  \label{3.24d}
\end{equation}

The St\"{a}ckel conditions for extended Hamiltonians $\widetilde{h}_{r}$
take the form (\ref{3.18bb})
\begin{equation*}
c_{1}\xi ^{(n+1)}+\widetilde{h}_{1}\xi ^{(n+1)-1}+...+\widetilde{h}%
_{n_{1}-1}\xi ^{(n+1)-(n_{1}-1)}\ \ \ \ \ \ \ \ \ \ \ \ \ \ \ \ \ \ \ \ \ \
\ \ \ \ \ \ \ \ \ \ \ \ \ \ \ \ \ \ \ \ \ \ \ \ \ \ \ \ \ \ \ \ \ \ \ \ \ \
\ \ \
\end{equation*}%
\begin{equation*}
+c_{2}\xi ^{(n+1)-n_{1}}+\widetilde{h}_{n_{1}+1}\xi ^{(n+1)-(n_{1}+1)}+...+%
\widetilde{h}_{n+1}\ \ \ \ \ \ \ \ \ \ \ \ \ \ \ \ \ \ \ \ \ \ \ \ \ \ \ \ \
\ \ \ \ \ \ \ \ \ \ \ \ \ \ \ \ \ \ \ \ \ \ \ \ \ \ \ \ \ \ \ \ \ \ \
\end{equation*}%
\begin{equation}
\ \ \ \ \ \ \ \ \ \ \ \ \ \ \ \ \ \ \ \ \ \ \ \ \ \ \ \ \ \ \ \ \ \ \ \ \ \
\ \ \ \ =\xi ^{n+2-n_{1}}\widetilde{h}^{(1)}(\xi )+\widetilde{h}^{(2)}(\xi
)\ =\frac{1}{2}f(\xi )\mu ^{2}+\gamma (\xi ),\   \tag{5.29}  \label{5.26}
\end{equation}%
where $(\xi ,\mu )=(\lambda _{i},\mu _{i}),\ i=1,...,n.$

\section{$k$-hole deformations of Benenti systems}

\subsection{Deformation procedure}

Here we extend the results of the previous section onto the general $k$-hole
case. Let us start with the separability condition
\begin{equation}
\widetilde{H}_{1}\xi ^{(n+k)-1}+\widetilde{H}_{2}\xi ^{(n+k)-2}+...+%
\widetilde{H}_{n+k}=\psi (\xi ,\mu ),  \tag{6.1}  \label{4.1}
\end{equation}%
with $k$-holes in $\xi ^{(n+k)-n_{1}},\xi ^{(n+k)-n_{2}},...,\xi
^{(n+k)-n_{k}}$, \ $1<n_{1}<...<n_{k}<n+k,$ $k\in \mathbb{N}$, i.e. $%
\widetilde{H}_{n_{1}}=\widetilde{H}_{n_{2}}=...=\widetilde{H}_{n_{k}}=0$,
and the separability condition for Benenti systems with the same $\psi $%
\begin{equation}
H_{1}\xi ^{n-1}+H_{2}\xi ^{n-2}+...+H_{n}=\psi (\xi ,\mu ).  \tag{6.2}
\label{4.2}
\end{equation}%
As for the basic potentials
\begin{equation*}
\xi ^{n+k}+V_{1}^{(n+k)}\xi ^{n-1}+...+V_{n}^{(n+k)}=0,
\end{equation*}%
substituting this relation to (\ref{4.1}) for $\xi ^{(n+k)-1},...,\xi ^{n}$
we get a deformation of the chain (\ref{4.1}) to the Benenti case (\ref{4.2}%
)
\begin{equation}
H_{r}=\widetilde{H}_{r+k}-V_{r}^{(n+k-1)}\widetilde{H}_{1}-V_{r}^{(n+k-2)}%
\widetilde{H}_{2}-...-V_{r}^{(n)}\widetilde{H}_{k},\ \ \ r=1,...,n,
\tag{6.3}  \label{4.3}
\end{equation}%
where $\widetilde{H}_{n_{1}}=...=\widetilde{H}_{n_{k}}=0$ and $V_{r}^{(m)}$
are appropriate basic Benenti potentials.

\begin{lemma}
Deformation of the Benenti case (\ref{4.2}) to the chain (\ref{4.1}), i.e.
the inverse formula to the (\ref{4.3}) one, is given by a following
determinant form
\begin{equation}
\widetilde{H}_{r}=\frac{\left\vert
\begin{array}{cccc}
H_{r-k} & \rho _{r-1} & \cdots & \rho _{r-k} \\
H_{n_{1}-k} & \rho _{n_{1}-1} & \cdots & \rho _{n_{1}-k} \\
\cdots & \cdots & \cdots & \cdots \\
H_{n_{k}-k} & \rho _{n_{k}-1} & \cdots & \rho _{n_{k}-k}%
\end{array}%
\right\vert }{\left\vert
\begin{array}{ccc}
\rho _{n_{1}-1} & \cdots & \rho _{n_{1}-k} \\
\cdots & \cdots & \cdots \\
\rho _{n_{k}-1} & \cdots & \rho _{n_{k}-k}%
\end{array}%
\right\vert }.  \tag{6.4}  \label{4.4}
\end{equation}
\end{lemma}

\begin{proof}
First, we select from (\ref{4.3}) $k$ equations containing $\widetilde{H}%
_{n_{1}},...,\widetilde{H}_{n_{k}}$%
\begin{eqnarray*}
H_{n_{1}-k} &=&-V_{n_{1}-k}^{(n+k-1)}\widetilde{H}_{1}-...-V_{n_{1}-k}^{(n)}%
\widetilde{H}_{k}, \\
&&\vdots \\
H_{n_{k}-k} &=&-V_{n_{k}-k}^{(n+k-1)}\widetilde{H}_{1}-...-V_{n_{k}-k}^{(n)}%
\widetilde{H}_{k}.
\end{eqnarray*}%
The solution with respect to $\widetilde{H}_{i},i=1,...,k$ is given by a
determinant form
\begin{equation*}
\widetilde{H}_{i}=\frac{W_{i}}{W},\ \ \ i=1,...,n,
\end{equation*}%
where
\begin{equation*}
W=(-1)^{k}\left\vert
\begin{array}{ccc}
V_{n_{1}-k}^{(n+k-1)} & \cdots & V_{n_{1}-k}^{(n)} \\
\cdots & \cdots & \cdots \\
V_{n_{k}-k}^{(n+k-1)} & \cdots & V_{n_{k}-k}^{(n)}%
\end{array}%
\right\vert
\end{equation*}%
and%
\begin{equation*}
W_{i}=(-1)^{k+i}\left\vert
\begin{array}{cccc}
H_{n_{1}-k} & V_{n_{1}-k}^{(n+k-1)} & \cdots & V_{n_{1}-k}^{(n)} \\
\cdots & \cdots & \cdots & \cdots \\
H_{n_{k}-k} & V_{n_{k}-k}^{(n+k-1)} & \cdots & V_{n_{k}-k}^{(n)}%
\end{array}%
\right\vert
\end{equation*}%
with the column $(V_{n_{1}-k}^{(n+k-i)},...,V_{n_{k}-k}^{(n+k-i)})^{T}$
missing. Substituting this result to (\ref{4.3}) we get
\begin{equation*}
\widetilde{H}_{r}=\frac{%
H_{r-k}W+V_{r-k}^{(n+k-1)}W_{1}+...+V_{r-k}^{(n)}W_{k}}{W}
\end{equation*}%
\begin{equation*}
\end{equation*}%
\begin{equation*}
=\frac{\left\vert
\begin{array}{cccc}
H_{r-k} & V_{r-k}^{(n+k-1)} & \cdots & V_{r-k}^{(n)} \\
H_{n_{1}-k} & V_{n_{1}-k}^{(n+k-1)} & \cdots & V_{n_{1}-k}^{(n)} \\
\cdots & \cdots & \cdots & \cdots \\
H_{n_{k}-k} & V_{n_{k}-k}^{(n+k-1)} & \cdots & V_{n_{k}-k}^{(n)}%
\end{array}%
\right\vert }{\left\vert
\begin{array}{ccc}
V_{n_{1}-k}^{(n+k-1)} & \cdots & V_{n_{1}-k}^{(n)} \\
\cdots & \cdots & \cdots \\
V_{n_{k}-k}^{(n+k-1)} & \cdots & V_{n_{k}-k}^{(n)}%
\end{array}%
\right\vert }
\end{equation*}%
\begin{equation*}
\end{equation*}%
\begin{equation*}
=\frac{\left\vert
\begin{array}{cccc}
H_{r-k} & \rho _{r-1} & \cdots & \rho _{r-k} \\
H_{n_{1}-k} & \rho _{n_{1}-1} & \cdots & \rho _{n_{1}-k} \\
\cdots & \cdots & \cdots & \cdots \\
H_{n_{k}-k} & \rho _{n_{k}-1} & \cdots & \rho _{n_{k}-k}%
\end{array}%
\right\vert }{\left\vert
\begin{array}{ccc}
\rho _{n_{1}-1} & \cdots & \rho _{n_{1}-k} \\
\cdots & \cdots & \cdots \\
\rho _{n_{k}-1} & \cdots & \rho _{n_{k}-k}%
\end{array}%
\right\vert }.
\end{equation*}%
The last step is valid due to the fact that $V_{i}^{(n)}=\rho _{i},$ the
form of the recursion formula for Benenti basic potentials (\ref{2.37}) and
the properties of determinants. It allow us to replace the arbitrary
potential $V^{(n+k-i)}$ in determinants by the $V^{(n)}=\rho $ one. For each
recursive step we have
\begin{equation*}
\left\vert
\begin{array}{cccc}
\cdots & V_{n_{1}-k}^{(n+k-i)} & \cdots & \rho _{n_{1}-k} \\
\cdots & \cdots & \cdots & \cdots \\
\cdots & V_{n_{k}-k}^{(n+k-i)} & \cdots & \rho _{n_{k}-k}%
\end{array}%
\right\vert \ \ \ \ \ \ \ \ \ \ \ \ \ \ \ \ \ \ \ \ \ \ \ \ \ \ \ \ \ \ \ \
\ \ \ \ \ \
\end{equation*}%
\begin{equation*}
\ \ \ \ \ \ \ \ \ \ \ \ \ \ \ \ \ \ \ \ \ \ \ \ \ =\left\vert
\begin{array}{cccc}
\cdots & V_{n_{1}-k+1}^{(n+k-i-1)}-\rho _{n_{1}-k}V_{1}^{(n+k-i-1)} & \cdots
& \rho _{n_{1}-k} \\
\cdots & \cdots & \cdots & \cdots \\
\cdots & V_{n_{k}-k+1}^{(n+k-i-1)}-\rho _{n_{k}-k}V_{1}^{(n+k-i-1)} & \cdots
& \rho _{n_{k}-k}%
\end{array}%
\right\vert
\end{equation*}%
\begin{equation*}
=\left\vert
\begin{array}{cccc}
\cdots & V_{n_{1}-k+1}^{(n+k-i-1)} & \cdots & \rho _{n_{1}-k} \\
\cdots & \cdots & \cdots & \cdots \\
\cdots & V_{n_{k}-k+1}^{(n+k-i-1)} & \cdots & \rho _{n_{k}-k}%
\end{array}%
\right\vert .
\end{equation*}
\end{proof}

The formula (\ref{4.4}) applies separately to the geodesic and the potential
parts.

\subsection{Deformed geodesic motion}

Let us first look onto $n$ geodesic Hamiltonians $\widetilde{E}_{r},\
r=1,...,n+k,\ r\neq n_{1},...,n_{k}.$ Introducing the abbreviation
\begin{equation}
\varphi (n_{1},...,n_{k})=\left\vert
\begin{array}{ccc}
\rho _{n_{1}-1} & \cdots & \rho _{n_{1}-k} \\
\cdots & \cdots & \cdots \\
\rho _{n_{k}-1} & \cdots & \rho _{n_{k}-k}%
\end{array}%
\right\vert ,  \tag{6.5}  \label{4.6}
\end{equation}%
one finds
\begin{equation}
\widetilde{E}_{r}=\frac{1}{\varphi }\left\vert
\begin{array}{cccc}
E_{r-k} & \rho _{r-1} & \cdots & \rho _{r-k} \\
E_{n_{1}-k} & \rho _{n_{1}-1} & \cdots & \rho _{n_{1}-k} \\
\cdots & \cdots & \cdots & \cdots \\
E_{n_{k}-k} & \rho _{n_{k}-1} & \cdots & \rho _{n_{k}-k}%
\end{array}%
\right\vert .  \tag{6.6}  \label{4.7}
\end{equation}%
Using the known relations for Benenti chain
\begin{equation*}
\rho _{r}I=K_{r+1}-LK_{r}
\end{equation*}%
and the property of determinants we get
\begin{eqnarray}
\widetilde{E}_{r} &=&\frac{1}{2}p^{T}\left\vert
\begin{array}{cccc}
K_{r-k} & \rho _{r-1} & \cdots & \rho _{r-k} \\
K_{n_{1}-k} & \rho _{n_{1}-1} & \cdots & \rho _{n_{1}-k} \\
\cdots & \cdots & \cdots & \cdots \\
K_{n_{k}-k} & \rho _{n_{k}-1} & \cdots & \rho _{n_{k}-k}%
\end{array}%
\right\vert \frac{1}{\varphi }Gp  \notag \\
&&  \notag \\
&=&\frac{1}{2}p^{T}\left\vert
\begin{array}{cccc}
K_{r-k} & K_{r} & \cdots & K_{r-k+1} \\
K_{n_{1}-k} & K_{n_{1}} & \cdots & K_{n_{1}-k+1} \\
\cdots & \cdots & \cdots & \cdots \\
K_{n_{k}-k} & K_{n_{k}} & \cdots & K_{n_{k}-k+1}%
\end{array}%
\right\vert \frac{1}{\varphi }Gp\ \ \ \ \ \ \ \ \ \ \ \ \ \ \ \ \ \ \ \ \ \
\ \   \TCItag{6.7}  \label{4.8}
\end{eqnarray}%
\begin{eqnarray*}
&=&(-1)^{k}\frac{1}{2}p^{T}\left\vert
\begin{array}{cccc}
K_{r} & \cdots & K_{r-k+1} & K_{r-k} \\
K_{n_{1}} & \cdots & K_{n_{1}-k+1} & K_{n_{1}-k} \\
\cdots & \cdots & \cdots & \cdots \\
K_{n_{k}} & \cdots & K_{n_{k}-k+1} & K_{n_{k}-k}%
\end{array}%
\right\vert \frac{1}{\varphi }Gp\ \  \\
&& \\
&=&(-1)^{k}\frac{1}{2}p^{T}(K_{r}D_{0}-K_{r-1}D_{1}+...+(-1)^{k}K_{r-k}D_{k})%
\frac{1}{\varphi }Gp,\ \ \
\end{eqnarray*}%
where
\begin{equation*}
D_{i}=\left\vert
\begin{array}{cccccc}
K_{n_{1}} & \cdots & K_{n_{1}-i+1} & K_{n_{1}-i-1} & \cdots & K_{n_{1}-k+1}
\\
\cdots & \cdots & \cdots & \cdots & \cdots & \cdots \\
K_{n_{k}} & \cdots & K_{n_{k}-i+1} & K_{n_{k}-i-1} & \cdots & K_{n_{k}-k+1}%
\end{array}%
\right\vert ,\ \ i=0,...,k
\end{equation*}%
and $K_{m}$ in determinant calculations are treated as symbols not matrices.

Then,
\begin{equation}
\widetilde{E}_{1}=\frac{1}{2}p^{T}\widetilde{G}p\ \ \ \Longrightarrow \ \
\widetilde{G}=(-1)^{k}\frac{1}{\varphi }D_{0}G,  \tag{6.8}  \label{4.9}
\end{equation}%
and
\begin{equation}
\widetilde{E}_{r}=\frac{1}{2}p^{T}\widetilde{K}_{r}\widetilde{G}p\
\Longrightarrow \widetilde{K}%
_{r}=K_{r}-K_{r-1}D_{1}D_{0}^{-1}+...+(-1)^{k}K_{r-k}D_{k}D_{0}^{-1},
\tag{6.9}  \label{4.10}
\end{equation}%
where
\begin{equation}
D_{0}=\left\vert
\begin{array}{ccc}
K_{n_{1}-1} & \cdots & K_{n_{1}-k} \\
\cdots & \cdots & \cdots \\
K_{n_{k}-1} & \cdots & K_{n_{k}-k}%
\end{array}%
\right\vert .  \tag{6.10}  \label{4.11}
\end{equation}

Again we know from the construction that $\widetilde{E}_{r}$ are in
involution, as they are St\"{a}ckel geodesics. Here we present the proof in
the coordinate free form but on the level of full Hamiltonians (geodesic
motion is included as a special case). Let us first consider basic
potentials.

\subsection{Basic deformed potentials}

From (\ref{4.4}) the deformed potentials are
\begin{equation}
\widetilde{V}_{r}^{(m)}=\frac{1}{\varphi (n_{1},...,n_{k})}\left\vert
\begin{array}{cccc}
V_{r-k}^{(m)} & \rho _{r-1} & \cdots & \rho _{r-k} \\
V_{n_{1}-k}^{(m)} & \rho _{n_{1}-1} & \cdots & \rho _{n_{1}-k} \\
\cdots & \cdots & \cdots & \cdots \\
V_{n_{k}-k}^{(m)} & \rho _{n_{k}-1} & \cdots & \rho _{n_{k}-k}%
\end{array}%
\right\vert ,  \tag{6.11}  \label{4.12}
\end{equation}%
so using the recursion formula for the Benenti potentials and the properties
of the determinants we have $\widetilde{V}_{r}^{(m)}=-\delta _{r-k,n-m},$ $%
m<n+k,$ $m\neq (n+k)-n_{i},$ $i=1,...,k,$
\begin{equation}
\widetilde{V}_{r}^{(n+k)-n_{i}}=(-1)^{i+1}\frac{1}{\varphi }\left\vert
\begin{array}{ccc}
\rho _{r-1} & \cdots & \rho _{r-k} \\
\rho _{n_{1}-1} & \cdots & \rho _{n_{1}-k} \\
\cdots & \cdots & \cdots \\
\rho _{n_{k}-1} & \cdots & \rho _{n_{k}-k}%
\end{array}%
\right\vert  \tag{6.12}  \label{4.13}
\end{equation}%
with the row $(\rho _{n_{i}-1},...,\rho _{n_{i}-k})$ missing,
\begin{equation}
\widetilde{V}_{r}^{(n+k)}=\frac{1}{\varphi }\left\vert
\begin{array}{cccc}
\rho _{r} & \rho _{r-1} & \cdots & \rho _{r-k} \\
\rho _{n_{1}} & \rho _{n_{1}-1} & \cdots & \rho _{n_{1}-k} \\
\cdots & \cdots & \cdots & \cdots \\
\rho _{n_{k}} & \rho _{n_{k}-1} & \cdots & \rho _{n_{k}-k}%
\end{array}%
\right\vert ,\ \ ....\ \ .  \tag{6.13}  \label{4.14}
\end{equation}%
Notice that $\widetilde{V}_{n_{i}}^{(m)}=0$ for arbitrary $m\geq n+k$ and $%
\widetilde{V}_{n_{i}}^{(n+k)-n_{j}}=\delta _{ij}$, $i,j=1,...,k.$

As in the 1-hole case, one can shaw that nontrivial basic potentials $%
\widetilde{V}_{r}^{(n+k-n_{i})},i=1,...,k,\ \widetilde{V}_{r}^{(n+k-1+l)}$
and $\widetilde{V}_{r}^{(-l)}$ $l=1,2,...$ fulfill the following generating
equations
\begin{equation}
\xi ^{n+k-1+l}+\widetilde{V}_{1}^{(n+k-1+l)}\xi ^{(n+k)-1}+...+\widetilde{V}%
_{n+k}^{(n+k-1+l)}=0,\   \tag{6.14a}  \label{4.15aa}
\end{equation}%
\begin{equation}
\xi ^{-l}+\widetilde{V}_{1}^{(-l)}\xi ^{(n+k)-1}+...+\widetilde{V}%
_{n+k}^{(-l)}=0,\ \   \tag{6.15b}  \label{4.15bb}
\end{equation}%
\begin{equation}
\xi ^{n+k-n_{i}}+\widetilde{V}_{1}^{(n+k-n_{i})}\xi ^{(n+k)-1}+...+%
\widetilde{V}_{n+k}^{(n+k-n_{i})}=0,\   \tag{6.14c}  \label{4.15cc}
\end{equation}%
and hence enter separation curve
\begin{equation}
\widetilde{H}_{1}\xi ^{(n+k)-1}+\widetilde{H}_{2}\xi ^{(n+k)-2}+...+%
\widetilde{H}_{n+k}=\frac{1}{2}f(\xi )\mu ^{2}+\gamma (\xi ),\ \ \widetilde{H%
}_{n_{i}}=0,  \tag{6.15}  \label{4.16aa}
\end{equation}%
as $\gamma (\xi )=-\xi ^{(n+k)-n_{i}},-\xi ^{n+k},-\xi ^{-k},$ where $%
i=1,...,k,\ \ l=1,2,...$ .

Also as in the 1-hole case, it is not difficult to shaw that basic
potentials $\widetilde{V}_{r}^{(m)},m>n+k$ can be constructed recursively by
the following recursion relation
\begin{equation}
\widetilde{V}_{r}^{(m+1)}=\widetilde{V}_{r+1}^{(m)}-\widetilde{V}_{r}^{(n+k)}%
\widetilde{V}_{1}^{(m)}-\sum_{i=1}^{k}\widetilde{V}_{r}^{(n+k-n_{i})}%
\widetilde{V}_{n_{i}+1}^{(m)},  \tag{6.16}  \label{4.17bb}
\end{equation}%
where $\widetilde{V}_{r}^{(n+k-n_{i})},i=1,...,k$ and $\widetilde{V}%
_{r}^{(n+k)}$ are given by (\ref{4.13}) and (\ref{4.14}).

\subsection{Deformed Hamiltonian functions}

For further applications we present now a recursion formula for deformed
Hamiltonian functions understood in the following sense. Having the
Hamiltonian functions for the $k$-hole case we construct the Hamiltonian
functions for $(k+1)$-hole case, respectively. Let us consider the
separability condition
\begin{equation}
\overline{H}_{1}\xi ^{(n+k+1)-1}+\overline{H}_{2}\xi ^{(n+k+1)-2}+...+%
\overline{H}_{n+k+1}=\psi (\xi ,\mu ),  \tag{6.17}  \label{4.17cc}
\end{equation}%
with $(k+1)$-holes in $\xi ^{(n+k+1)-(n_{1}+1)}$, $...$, $\xi
^{(n+k+1)-(n_{k}+1)}$, $\xi ^{(n+k+1)-n_{k+1}}$, $\
1<n_{1}<...<n_{k+1}<n+k+1 $, $k\in \mathbb{N}$, i.e. $\overline{H}%
_{n_{1}+1}=...=\overline{H}_{n_{k}+1} $ $=\overline{H}_{n_{k+1}}=0$, and the
separability condition \
\begin{equation}
\widetilde{H}_{1}\xi ^{(n+k)-1}+\widetilde{H}_{2}\xi ^{(n+k)-2}+...+%
\widetilde{H}_{n+k}=\psi (\xi ,\mu ),  \tag{6.18}  \label{4.18cc}
\end{equation}%
with $k$-holes in $\xi ^{(n+k)-n_{1}},\xi ^{(n+k)-n_{2}},...,\xi
^{(n+k)-n_{k}},$ $\ 1<n_{1}<...<n_{k}<n+k,$ $k\in \mathbb{N}$, i.e. $%
\widetilde{H}_{n_{1}}=\widetilde{H}_{n_{2}}=...=\widetilde{H}_{n_{k}}=0$,
and the same $\psi .$ It means that the first $k$ holes are fixed. From (\ref%
{4.15aa}) with $l=1$ we have
\begin{equation}
\xi ^{n+k}+\widetilde{V}_{1}^{(n+k)}\xi ^{(n+k)-1}+...+\widetilde{V}%
_{n+k}^{(n+k)}=0.  \tag{6.19}  \label{4.19cc}
\end{equation}%
Substituting (\ref{4.19cc}) to the first term of (\ref{4.17cc}) we get
\begin{equation*}
(\overline{H}_{2}-\widetilde{V}_{1}^{(n+k)}\overline{H}_{1})\xi
^{(n+k)-1}+...-\widetilde{V}_{1}^{(n+k)}\overline{H}_{1}\xi
^{(n+k)-n_{k+1}}+...~\ \ \ \ \ \ \ \ \ \ \ \ \ \ \ \ \ \ \ \
\end{equation*}%
\begin{equation}
+(\overline{H}_{i+1}-\widetilde{V}_{i}^{(n+k)}\overline{H}_{1})\xi
^{(n+k)-i}+...+(\overline{H}_{n+k+1}-\widetilde{V}_{n+k}^{(n+k)}\overline{H}%
_{1})=\psi (\xi ,\mu )  \tag{6.20}  \label{4.20cc}
\end{equation}%
and hence
\begin{equation}
\widetilde{H}_{r}=\overline{H}_{r+1}-\widetilde{V}_{r}^{(n+k)}\overline{H}%
_{1},\ \ \ r=1,...,n+k,  \tag{6.21}  \label{4.21cc}
\end{equation}%
where $\overline{H}_{n_{1}+1}=...=\overline{H}_{n_{k}+1}=\overline{H}%
_{n_{k+1}}=0,\ \widetilde{V}_{n_{1}}^{(n+k)}=...=\widetilde{V}%
_{n_{k}}^{(n+k)}=0.$ The inverse of (\ref{4.21cc})
\begin{equation}
\overline{H}_{r+1}=\widetilde{H}_{r}-\frac{\widetilde{V}_{r}^{(n+k)}}{%
\widetilde{V}_{n_{k+1}-1}^{(n+k)}}\widetilde{H}_{n_{k+1}-1}  \tag{6.22}
\label{4.22cc}
\end{equation}%
is our final recursion formula. The formula is applicable separately for
geodesic Hamiltonian functions $\overline{E}_{r}$ and potentials $\overline{V%
}_{r}.$

\begin{lemma}
Functions $\overline{H}_{r}$ are in involution with respect to the Poisson
tensor $\theta _{0}.$
\end{lemma}

\begin{proof}
The proof is inductive. The 1-hole case was proved in Section 3. From the
involutivity of $k$-holes Hamiltonians $\widetilde{H}_{r}$ we prove the
involutivity of $\overline{H}_{r}$ Hamiltonians from $(k+1)$-holes case.
From involutivity of $\widetilde{H}_{r}$ Hamiltonians we have
\begin{eqnarray*}
0 &=&\{\widetilde{H}_{r},\widetilde{H}_{s}\}_{\theta _{0}}=\{\widetilde{E}%
_{r}+\widetilde{V}_{r},\widetilde{E}_{s}+\widetilde{V}_{s}\}_{\theta _{0}}=\{%
\widetilde{E}_{r},\widetilde{V}_{s}\}_{\theta _{0}}+\{\widetilde{V}_{r},%
\widetilde{E}_{s}\}_{\theta _{0}} \\
&& \\
&\Longrightarrow &\{\widetilde{H}_{r},\widetilde{V}_{s}\}_{\theta _{0}}=\{%
\widetilde{H}_{s},\widetilde{V}_{r}\}_{\theta _{0}}\Longrightarrow \{%
\widetilde{H}_{r},\widetilde{V}_{s}^{(n+k)}\}_{\theta _{0}}=\{\widetilde{H}%
_{s},\widetilde{V}_{r}^{(n+k)}\}_{\theta _{0}}.
\end{eqnarray*}%
Then, from (\ref{4.22cc}) we have
\begin{eqnarray*}
\{\overline{H}_{r+1},\overline{H}_{s+1}\}_{\theta _{0}} &=&\left\{
\widetilde{H}_{r},-\frac{\widetilde{V}_{s}^{(n+k)}}{\widetilde{V}%
_{n_{k+1}-1}^{(n+k)}}\widetilde{H}_{n_{k+1}-1}\right\} _{\theta
_{0}}+\left\{ -\frac{\widetilde{V}_{r}^{(n+k)}}{\widetilde{V}%
_{n_{k+1}-1}^{(n+k)}}\widetilde{H}_{n_{k+1}-1},\widetilde{H}_{s}\right\}
_{\theta _{0}} \\
&&+\left\{ \frac{\widetilde{V}_{r}^{(n+k)}}{\widetilde{V}_{n_{k+1}-1}^{(n+k)}%
}\widetilde{H}_{n_{k+1}-1},\frac{\widetilde{V}_{s}^{(n+k)}}{\widetilde{V}%
_{n_{k+1}-1}^{(n+k)}}\widetilde{H}_{n_{k+1}-1}\right\} _{\theta _{0}} \\
&=&\{\widetilde{H}_{r},\widetilde{V}_{n_{k+1}-1}^{(n+k)}\}_{\theta _{0}}%
\frac{\widetilde{V}_{s}^{(n+k)}}{\left( \widetilde{V}_{n_{k+1}-1}^{(n+k)}%
\right) ^{2}}\widetilde{H}_{n_{k+1}-1} \\
&&+\{\widetilde{V}_{n_{k+1}-1}^{(n+k)},\widetilde{H}_{s}\}_{\theta _{0}}%
\frac{\widetilde{V}_{r}^{(n+k)}}{\widetilde{V}_{n_{k+1}-1}^{(n+k)}}%
\widetilde{H}_{n_{k+1}-1} \\
&&+\{\widetilde{V}_{r}^{(n+k)},\widetilde{H}_{n_{k+1}-1}\}_{\theta _{0}}%
\frac{\widetilde{V}_{s}^{(n+k)}}{\left( \widetilde{V}_{n_{k+1}-1}^{(n+k)}%
\right) ^{2}}\widetilde{H}_{n_{k+1}-1}+ \\
&&\{\widetilde{H}_{n_{k+1}-1},\widetilde{V}_{s}^{(n+k)}\}_{\theta _{0}}\frac{%
\widetilde{V}_{r}^{(n+k)}}{\widetilde{V}_{n_{k+1}-1}^{(n+k)}}\widetilde{H}%
_{n_{k+1}-1} \\
&=&0.
\end{eqnarray*}
\end{proof}

\subsection{Quasi-bi-Hamiltonian representation}

Now we demonstrate that the subspace span by $\ (d\widetilde{H}_{1},...,d%
\widetilde{H}_{n+k})$ is invariant with respect to $N^{\ast }$ (Theorem 3).

\begin{theorem}
Hamiltonian functions $\widetilde{H}_{r}$ belong to the following quasi-bi-
Hamiltonian chain
\begin{equation*}
d\widetilde{H}_{r+1}=N^{\ast }d\widetilde{H}_{r}+\alpha _{r}^{0}d\widetilde{H%
}_{1}+\sum_{i=1}^{k}\alpha _{r}^{n_{i}}d\widetilde{H}_{n_{i}+1}
\end{equation*}%
\begin{equation}
\Updownarrow  \tag{6.23}  \label{4.15}
\end{equation}%
\begin{equation*}
\theta _{0}d\widetilde{H}_{r+1}=\theta _{1}d\widetilde{H}_{r}+\alpha
_{r}^{0}\theta _{0}d\widetilde{H}_{1}+\sum_{i=1}^{k}\alpha
_{r}^{n_{i}}\theta _{0}d\widetilde{H}_{n_{i}+1},
\end{equation*}%
where $\alpha _{r}^{s}=\widetilde{V}_{r}^{(n+k)-s}.$Of course formula (\ref%
{4.15}) works separately for the geodesic Hamiltonians $\widetilde{E}_{r}$
and potentials $\widetilde{V}_{r}$
\begin{equation}
d\widetilde{E}_{r+1}=N^{\ast }d\widetilde{E}_{r}+\alpha _{r}^{0}d\widetilde{E%
}_{1}+\sum_{i=1}^{k}\alpha _{r}^{ni}d\widetilde{E}_{n_{i}+1},  \tag{6.24a}
\label{4.16a}
\end{equation}%
\begin{equation}
d\widetilde{V}_{r+1}=L^{\ast }d\widetilde{V}_{r}+\alpha _{r}^{0}d\widetilde{V%
}_{1}+\sum_{i=1}^{k}\alpha _{r}^{n_{i}}d\widetilde{V}_{n_{i}+1}.  \tag{6.24b}
\label{4.16b}
\end{equation}
\end{theorem}

\begin{proof}
The proof is inductive. Assuming that (\ref{4.15}) and (\ref{4.16b}) are
valid for the $k$-hole case, with holes at positions $\xi
^{(n+k)-n_{i}},i=1,...,k,$ we prove the validity of (\ref{4.15}) for the $%
(k+1)$-hole case with an extra hole at the position $\xi ^{(n+k+1)-n_{k+1}}.$
We take the $k$-hole and $(k+1)$-hole separability conditions as in (\ref%
{4.17cc}) and (\ref{4.18cc}). Our assumptions are as follows
\begin{equation}
d\widetilde{H}_{r+1}=N^{\ast }d\widetilde{H}_{r}+\alpha _{r}^{0}d\widetilde{H%
}_{1}+\sum_{i=1}^{k}\alpha _{r}^{n_{i}}d\widetilde{H}_{n_{i}+1},  \tag{6.25a}
\label{4.26cc}
\end{equation}%
\begin{equation}
d\alpha _{r+1}^{n_{i}}=L^{\ast }d\alpha _{r}^{n_{i}}+\alpha _{r}^{0}d\alpha
_{1}^{n_{i}}+\sum_{j=1}^{k}\alpha _{r}^{n_{j}}d\alpha _{n_{j}+1}^{n_{i}},
\tag{6.25b}  \label{4.26dd}
\end{equation}%
where $i,j=0,...,k,n_{0}=0.$ Notice that (\ref{4.26dd}) is a particular case
of the condition (\ref{0.22}) and follows from the fact that in a St\"{a}%
ckel case $\alpha _{ij}$ are particular basic separable potentials. From (%
\ref{4.22cc}) and (\ref{4.17bb}) we find the following useful relations
\begin{equation}
\overline{H}_{r+1}=\widetilde{H}_{r}-\frac{\alpha _{r}^{0}}{\alpha
_{n_{k+1}-1}^{0}}\widetilde{H}_{n_{k+1}-1},\ \ \overline{H}_{1}=-\frac{1}{%
\alpha _{n_{k+1}-1}^{0}}\widetilde{H}_{n_{k+1}-1},\ \ r=1,...,n+k,
\tag{6.26a}  \label{4.27cc}
\end{equation}%
\begin{equation}
\overline{\alpha }_{r+1}^{n_{i}+1}=\alpha _{r}^{n_{i}}-\frac{\alpha _{r}^{0}%
}{\alpha _{n_{k+1}-1}^{0}}\alpha _{n_{k+1}-1}^{n_{i}},\ \ \ \ r=1,...,n+k,
\tag{6.26b}  \label{4.27bb}
\end{equation}%
\begin{equation}
\overline{\alpha }_{r+1}^{n_{k+1}}=\frac{\alpha _{r}^{0}}{\alpha
_{n_{k+1}-1}^{0}},  \tag{6.26c}  \label{4.27aa}
\end{equation}%
\begin{equation}
\overline{\alpha }_{r+1}^{0}=\alpha _{r+1}^{0}+\frac{\alpha _{r}^{0}}{\alpha
_{n_{k+1}-1}^{0}}\sum_{i=1}^{k}\alpha _{n_{k+1}-1}^{n_{i}}\alpha
_{n_{i}+1}^{0}-\frac{\alpha _{r}^{0}}{\alpha _{n_{k+1}-1}^{0}}\alpha
_{n_{k+1}}^{0}-\sum_{i=1}^{k}\alpha _{r}^{n_{i}}\alpha _{n_{i}+1}^{0},
\tag{6.26d}  \label{4.27dd}
\end{equation}%
where $\widetilde{H}_{n_{i}}=\alpha _{n_{i}}^{0}=0,i=1,...,k,\alpha
_{n_{i}}^{n_{j}}=\delta _{ij}.$ Expressing $\overline{H},\overline{\alpha }$
by $\widetilde{H},\alpha $ from (\ref{4.27cc})-(\ref{4.27dd}) and using
assumptions (\ref{4.26cc}), (\ref{4.26dd}) we find, after long but
straightforward calculations, that
\begin{eqnarray*}
&&N^{\ast }d\overline{H}_{r}+\overline{\alpha }_{r}^{0}d\overline{H}%
_{1}+\sum_{i=1}^{k}\overline{\alpha }_{r}^{n_{i}+1}d\overline{H}_{n_{i}+2}+%
\overline{\alpha }_{r}^{n_{k+1}}d\overline{H}_{n_{k+1}+1} \\
&=&d\widetilde{H}_{r}-\frac{\widetilde{H}_{n_{k+1}-1}}{\alpha
_{n_{k+1}-1}^{0}}d\alpha _{r}^{0}-\alpha _{r}^{0}d\left( \frac{\widetilde{H}%
_{n_{k+1}-1}}{\alpha _{n_{k+1}-1}^{0}}\right) \\
&=&d\left( \widetilde{H}_{r}-\frac{\alpha _{r}^{0}}{\alpha _{n_{k+1}-1}^{0}}%
\widetilde{H}_{n_{k+1}-1}\right) =d\overline{H}_{r+1}.
\end{eqnarray*}
\end{proof}

Using the known relation for the Benenti potentials
\begin{equation*}
K_{r+1}=\sum_{k=0}^{r}\rho _{k}L^{r-k},
\end{equation*}%
one finds from (\ref{4.26dd}) that
\begin{eqnarray}
d\widetilde{V}_{r+1} &=&[(L^{\ast })^{r}+\alpha _{1}^{0}(L^{\ast
})^{r-1}+...+\alpha _{r}^{0}]d\widetilde{V}_{1}+[\alpha _{1}^{n_{1}}(L^{\ast
})^{r-1}+...+\alpha _{r}^{n_{1}}]  \notag \\
&&d\widetilde{V}_{n_{1}+1}+...+[\alpha _{1}^{n_{k}}(L^{\ast
})^{r-1}+...+\alpha _{r}^{n_{k}}]d\widetilde{V}_{n_{k}+1}  \notag \\
&=&A_{r}^{0}d\widetilde{V}_{1}+A_{r}^{n_{1}}d\widetilde{V}%
_{n_{1}+1}+...+A_{r}^{n_{k}}d\widetilde{V}_{n_{k}+1},  \TCItag{6.27}
\label{4.17}
\end{eqnarray}%
where
\begin{equation*}
A_{r}^{0}=\frac{1}{\varphi }\left\vert
\begin{array}{cccc}
K_{r} & K_{r-1} & \cdots & K_{r-k} \\
\rho _{n_{1}} & \rho _{n_{1}-1} & \cdots & \rho _{n_{1}-k} \\
\cdots & \cdots & \cdots & \cdots \\
\rho _{n_{k}} & \rho _{n_{k}-1} & \cdots & \rho _{n_{k}-k}%
\end{array}%
\right\vert ,
\end{equation*}%
\begin{equation*}
A_{r}^{n_{i}}=(-1)^{i+1}\frac{1}{\varphi }\left\vert
\begin{array}{ccc}
K_{r-1} & \cdots & K_{r-k} \\
\rho _{n_{1}-1} & \cdots & \rho _{n_{1}-k} \\
\cdots & \cdots & \cdots \\
\rho _{n_{k}-1} & \cdots & \rho _{n_{k}-k}%
\end{array}%
\right\vert
\end{equation*}%
with the row $(\rho _{n_{i}-1},...,\rho _{n_{i}-k})$ missing in $%
A_{r}^{n_{i}}$ and again $K_{i}$ in the calculation of determinants are
treated as symbols not matrices.

As in the one-hole case (\ref{3.17c}), the $d_{p}$ part of of (\ref{3.16a})
gives us the analog of formulae (\ref{2.7b}) for the $k$-hole case
\begin{equation}
\widetilde{K}_{r+1}=L\widetilde{K}_{r}+\alpha _{r}^{0}I+\sum_{i=1}^{k}\alpha
_{r}^{n_{i}}\widetilde{K}_{n_{i}+1}.  \tag{6.28}  \label{4.17b}
\end{equation}%
Then, from (\ref{4.17b}) we find that for $\widetilde{L}$ function (\ref%
{3.17d}) the following relation holds
\begin{equation}
\widetilde{L}=\widetilde{E}_{2}-\alpha _{1}^{0}\widetilde{E}%
_{1}-\sum_{i=1}^{k}\alpha _{r}^{n_{i}}\widetilde{E}_{n_{i}+1},  \tag{6.29}
\label{6.29}
\end{equation}%
hence
\begin{eqnarray}
\{\widetilde{L},\widetilde{E}_{1}\}_{\theta _{0}} &=&\{\widetilde{E}%
_{1},\alpha _{1}^{0}\}_{\theta _{0}}\widetilde{E}_{1}+\sum_{i=1}^{k}\{%
\widetilde{E}_{1},\alpha _{1}^{n_{i}}\}_{\theta _{0}}\widetilde{E}_{n_{1}+1}
\notag \\
&=&\kappa ^{0}\widetilde{E}_{1}+\sum_{i=1}^{k}\kappa ^{n_{i}}\widetilde{E}%
_{n_{i}+1}.  \TCItag{6.30}  \label{6.30}
\end{eqnarray}%
so obviously, $L$ is not a conformal Killing tensor for $\widetilde{G}$
given by (\ref{4.9}).

Notice, that although the number $k$ can be arbitrary large, nevertheless,
the maximal number of nonvanishing terms $\alpha _{r}^{n_{i}}d\widetilde{H}%
_{n_{i}+1}$ in (\ref{4.15}) is lower or equal to $n$. In fact, if $n_{i}$
and $n_{i+1}$ are two successive numbers, i.e. $n_{i+1}=n_{i}+1,$ then $%
\alpha _{r}^{n_{i}}d\widetilde{H}_{n_{i}+1}=\alpha _{r}^{n_{i}}d\widetilde{H}%
_{n_{i+1}}=0,$ as from construction $\widetilde{H}_{n_{i+1}}=0.$ Hence, for
a string of successive numbers $n_{i}-s_{i}+1,n_{i}-s_{i}+2,...,n_{i}$, only
the term with $n_{i}$ is nonzero in formula (\ref{4.15}), as
\begin{equation*}
\widetilde{H}_{n_{i}-s_{i}+1}=\widetilde{H}_{n_{i}-s_{i}+2}=...=\widetilde{H}%
_{n_{i}}=0.
\end{equation*}

Thus, assume that in the sequence
\begin{equation*}
\widetilde{H}_{1}\xi ^{(n+k)-1}+\widetilde{H}_{2}\xi ^{(n+k)-2}+...+%
\widetilde{H}_{n+k}
\end{equation*}%
we have $l$ strings of holes, where the $i$-th string has $s_{i}$ holes and $%
s_{1}+...+s_{l}=k.$ Then, the quasi-bi-Hamiltonian chain (\ref{4.15}) takes
the form
\begin{equation}
d\widetilde{H}_{r+1}=N^{\ast }d\widetilde{H}_{r}+\alpha _{r}^{n_{0}}d%
\widetilde{H}_{1}+\alpha _{r}^{n_{1}}d\widetilde{H}_{n_{1}+1}+...+\alpha
_{r}^{n_{l}}d\widetilde{H}_{n_{l}+1},  \tag{6.31}  \label{4.18}
\end{equation}%
where $n_{0}=0.$

\subsection{\protect\bigskip Gel'fand-Zakharevich bi-Hamiltonian
representation}

Now we lift our quasi-bi-Hamiltonian representation, constructed in a
previous subsection, into a GZ form. Adding $l+1$ Casimir coordinates $c_{i}$
(with respect to $\theta _{0}$) and extending Hamiltonians $\widetilde{H}%
_{r} $ to the form affine in $c_{i}$
\begin{equation}
\widetilde{h}_{r}=\widetilde{H}_{r}+\sum_{i=1}^{l+1}\alpha
_{r}^{n_{i-1}}c_{i},  \tag{6.32}  \label{4.19}
\end{equation}%
\ one can transform the quasi-bi-Hamiltonian chain on $T^{\ast }Q$ into a
bi-Hamiltonian chain on $T^{\ast }Q\times \underset{l+1}{\underbrace{\mathbb{%
R}\times ...\times \mathbb{R}}},$ being a composition of $l+1$
bi-Hamiltonian sub-chains
\begin{equation}
\begin{array}{l}
\ \ \ \ \ \ \ \ \widetilde{\pi }_{0}d\widetilde{h}_{n_{i}}=0 \\
\ \ \ \ \ \widetilde{\pi }_{0}d\widetilde{h}_{n_{i}+1}=\widetilde{\pi }_{1}d%
\widetilde{h}_{n_{i}} \\
\,\,\,\,\,\,\,\,\,\,\,\ \ \ \ \
\,\,\,\,\,\,\,\,\,\,\,\,\,\,\,\,\,\,\,\,\,\vdots \\
\widetilde{\pi }_{0}d\widetilde{h}_{n_{i+1}-s_{i+1}}=\widetilde{\pi }_{1}d%
\widetilde{h}_{n_{i+1}-s_{i+1}-1} \\
\qquad \qquad \,\,\ \,\ \,0=\widetilde{\pi }_{1}d\widetilde{h}%
_{n_{i+1}-s_{i+1}},%
\end{array}
\tag{6.33}  \label{4.20b}
\end{equation}%
where $i=0,1,...,l,\ \widetilde{h}_{0}=c_{1},\ \ \widetilde{h}%
_{n_{i}}=c_{i+1},\ \ \widetilde{h}_{n_{l+1}-s_{l+1}}=\ \widetilde{h}%
_{n_{l+1}}=\widetilde{h}_{n+k}$ and
\begin{eqnarray*}
\widetilde{\pi }_{0} &=&\left(
\begin{array}{c|c}
\theta _{0} & 0\ \ 0\ \ ...\ \ 0 \\ \hline
\begin{array}{c}
0 \\
0 \\
... \\
0%
\end{array}
& 0%
\end{array}%
\right) \text{, } \\
&& \\
\text{\ }\widetilde{\pi }_{1} &=&\left(
\begin{array}{c|c}
\theta _{1} & \theta _{0}d\widetilde{h}_{1}\ \ \ \theta _{0}d\widetilde{h}%
_{n_{1}+1}\ \ ...\ \ \theta _{0}d\widetilde{h}_{n_{l}+1} \\ \hline
\begin{array}{c}
-(\theta _{0}d\widetilde{h}_{1})^{T} \\
-(\theta _{0}d\widetilde{h}_{n_{1}+1})^{T} \\
... \\
-(\theta _{0}d\widetilde{h}_{n_{l}+1})^{T}%
\end{array}
& 0%
\end{array}%
\right) .
\end{eqnarray*}%
Again $\widetilde{\pi }_{0}$ and $\widetilde{\pi }_{1}$ are compatible
Poisson structures as, according to Lemma \ref{l3}, $\widetilde{\pi }_{1}$
takes the form
\begin{equation*}
\widetilde{\pi }_{1}=\widetilde{\pi }_{1D}+\sum_{i=1}^{l+1}X_{i}\wedge Z_{i},
\end{equation*}%
where $Z_{i}=\frac{\partial }{\partial c_{i}},X_{i}=\widetilde{\pi }_{0}d%
\widetilde{h}_{n_{i}+1},L_{Z_{i}}\widetilde{\pi }_{0}=0.$ The reduction of $%
l+1$ chains onto the symplectic leaf $c_{1}=...=c_{l+1}=0$ of $\widetilde{%
\pi }_{0}$ along the distribution $\mathcal{Z}=(\frac{\partial }{\partial
c_{1}},...,\frac{\partial }{\partial c_{l+1}})$ reconstructs immediately the
quasi-bi-Hamiltonian chain (\ref{4.15}).

If we introduce the Poisson pencil $\widetilde{\pi }_{\xi }=\widetilde{\pi }%
_{1}-\xi \widetilde{\pi }_{0},$ $l+1$ chains (\ref{4.20b}) can be written in
a compact form
\begin{equation}
\widetilde{\pi }_{\xi }d\widetilde{h}(\xi )=0,\ \ \widetilde{h}(\xi )=(%
\widetilde{h}^{(1)}(\xi ),\widetilde{h}^{(2)}(\xi ),...,\widetilde{h}%
^{(l+1)}(\xi ))^{T},  \tag{6.34}  \label{4.23}
\end{equation}%
where
\begin{equation}
\ \widetilde{h}^{(i)}(\xi )=\widetilde{h}_{n_{i}}\xi
^{n_{i+1}-n_{i}-s_{i+1}}+...+\widetilde{h}_{n_{i+1}-s_{i+1}},\ \ \
i=1,2,...,l+1.  \tag{6.34a}  \label{4.23a}
\end{equation}%
The quasi-bi-Hamiltonian chain (\ref{3.14a}) takes the form
\begin{equation*}
\widetilde{\theta }_{\xi }d\widetilde{H}(\xi )+\widetilde{\alpha }_{\xi }%
\widetilde{\theta }_{0}\widetilde{H}_{(1)}(q,p)=0
\end{equation*}%
\begin{equation}
\Updownarrow  \tag{6.35}  \label{4.24}
\end{equation}%
\begin{equation*}
(N^{\ast }-\xi I)d\widetilde{H}(\xi )+\widetilde{\alpha }(\xi )d\widetilde{H}%
_{(1)}(q,p)=0,
\end{equation*}%
where%
\begin{equation}
\begin{array}{l}
\widetilde{H}^{(i)}(\xi )=\widetilde{H}_{n_{i}+1}\xi
^{n_{i+1}-n_{i}-s_{i+1}+1}+...+\widetilde{H}_{n_{i+1}-s_{i+1}} \\
\\
\widetilde{H}_{(1)}=(\widetilde{H}_{1},\widetilde{H}_{n_{1}+1},...,%
\widetilde{H}_{n_{l}+1})^{T}%
\end{array}%
\ \ \ \ i=1,2,...,l+1,  \tag{6.35a}  \label{4.34}
\end{equation}%
and
\begin{equation}
\widetilde{\alpha }(\xi )_{j}^{i}=\frac{\partial \widetilde{h}^{(j)}(\xi )}{%
\partial c_{i}},\ \ i,j=1,...,l+1.  \tag{6.35b}  \label{4.24b}
\end{equation}

The St\"{a}ckel separation conditions for extended Hamiltonians $\widetilde{h%
}_{r}$ take the form (\ref{3.18bb})
\begin{equation*}
c_{1}\xi ^{(n+k)}+\widetilde{h}_{1}\xi ^{(n+k)-1}+...+\widetilde{h}%
_{n_{1}-s_{1}-1}\xi ^{(n+k)-(n_{1}-s_{1}-1)}\ \ \ \ \ \ \ \ \ \ \ \ \ \ \ \
\ \ \ \ \ \ \ \ \ \ \ \ \ \ \ \ \ \ \ \ \ \ \ \ \ \
\end{equation*}%
\begin{equation}
\ \ \ \ +c_{2}\xi ^{(n+k)-n_{1}}+\widetilde{h}_{n_{1}+1}\xi
^{(n+k)-(n_{1}+1)}+...+\widetilde{h}_{n+1}\ \ \ \ \ \ \ \ \ \ \ \ \ \ \ \ \
\ \ \ \ \ \ \ \ \ \ \ \ \ \ \ \ \ \ \ \ \ \ \ \ \ \ \ \ \ \ \ \ \ \ \ \ \ \
\   \tag{6.36}  \label{4.25}
\end{equation}%
\begin{equation*}
=\sum_{i=1}^{l+1}\xi ^{(n+k)-n_{i+1}-s_{i+1}}\ \widetilde{h}^{(i)}(\xi )=%
\frac{1}{2}f(\xi )\mu ^{2}+\gamma (\xi ),\ \ \ \ (\xi ,\mu )=(\lambda
_{i},\mu _{i}),\ i=1,...,n.
\end{equation*}

\begin{remark}
Systems considered in this paper, although obtained through the deformation
procedure on the level of Hamiltonian functions, are far from being trivial
generalizations of Benenti systems. There is no obvious relations between
solutions of a given Benenti system and all its deformations. In each case
we have a different inverse Jacobi problem to solve. Notice, that the common
feature of appropriate deformed systems is the same set of separated
coordinates, determined by the related Benenti system.
\end{remark}

\section{Examples}

\subsection{Henon-Heiles system}

Let us consider the integrable case of the Henon-Heiles system
\begin{eqnarray}
(q^{1})_{tt} &=&-3(q^{1})^{2}-\frac{1}{2}(q^{2})^{2}+c  \notag \\
(q^{2})_{tt} &=&-q^{1}q^{2},  \TCItag{7.1}  \label{5.0}
\end{eqnarray}%
with the corresponding Lagrangian
\begin{equation}
L=\frac{1}{2}(q^{1})_{t}^{2}+\frac{1}{2}(q^{2})_{t}^{2}-(q^{1})^{3}-\frac{1}{%
2}q^{1}(q^{2})^{2}+cq^{1}.  \tag{7.1a}  \label{7.1a}
\end{equation}%
The bi-Hamiltonian chain is of the following form \cite{bw}
\begin{equation}
\begin{array}{l}
\pi _{0}dh_{0}=0 \\
\pi _{0}dh_{1}=X_{1}=\pi _{1}dh_{0} \\
\pi _{0}dh_{2}=X_{2}=\pi _{1}dh_{1} \\
\qquad \qquad \,\,\,\,\,0=\pi _{1}dh_{2,}\,\,%
\end{array}
\tag{7.2}  \label{5.1}
\end{equation}%
where
\begin{eqnarray*}
h_{0} &=&c, \\
h_{1} &=&\frac{1}{2}p_{1}^{2}+\frac{1}{2}p_{2}^{2}+(q^{1})^{3}+\frac{1}{2}%
q^{1}(q^{2})^{2}-cq^{1} \\
&=&E_{1}+V_{1}(q)+\rho _{1}(q)\ c=H_{1}+\rho _{1}(q)c, \\
h_{2} &=&\frac{1}{2}q^{2}p_{1}p_{2}-\frac{1}{2}q^{1}p_{2}^{2}+\frac{1}{16}%
(q^{2})^{4}+\frac{1}{4}(q^{1})^{2}(q^{2})^{2}-\frac{1}{4}c(q^{2})^{2} \\
&=&E_{2}+V_{2}(q)+\rho _{2}(q)\ c=H_{2}+\rho _{2}(q)c,
\end{eqnarray*}%
\begin{equation*}
\pi _{0}=\left(
\begin{array}{c|c}
\theta _{0} & 0 \\ \hline
0 & 0%
\end{array}%
\right) \text{ \ , \ }\pi _{1}=\left(
\begin{array}{c|c}
\theta _{1} & \theta _{0}dh_{1} \\ \hline
-(\theta _{0}dh_{1})^{T} & 0%
\end{array}%
\right) ,
\end{equation*}%
\begin{equation*}
\theta _{0}=\left(
\begin{array}{cccc}
0 & 0 & 1 & 0 \\
0 & 0 & 0 & 1 \\
-1 & 0 & 0 & 0 \\
0 & -1 & 0 & 0%
\end{array}%
\right) ,\ \theta _{1}=\left(
\begin{array}{cccc}
0 & 0 & q^{1} & \frac{1}{2}q^{2} \\
0 & 0 & \frac{1}{2}q^{2} & 0 \\
-q^{1} & -\frac{1}{2}q^{2} & 0 & \frac{1}{2}p_{2} \\
-\frac{1}{2}q^{2} & 0 & -\frac{1}{2}p_{2} & 0%
\end{array}%
\right) .
\end{equation*}%
Notice that
\begin{equation*}
G=\left(
\begin{array}{cc}
1 & 0 \\
0 & 1%
\end{array}%
\right) ,\ L=\left(
\begin{array}{cc}
q^{1} & \frac{1}{2}q^{2} \\
\frac{1}{2}q^{2} & 0%
\end{array}%
\right) ,\ K_{2}=\left(
\begin{array}{cc}
0 & \frac{1}{2}q^{2} \\
\frac{1}{2}q^{2} & -q^{1}%
\end{array}%
\right) .
\end{equation*}%
The quasi-bi-Hamiltonian chain is
\begin{equation}
dH_{r+1}=N^{\ast }dH_{r}+\rho _{r}dH_{1},\ \ \rho _{r}=\frac{\partial h_{r}}{%
\partial c},\ \ \ r=1,2,  \tag{7.3}  \label{5.6}
\end{equation}%
where $N^{\ast }=\theta _{0}^{-1}\theta _{1}.$

The transformation to separated coordinates $(\lambda ,\mu )$ takes the form
\begin{equation*}
q^{1}=\lambda ^{1}+\lambda ^{2},\ \ \ q^{2}=2\sqrt{-\lambda ^{1}\lambda ^{2}}%
,
\end{equation*}%
\begin{equation}
p_{1}=\frac{\lambda ^{1}\mu _{1}}{\lambda ^{1}-\lambda ^{2}}+\frac{\lambda
^{2}\mu _{2}}{\lambda ^{2}-\lambda ^{1}},\ \ \ p_{2}=\sqrt{-\lambda
^{1}\lambda ^{2}}\left( \frac{\mu _{1}}{\lambda ^{1}-\lambda ^{2}}+\frac{\mu
_{2}}{\lambda ^{2}-\lambda ^{1}}\right) ,  \tag{7.4}  \label{5.7}
\end{equation}%
and the separability conditions for $H_{i}$ respectively $h_{i}$ are
reconstructed from separation curves
\begin{equation}
H_{1}\xi +H_{2}=\frac{1}{2}\xi \mu ^{2}+\xi ^{4},  \tag{7.5a}  \label{5.8a}
\end{equation}%
\begin{equation}
c\xi ^{2}+h_{1}\xi +h_{2}=\frac{1}{2}\xi \mu ^{2}+\xi ^{4}.  \tag{7.5b}
\label{5.8b}
\end{equation}

\textbf{1-hole deformation of the Henon-Heiles.}

The only possibility is $n_{1}=2.$ Then,
\begin{equation*}
\widetilde{G}=-\frac{1}{\rho _{1}}G=\left(
\begin{array}{cc}
\frac{1}{q^{1}} & 0 \\
0 & \frac{1}{q^{1}}%
\end{array}%
\right) ,\widetilde{K}_{2}=0,\widetilde{K}_{3}=-K_{2}^{2}=\left(
\begin{array}{cc}
-\frac{1}{4}(q^{2})^{2} & \frac{1}{2}q^{1}q^{2} \\
\frac{1}{2}q^{1}q^{2} & -\frac{1}{4}(q^{2})^{2}-(q^{1})^{2}%
\end{array}%
\right) ,
\end{equation*}%
\begin{equation*}
\widetilde{V}_{1}=-\frac{1}{\rho _{1}}V_{1},\ \ \widetilde{V}_{2}=0,\ \
\widetilde{V}_{3}=V_{2}-\frac{\rho _{2}}{\rho _{1}}V_{1},
\end{equation*}%
and%
\begin{equation*}
\alpha _{r}=\rho _{r}-\rho _{r-1}\rho _{2}\rho _{1}^{-1},\ \ \ \beta
_{r}=\rho _{r-1}\rho _{1}^{-1}.
\end{equation*}%
The quasi-bi-Hamiltonian chain takes the form
\begin{equation}
d\widetilde{H}_{r+1}=N^{\ast }d\widetilde{H}_{r}+\alpha _{r}d\widetilde{H}%
_{1}+\beta _{r}d\widetilde{H}_{n_{1}+1},\ r=1,2,3.  \tag{7.6}  \label{5.10}
\end{equation}%
while two bi-Hamiltonian sub-chains are
\begin{equation}
\begin{array}{ccc}
\begin{array}{l}
\widetilde{\pi }_{0}d\widetilde{h}_{0}=0 \\
\widetilde{\pi }_{0}d\widetilde{h}_{1}=\widetilde{X}_{1}=\widetilde{\pi }%
_{1}d\widetilde{h}_{0} \\
\,\ \ \ \ \ \ \ \ \ \ \ \ \ \ \,\,0=\widetilde{\pi }_{1}d\widetilde{h}%
_{1}\,\,%
\end{array}
&  &
\begin{array}{l}
\widetilde{\pi }_{0}d\widetilde{h}_{2}=0 \\
\widetilde{\pi }_{0}d\widetilde{h}_{3}=\widetilde{X}_{3}=\widetilde{\pi }%
_{1}d\widetilde{h}_{2} \\
\qquad \qquad \,\,\,\ 0=\widetilde{\pi }_{1}d\widetilde{h}_{3}\,\,,%
\end{array}%
\end{array}
\tag{7.7}  \label{5.9}
\end{equation}%
where
\begin{eqnarray*}
\widetilde{h}_{0} &=&c_{1}, \\
\widetilde{h}_{1} &=&\frac{1}{2}\frac{1}{q^{1}}p_{1}^{2}+\frac{1}{2}\frac{1}{%
q^{1}}p_{2}^{2}+(q^{1})^{2}+\frac{1}{2}(q^{2})^{2}-c_{1}[q_{1}+\frac{1}{4}%
\frac{1}{q^{1}}(q^{2})^{2}]-c_{2}\frac{1}{q^{1}} \\
&=&\widetilde{E}_{1}+\widetilde{V}q_{1}^{1}(q)+\alpha _{1}(q)\ c_{1}+\beta
_{1}(q)c_{2}, \\
\widetilde{h}_{2} &=&c_{2}, \\
\widetilde{h}_{3} &=&-\frac{1}{8}\frac{q_{2}^{2}}{q^{1}}p_{1}^{2}+\frac{1}{2}%
q^{2}p_{1}p_{2}-\frac{1}{8}\frac{q_{2}^{2}}{q^{1}}p_{2}^{2}-\frac{1}{16}%
(q^{2})^{4}+\frac{1}{16}c_{1}\frac{(q^{2})^{4}}{q^{1}}+\frac{1}{4}c_{2}\frac{%
(q^{2})^{2}}{q^{1}} \\
&=&\widetilde{E}_{3}+\widetilde{V}_{3}(q)+\alpha _{3}(q)\ c_{1}+\beta
_{3}(q)c_{2},
\end{eqnarray*}%
and%
\begin{equation*}
\widetilde{\pi }_{0}=\left(
\begin{array}{c|c}
\theta _{0} & 0\ \ 0 \\ \hline
\begin{array}{c}
0 \\
0%
\end{array}
& 0%
\end{array}%
\right) \text{ \ , \ }\widetilde{\pi }_{1}=\left(
\begin{array}{c|c}
\theta _{1} & \theta _{0}d\widetilde{h}_{1}\ \ \ \theta _{0}d\widetilde{h}%
_{3} \\ \hline
\begin{array}{c}
-(\theta _{0}d\widetilde{h}_{1})^{T} \\
-(\theta _{0}d\widetilde{h}_{3})^{T}%
\end{array}
& 0%
\end{array}%
\right) .
\end{equation*}%
The transformation to separated coordinates is given by (\ref{5.7}) and the
separability conditions for $H_{i}$ respectively $h_{i}$ are represented by
separation curves
\begin{equation}
\widetilde{H}_{1}\xi ^{2}+\widetilde{H}_{3}=\frac{1}{2}\xi \mu ^{2}+\xi ^{4},
\tag{7.8a}  \label{7.7a}
\end{equation}%
\begin{equation}
c_{1}\xi ^{3}+\widetilde{h}_{1}\xi ^{2}+c_{2}\xi +\widetilde{h}_{3}=\frac{1}{%
2}\xi \mu ^{2}+\xi ^{4}.  \tag{7.8b}  \label{5.11}
\end{equation}%
The related Euler-Lagrange equations (\ref{1.2}) are
\begin{equation}
q^{1}q_{tt}^{1}+\frac{1}{2}(q_{t}^{1})^{2}-\frac{1}{2}%
(q_{t}^{2})^{2}=-2q^{1}+[1-\frac{1}{4}%
(q^{1})^{-2}(q^{2})^{2}]c_{1}-(q^{1})^{-2}c_{2},  \tag{7.9}  \label{7.8}
\end{equation}%
\begin{equation*}
q^{1}q_{tt}^{2}+q_{t}^{1}q_{t}^{2}=-q^{2}+\frac{1}{2}(q^{1})^{-1}q^{2}c_{1}.
\end{equation*}

\textbf{2-hole deformation of the Henon-Heiles.}

The only possibility is 2-hole string $n_{1}=2,n_{2}=n_{1}+1=3.$ Then,
\begin{equation*}
\widetilde{G}=\frac{1}{\rho _{1}^{2}-\rho _{2}}G=\left(
\begin{array}{cc}
\frac{4}{(q^{1})^{2}+q^{2}} & 0 \\
0 & \frac{4}{(q^{1})^{2}+q^{2}}%
\end{array}%
\right) ,\widetilde{K}_{2}=\widetilde{K}_{3}=0,
\end{equation*}%
\begin{equation*}
\widetilde{K}_{4}=K_{2}^{3}=\left(
\begin{array}{cc}
-\frac{1}{4}q^{1}(q^{2})^{2} & \frac{1}{8}(q^{2})^{3}+\frac{1}{2}%
(q^{1})^{2}+q^{2} \\
\frac{1}{8}(q^{2})^{3}+\frac{1}{2}(q^{1})^{2}+q^{2} & -\frac{1}{4}%
q^{1}(q^{2})^{2}-(q^{1})^{3}%
\end{array}%
\right) ,
\end{equation*}%
\begin{equation*}
\widetilde{V}_{1}=-\frac{1}{\rho _{1}^{2}-\rho _{2}}V_{1},\ \ \widetilde{V}%
_{2}=\widetilde{V}_{3}=0,\ \ \widetilde{V}_{4}=V_{2}-\frac{\rho _{1}\rho _{2}%
}{\rho _{1}^{2}-\rho _{2}}V_{1},
\end{equation*}%
\begin{equation*}
\alpha _{1}=\rho _{1}-\frac{\rho _{2}\rho _{1}}{\rho _{1}^{2}-\rho _{2}},\ \
\alpha _{2}=\alpha _{3}=0,\ \ \alpha _{4}=\frac{\rho _{2}^{3}}{\rho
_{1}^{2}-\rho _{2}},
\end{equation*}%
\begin{equation*}
\beta _{1}=-\frac{1}{\rho _{1}^{2}-\rho _{2}},\ \ \beta _{2}=\beta _{3}=0,\
\ \beta _{4}=\frac{\rho _{2}\rho _{1}}{\rho _{1}^{2}-\rho _{2}}
\end{equation*}%
and two bi-Hamiltonian chain are
\begin{equation}
\begin{array}{ccc}
\begin{array}{l}
\widetilde{\pi }_{0}d\widetilde{h}_{0}=0 \\
\widetilde{\pi }_{0}d\widetilde{h}_{1}=\widetilde{X}_{1}=\widetilde{\pi }%
_{1}d\widetilde{h}_{0} \\
\qquad \qquad \,\,\,\,0=\widetilde{\pi }_{1}d\widetilde{h}_{1}\,\,%
\end{array}
&  &
\begin{array}{l}
\widetilde{\pi }_{0}d\widetilde{h}_{3}=0 \\
\widetilde{\pi }_{0}d\widetilde{h}_{4}=\widetilde{X}_{4}=\widetilde{\pi }%
_{1}d\widetilde{h}_{3} \\
\qquad \qquad \,\,\,\,0=\widetilde{\pi }_{1}d\widetilde{h}_{4}\,\,,%
\end{array}%
\end{array}
\tag{7.10}  \label{5.12}
\end{equation}%
where
\begin{eqnarray*}
\widetilde{h}_{0} &=&c_{1}, \\
\widetilde{h}_{1} &=&\frac{1}{4(q^{1})^{2}+q^{2}}%
[2p_{1}^{2}+2p_{2}^{2}+4(q^{1})^{3}+2q^{1}(q^{2})^{2}]-c_{1}q_{1}-\frac{%
4c_{2}}{4(q^{1})^{2}+q^{2}}, \\
\widetilde{h}_{2} &=&0, \\
\widetilde{h}_{3} &=&c_{2}, \\
\widetilde{h}_{4} &=&\frac{1}{4(q^{1})^{2}+q^{2}}[-\frac{1}{2}%
q^{1}(q^{2})^{2}p_{1}^{2}+(2(q^{1})^{2}q^{2}+\frac{1}{2}%
(q^{2})^{3})p_{1}p_{2}-(q^{1}(q^{2})^{2} \\
&&+2(q^{1})^{3})p_{2}^{2}+\frac{1}{16}(q^{2})^{6}-\frac{1}{16}%
c_{1}(q^{2})^{6}+c_{2}q^{1}(q^{2})^{2}],
\end{eqnarray*}%
\begin{equation*}
\widetilde{\pi }_{0}=\left(
\begin{array}{c|c}
\theta _{0} & 0\ \ 0 \\ \hline
\begin{array}{c}
0 \\
0%
\end{array}
& 0%
\end{array}%
\right) \text{,\ }\widetilde{\pi }_{1}=\left(
\begin{array}{c|c}
\theta _{1} & \theta _{0}d\widetilde{h}_{1}\ \ \ \theta _{0}d\widetilde{h}%
_{4} \\ \hline
\begin{array}{c}
-(\theta _{0}d\widetilde{h}_{1})^{T} \\
-(\theta _{0}d\widetilde{h}_{4})^{T}%
\end{array}
& 0%
\end{array}%
\right) .
\end{equation*}%
The transformation to separated coordinates is given by (\ref{5.7}) and the
separation curves for $H_{i}$ respectively $h_{i}$ are
\begin{equation}
\widetilde{H}_{1}\xi ^{3}+\widetilde{H}_{4}=\frac{1}{2}\xi \mu ^{2}+\xi ^{4},
\tag{7.11a}  \label{7.10a}
\end{equation}%
\begin{equation}
c_{1}\xi ^{4}+\widetilde{h}_{1}\xi ^{3}+c_{2}\xi +\widetilde{h}_{4}=\frac{1}{%
2}\xi \mu ^{2}+\xi ^{4}.  \tag{7.11b}  \label{5.14}
\end{equation}

\subsection{7-th order stationary KdV}

Let us consider the so-called first Newton representation of the
seventh-order stationary flow of the KdV hierarchy \cite{myPhysicaA},\cite{1}%
. It is a Lagrangian system of second order Newton equations
\begin{align}
q_{tt}^{1}& =-10(q^{1})^{2}+4q^{2}  \notag \\
q_{tt}^{2}& =-16q^{1}q^{2}+10(q^{1})^{3}+4q^{3}  \tag{7.12}  \label{7.11} \\
q_{tt}^{3}& =-20q^{1}q^{3}-8(q^{2})^{2}+30(q^{1})^{2}q^{2}-15(q^{1})^{4}+c
\notag
\end{align}%
with the corresponding Lagrangian
\begin{equation}
\mathcal{L}=q_{t}^{1}q_{t}^{3}+\tfrac{1}{2}%
(q_{t}^{2})^{2}+4q^{2}q^{3}-10(q^{1})^{2}q^{3}-8q^{1}(q^{2})^{2}+10(q^{1})^{3}q^{2}-3(q^{1})^{5}+cq^{1},
\tag{7.12a}  \label{7.11a}
\end{equation}%
so that it can be cast in a Hamiltonian form. In fact, the above system can
be represented as a bi-Hamiltonian vector field $X_{1}$ belonging to a
bi-Hamiltonian chain of the form
\begin{equation}
\begin{array}{l}
\pi _{0}dh_{0}=0 \\
\pi _{0}dh_{1}=X_{1}=\pi _{1}dh_{0} \\
\pi _{0}dh_{2}=X_{2}=\pi _{1}dh_{1} \\
\pi _{0}dh_{3}=X_{3}=\pi _{1}dh_{2} \\
\qquad \qquad \,\,\,\,\,0=\pi _{1}dh_{3,}\,\,%
\end{array}
\tag{7.13}  \label{7.15}
\end{equation}

\begin{align*}
h_{0}& =c \\
h_{1}& =p_{1}p_{3}+\tfrac{1}{2}%
p_{2}^{2}+10(q^{1})^{2}q^{3}-4q^{2}q^{3}+8q^{1}(q^{2})^{2}-10(q^{1})^{3}q^{2}+3(q^{1})^{5}-cq^{1}
\\
& =E_{1}+V_{1}(q)+\rho _{1}(q)\ c=H_{1}+\rho _{1}(q)c \\
h_{2}& =\tfrac{1}{2}q^{3}p_{3}^{2}-\tfrac{1}{2}q^{1}p_{2}^{2}+\tfrac{1}{2}%
q^{2}p_{2}p_{3}-\tfrac{1}{2}p_{1}p_{2}-\tfrac{1}{2}%
q^{1}p_{1}p_{3}+2(q^{1})^{2}(q^{2})^{2}+\tfrac{5}{2}(q^{1})^{4}q^{2} \\
& -\tfrac{5}{4}(q^{1})^{6}-2(q^{2})^{3}+(q^{3})^{2}-6q^{1}q^{2}q^{3}+\tfrac{1%
}{4}c(q^{1})^{2}+\tfrac{1}{2}cq^{2} \\
& =E_{2}+V_{2}(q)+\rho _{2}(q)c=H_{2}+\rho _{2}(q)c \\
h_{3}& =\tfrac{1}{8}(q^{2})^{2}p_{3}^{2}+\tfrac{1}{8}(q^{1})^{2}p_{2}^{2}+%
\tfrac{1}{8}p_{1}^{2}+\tfrac{1}{4}q^{1}p_{1}p_{2}+\tfrac{1}{4}%
q^{2}p_{1}p_{3}-\tfrac{1}{4}q^{1}q^{2}p_{2}p_{3} \\
& -\tfrac{1}{2}q^{3}p_{2}p_{3}-3(q^{1})^{3}(q^{2})^{2}+q^{1}(q^{2})^{3}+%
\tfrac{5}{4}(q^{1})^{5}q^{2}+2q^{1}(q^{3})^{2} \\
& +\tfrac{5}{4}(q^{1})^{4}q^{3}+(q^{2})^{2}q^{3}-(q^{1})^{2}q^{2}q^{3}-%
\tfrac{1}{4}cq^{1}q^{2}-\tfrac{1}{4}cq^{3} \\
& =E_{3}+V_{3}(q)+\rho _{3}(q)c=H_{3}+\rho _{3}(q)c,
\end{align*}%
with the corresponding operators $\pi _{0}$ and $\pi _{1}$
\begin{equation*}
\Pi _{0}=\left(
\begin{array}{c|c}
\theta _{0} & 0_{6\times 1} \\ \hline
0_{1\times 6} & 0%
\end{array}%
\right) \text{ \ \ , \ \ }\theta _{0}=\left(
\begin{array}{cc}
0_{3} & I_{3} \\
-I_{3} & 0_{3}%
\end{array}%
\right)
\end{equation*}%
\begin{equation*}
\Pi _{1}=\frac{1}{2}\left(
\begin{array}{c|c}
\begin{array}{cccccc}
0 & 0 & 0 & q^{1} & -1 & 0 \\
0 & 0 & 0 & q^{2} & 0 & -1 \\
0 & 0 & 0 & 2q^{3} & q^{2} & q^{1} \\
-q^{1} & -q^{2} & -2q^{3} & 0 & p_{2} & p_{3} \\
1 & 0 & -q^{2} & -p_{2} & 0 & 0 \\
0 & 1 & -q^{1} & -p_{3} & 0 & 0%
\end{array}
& 2X_{1} \\ \hline
-2X_{1}^{T} & 0%
\end{array}%
\right)
\end{equation*}%
and $X_{1}=\theta _{0}dh_{1}$, $d=\left( \frac{\partial }{\partial q},\frac{%
\partial }{\partial p}\right) ^{T}$, so that the operator $\theta _{1}$ in
the corresponding quasi-bi-Hamiltonian chain is of the form
\begin{equation*}
\theta _{1}=\frac{1}{2}\left(
\begin{array}{cccccc}
0 & 0 & 0 & q^{1} & -1 & 0 \\
0 & 0 & 0 & q^{2} & 0 & -1 \\
0 & 0 & 0 & 2q^{3} & q^{2} & q^{1} \\
-q^{1} & -q^{2} & -2q^{3} & 0 & p_{2} & p_{3} \\
1 & 0 & -q^{2} & -p_{2} & 0 & 0 \\
0 & 1 & -q^{1} & -p_{3} & 0 & 0%
\end{array}%
\right) .
\end{equation*}%
From the form of $h_{1}$ one can directly see that the inverse metric tensor
$G$ expressed in $(q,p)$ variables has in this example an anti-diagonal form
\begin{equation*}
G=\left(
\begin{array}{ccc}
0 & 0 & 1 \\
0 & 1 & 0 \\
1 & 0 & 0%
\end{array}%
\right)
\end{equation*}%
while the conformal Killing tensor $L$ has the form
\begin{equation*}
L=\frac{1}{2}\left(
\begin{array}{ccc}
q^{1} & -1 & 0 \\
q^{2} & 0 & -1 \\
2q^{3} & q^{2} & q^{1}%
\end{array}%
\right)
\end{equation*}%
and hence $K_{1}=I,\ A_{1}=G,\ \ $
\begin{equation*}
K_{2}=\frac{1}{2}\left(
\begin{array}{ccc}
-q^{1} & -1 & 0 \\
q^{2} & -2q^{1} & -1 \\
2q^{3} & q^{2} & -q^{1}%
\end{array}%
\right) ,\ K_{3}=\frac{1}{4}\left(
\begin{array}{ccc}
q^{2} & q^{1} & 1 \\
-q^{1}q^{2}-2q^{3} & (q^{1})^{2} & q^{1} \\
(q^{2})^{2} & -q^{1}q^{2}-2q^{3} & q^{2}%
\end{array}%
\right) ,
\end{equation*}%
\begin{equation*}
A_{2}=\frac{1}{2}\left(
\begin{array}{ccc}
0 & -1 & q^{1} \\
-1 & -2q^{1} & q^{2} \\
-q^{1} & q^{2} & 2q^{3}%
\end{array}%
\right) ,\ A_{3}=\frac{1}{4}\left(
\begin{array}{ccc}
1 & q^{1} & q^{2} \\
q^{1} & (q^{1})^{2} & -q^{1}q^{2}-2q^{3} \\
q^{2} & -q^{1}q^{2}-2q^{3} & (q^{2})^{2}%
\end{array}%
\right) .
\end{equation*}%
The quasi-bi-Hamiltonian chain is given by (\ref{5.6}) with $r=1,2,3,$ where
\begin{equation*}
\rho _{1}=-q^{1},\ \ \ \rho _{2}=\frac{1}{4}(q^{1})^{2}+\frac{1}{2}q^{2},\ \
\ \rho _{3}=-\frac{1}{4}q^{1}q^{2}-\frac{1}{4}q^{3}.
\end{equation*}%
The transformation $(\lambda ,\mu )\rightarrow (q,p)$ is constructed from
the relations
\begin{equation*}
q^{1}=\lambda ^{1}+\lambda ^{2}+\lambda ^{3},\ \ \frac{1}{4}(q^{1})^{2}+%
\frac{1}{2}q^{2}=\lambda ^{1}\lambda ^{2}+\lambda ^{1}\lambda ^{3}+\lambda
^{2}\lambda ^{3},\ \ \frac{1}{4}q^{1}q^{2}+\frac{1}{4}q^{3}=\lambda
^{1}\lambda ^{2}\lambda ^{3},
\end{equation*}%
(the explicit formulas are given in \cite{1}) and the separation curves for $%
H_{i}$ respectively $h_{i}$ are
\begin{equation}
H_{1}\xi ^{2}+H_{2}\xi +H_{3}=\frac{1}{8}\mu ^{2}+16\xi ^{7},  \tag{7.14a}
\label{7.21a}
\end{equation}%
\begin{equation}
c\xi ^{3}+h_{1}\xi ^{2}+h_{2}\xi +h_{3}=\frac{1}{8}\mu ^{2}+16\xi ^{7}.
\tag{7.14b}  \label{7.21b}
\end{equation}%
\begin{equation*}
\end{equation*}

\textbf{1-hole deformation.}

There are two admissible cases of one-hole deformation: $n_{1}=2,3.$ Here we
illustrate the case $n_{1}=3.$ The deformed Hamiltonians are the following
\begin{eqnarray}
\widetilde{H}_{1} &=&-\frac{1}{\rho _{2}}H_{2}=\frac{1}{2}p^{T}\widetilde{G}%
\ p+\widetilde{V}_{1},  \notag \\
\widetilde{H}_{2} &=&H_{1}-\frac{\rho _{1}}{\rho _{2}}H_{2}=\frac{1}{2}p^{T}%
\widetilde{A}_{2}\ p+\widetilde{V}_{2},  \notag \\
\widetilde{H}_{3} &=&0,  \TCItag{7.15}  \label{7.22} \\
\widetilde{H}_{4} &=&H_{3}-\frac{\rho _{3}}{\rho _{2}}H_{2}=\frac{1}{2}p^{T}%
\widetilde{A}_{4}\ p+\widetilde{V}_{4},  \notag
\end{eqnarray}%
where
\begin{equation*}
\widetilde{G}=-\frac{1}{\rho _{2}}K_{2}G=\frac{2}{(q^{1})^{2}+2q^{2}}\left(
\begin{array}{ccc}
0 & 1 & q^{1} \\
1 & \frac{1}{2}q^{1} & -q^{2} \\
q^{1} & -q^{2} & -\frac{1}{2}q^{3}%
\end{array}%
\right) ,\ \ \ \widetilde{A}_{1}=\widetilde{G},
\end{equation*}%
\begin{eqnarray*}
\widetilde{A}_{2} &=&-\frac{1}{\rho _{2}}(K_{2}^{2}-K_{1}K_{3})G \\
&=&-\frac{1}{(q^{1})^{2}+2q^{2}}\left(
\begin{array}{ccc}
0 & 2q^{1} & (q^{1})^{2}-2q^{2} \\
2q^{1} & 3(q^{1})^{2}-2q^{2} & -2q^{1}q^{2} \\
(q^{1})^{2}-2q^{2} & -2q^{1}q^{2} & -4q^{1}q^{3}%
\end{array}%
\right) ,\ \ \  \\
\widetilde{A}_{3} &=&0, \\
\ \widetilde{A}_{4} &=&\frac{1}{\rho _{2}}K_{3}^{2}G=\frac{1}{4}\frac{1}{%
(q^{1})^{2}+2q^{2}}
\end{eqnarray*}%
\begin{equation*}
\left(
\begin{array}{ccc}
(q^{1})^{2}+2q^{2} & (q^{1})^{3}-2q^{3} &
2(q^{2})^{2}-(q^{1})^{2}q^{2}-2q^{1}q^{3} \\
\ast & q^{1}[(q^{1})^{3}-2q^{1}q^{2}-4q^{3}] &
-(q^{1})^{3}q^{2}-2(q^{1})^{2}q^{3}-2q^{2}q^{3} \\
\ast & \ast &
2(q^{2})^{3}+(q^{1})^{2}(q^{2})^{2}+4q^{1}q^{2}q^{3}+4(q^{3})^{2}%
\end{array}%
\right)
\end{equation*}%
the respective potentials are
\begin{eqnarray*}
\widetilde{V}_{1} &=&\frac{%
5(q^{1})^{6}-8(q^{1})^{2}(q^{2})^{2}-10(q^{1})^{4}q^{2}+8(q^{2})^{3}-4(q^{3})^{2}+24q^{1}q^{2}q^{3}%
}{(q^{1})^{2}+2q^{2}}, \\
&& \\
\widetilde{V}_{2}
&=&[(q^{1})^{7}-3(q^{1})^{5}q^{2}-5(q^{1})^{4}q^{3}+2(q^{1})^{3}(q^{2})^{2}+4(q^{1})^{2}q^{2}q^{3}-4q^{1}(q^{2})^{3}
\\
&&-2q^{1}(q^{3})^{2}+4(q^{2})^{2}q^{3}]/[(q^{1})^{2}+2q^{2}], \\
\widetilde{V}_{3} &=&0, \\
\widetilde{V}_{4}
&=&[2(q^{1})^{5}(q^{2})^{2}-3(q^{1})^{3}(q^{2})^{3}+2(q^{1})^{3}(q^{3})^{2}-q^{1}q^{2}(q^{3})^{2}+4(q^{1})^{4}q^{2}q^{3}
\\
&&-5(q^{1})^{2}(q^{2})^{2}q^{3}+(q^{3})^{3}]/[(q^{1})^{2}+2q^{2}],
\end{eqnarray*}%
and (*) are elements making matrix symmetric. The quasi-bi-Hamiltonian chain
takes the form (\ref{5.10}) with $r=1,...,4$ and
\begin{equation*}
\alpha _{r}=\rho _{r}-\frac{\rho _{r-1}\rho _{3}}{\rho _{2}},\ \ \ \beta
_{r}=\frac{\rho _{r-1}}{\rho _{2}},
\end{equation*}%
while the related bi-Hamiltonian chain splits onto two sub-chains of the
form
\begin{equation}
\begin{array}{ccc}
\begin{array}{l}
\widetilde{\pi }_{0}d\widetilde{h}_{0}=0 \\
\widetilde{\pi }_{0}d\widetilde{h}_{1}=\widetilde{X}_{1}=\widetilde{\pi }%
_{1}d\widetilde{h}_{0} \\
\widetilde{\pi }_{0}d\widetilde{h}_{2}=\widetilde{X}_{2}=\widetilde{\pi }%
_{1}d\widetilde{h}_{1} \\
\qquad \qquad \,\,\,\,0=\widetilde{\pi }_{1}d\widetilde{h}_{2}\,\,%
\end{array}
&  &
\begin{array}{l}
\widetilde{\pi }_{0}d\widetilde{h}_{3}=0 \\
\widetilde{\pi }_{0}d\widetilde{h}_{4}=\widetilde{X}_{4}=\widetilde{\pi }%
_{1}d\widetilde{h}_{3} \\
\qquad \qquad \,\,\,\,0=\widetilde{\pi }_{1}d\widetilde{h}_{4}\,\,,%
\end{array}%
\end{array}
\tag{7.16}  \label{7.15a}
\end{equation}%
where
\begin{equation*}
\widetilde{h}_{0}=c_{1},\ \ \widetilde{h}_{3}=c_{2},\ \widetilde{h}_{i}=%
\frac{1}{2}p^{T}\widetilde{A}_{i}\ p+\widetilde{V}_{i}+\alpha
_{i}c_{1}+\beta _{i}c_{2},\ \ \ i=1,2,4.
\end{equation*}%
Separation curves for respective Hamiltonian functions are
\begin{equation*}
\widetilde{H}_{1}\xi ^{3}+\widetilde{H}_{2}\xi ^{2}+\widetilde{H}_{4}=\frac{1%
}{8}\mu ^{2}+16\xi ^{7},
\end{equation*}%
\begin{equation*}
c_{1}\xi ^{4}+\widetilde{h}_{1}\xi ^{3}+\widetilde{h}_{2}\xi ^{2}+c_{2}\xi +%
\widetilde{h}_{4}=\frac{1}{8}\mu ^{2}+16\xi ^{7}.
\end{equation*}%
\begin{equation*}
\end{equation*}

\textbf{2-hole deformation.}

There are three admissible cases of $2$-hole deformation. One related to
3-Casimir Poisson pencil: $n_{1}=2,n_{2}=4$ and two cases related to
2-Casimir Poisson pencils, i.e. $n_{1}=2,n_{2}=3$ and $n_{1}=3,n_{2}=4.$
Here we present the first case. The deformed Hamiltonians are the following
\begin{eqnarray}
\widetilde{H}_{1} &=&\frac{1}{\rho _{1}\rho _{2}-\rho _{3}}H_{2}=\frac{1}{2}%
p^{T}\widetilde{G}\ p+\widetilde{V}_{1},  \notag \\
\widetilde{H}_{2} &=&0,  \notag \\
\widetilde{H}_{3} &=&H_{1}+\frac{\rho _{2}-\rho _{1}^{2}}{\rho _{1}\rho
_{2}-\rho _{3}}H_{2}=\frac{1}{2}p^{T}\widetilde{A}_{3}\ p+\widetilde{V}_{3},
\TCItag{7.17}  \label{7.16} \\
\widetilde{H}_{4} &=&0,  \notag \\
\widetilde{H}_{5} &=&H_{3}+\frac{\rho _{1}\rho _{3}}{\rho _{1}\rho _{2}-\rho
_{3}}H_{2}=\frac{1}{2}p^{T}\widetilde{A}_{5}\ p+\widetilde{V}_{5},  \notag
\end{eqnarray}%
where
\begin{equation*}
\widetilde{A}_{1}=\widetilde{G}=\frac{1}{\rho _{1}\rho _{2}-\rho _{3}}%
K_{1}K_{2}G=\frac{2}{q_{1}^{3}-q_{1}q_{2}+q_{3}}\left(
\begin{array}{ccc}
0 & 1 & q^{1} \\
1 & \frac{1}{2}q^{1} & -q^{2} \\
q^{1} & -q^{2} & -\frac{1}{2}q^{3}%
\end{array}%
\right) ,
\end{equation*}%
\begin{equation*}
\widetilde{A}_{2}=0,\ \ \ \widetilde{A}_{3}=\frac{1}{\rho _{1}\rho _{2}-\rho
_{3}}K_{2}(K_{3}-K_{2}^{2})G,\ \ \ \widetilde{A}_{4}=0,\ \ \ \widetilde{A}%
_{5}=\frac{1}{\rho _{1}\rho _{2}-\rho _{3}}K_{2}K_{3}^{2}G,
\end{equation*}%
\begin{equation*}
\widetilde{V}_{1}=\frac{4}{q_{3}-q_{1}q_{2}-q_{1}^{3}}V_{2},\ \widetilde{V}%
_{2}=0,\ \widetilde{V}_{3}=V_{1}+\frac{2q_{2}-3q_{1}^{2}}{%
q_{3}-q_{1}q_{2}-q_{1}^{3}}V_{2},\
\end{equation*}%
\begin{equation*}
\widetilde{V}_{4}=0,\ \widetilde{V}_{5}=V_{3}+\frac{q_{1}(q_{1}q_{2}+q_{3})}{%
q_{3}-q_{1}q_{2}-q_{1}^{3}}V_{2}.
\end{equation*}%
The quasi-bi-Hamiltonian chain takes the form
\begin{equation}
d\widetilde{H}_{r+1}=N^{\ast }d\widetilde{H}_{r}+\alpha _{r}^{0}d\widetilde{H%
}_{1}+\alpha _{r}^{2}d\widetilde{H}_{3}+\alpha _{r}^{4}d\widetilde{H}_{5},\
r=1,...,5,  \tag{7.18}  \label{7.17}
\end{equation}%
\begin{equation*}
\alpha _{r}^{0}=\rho _{r}+\frac{\rho _{r-2}\rho _{2}\rho _{3}-\rho
_{r-1}\rho _{2}^{2}}{\rho _{1}\rho _{2}-\rho _{3}},\ \ \alpha _{r}^{2}=\frac{%
\rho _{r-1}\rho _{2}-\rho _{r-2}\rho _{3}}{\rho _{1}\rho _{2}-\rho _{3}},\ \
a_{r}^{4}=\frac{\rho _{r-2}\rho _{1}-\rho _{r-1}}{\rho _{1}\rho _{2}-\rho
_{3}}
\end{equation*}%
and the respective bi-Hamiltonian chain splits into three sub-chains
\begin{equation}
\begin{array}{ccc}
\begin{array}{l}
\widetilde{\pi }_{0}d\widetilde{h}_{0}=0 \\
\widetilde{\pi }_{0}d\widetilde{h}_{1}=\widetilde{X}_{1}=\widetilde{\pi }%
_{1}d\widetilde{h}_{0} \\
\qquad \qquad \,\,\,\,0=\widetilde{\pi }_{1}d\widetilde{h}_{1}\,\,%
\end{array}
&
\begin{array}{l}
\widetilde{\pi }_{0}d\widetilde{h}_{2}=0 \\
\widetilde{\pi }_{0}d\widetilde{h}_{3}=\widetilde{X}_{3}=\widetilde{\pi }%
_{1}d\widetilde{h}_{2} \\
\qquad \qquad \,\,\,\,0=\widetilde{\pi }_{1}d\widetilde{h}_{3}\,\,%
\end{array}
&
\begin{array}{l}
\widetilde{\pi }_{0}d\widetilde{h}_{4}=0 \\
\widetilde{\pi }_{0}d\widetilde{h}_{5}=\widetilde{X}_{5}=\widetilde{\pi }%
_{1}d\widetilde{h}_{4} \\
\qquad \qquad \,\,\,\,0=\widetilde{\pi }_{1}d\widetilde{h}_{5,}\,\,%
\end{array}%
\end{array}
\tag{7.19}  \label{7.18}
\end{equation}%
where%
\begin{equation*}
\widetilde{\pi }_{0}=\left(
\begin{array}{c|c}
\theta _{0} & 0\ \ 0\ \ \ 0 \\ \hline
\begin{array}{c}
0 \\
0 \\
0%
\end{array}
& 0%
\end{array}%
\right) \text{ \ , \ }\widetilde{\pi }_{1}=\left(
\begin{array}{c|c}
\theta _{1} & \theta _{0}d\widetilde{h}_{1}\ \ \ \theta _{0}d\widetilde{h}%
_{3}\ \ \ \theta _{0}d\widetilde{h}_{5} \\ \hline
\begin{array}{c}
-(\theta _{0}d\widetilde{h}_{1})^{T} \\
-(\theta _{0}d\widetilde{h}_{3})^{T} \\
-(\theta _{0}d\widetilde{h}_{5})^{T}%
\end{array}
& 0%
\end{array}%
\right) .
\end{equation*}%
and
\begin{equation*}
\widetilde{h}_{0}=c_{1},\ \ \widetilde{h}_{2}=c_{2}\ \ \widetilde{h}%
_{4}=c_{3},\ \ \widetilde{h}_{i}=\widetilde{H}_{i}+\alpha
_{i}^{0}c_{1}+\alpha _{i}^{2}c_{2}+\alpha _{i}^{4}c_{3},\ \ i=1,3,5.
\end{equation*}%
The respective separation curves are
\begin{equation}
\widetilde{H}_{1}\xi ^{4}+\widetilde{H}_{3}\xi ^{2}+\widetilde{H}_{5}=\frac{1%
}{8}\mu ^{2}+16\xi ^{7},  \tag{7.20a}  \label{7.19a}
\end{equation}%
\begin{equation}
c_{1}\xi ^{5}+\widetilde{h}_{1}\xi ^{4}+c_{2}\xi ^{3}+\widetilde{h}_{3}\xi
^{2}+c_{3}\xi +\widetilde{h}_{5}=\frac{1}{8}\mu ^{2}+16\xi ^{7}.  \tag{7.20b}
\label{7.19b}
\end{equation}

\section{Summary}

We have presented a geometric separability theory of Liouville integrable
systems with $n$ quadratic in momenta constants of motion
\begin{equation}
H_{i}(q,p)=p^{T}A_{i}(q)p+V_{i}(q),\ \ \ i=1,...,n  \tag{8.1}  \label{8.1}
\end{equation}%
and with separation curves of polynomial type%
\begin{equation}
H_{1}\xi ^{m_{1}}+...+H_{n}\xi ^{m_{n}}=\frac{1}{2}f(\xi )\mu ^{2}+\gamma
(\xi ),\ \ \ m_{n}=0<m_{n-1}<...<m_{1}\in \mathbb{N},  \tag{8.2}  \label{8.2}
\end{equation}%
where $f(\xi ),$ and $\gamma (\xi )$ are Laurent polynomials of $\xi $. Such
systems can be lifted to the class of Gel'fand-Zakharevich bi-Hamiltonian
systems, defined by linear Poisson pencils and their Casimirs: polynomial
functions in pencil parameter. First we briefly summarized the geometric
separability theory of all GZ bi-Hamiltonian systems. Then, we reviewed with
details a special class of such systems introduced by Benenti and
characterized by separation curves of the form (\ref{8.2}) with $m_{k}=n-k.$
Finally, we proved in a very systematic way that all remaining separable
systems, characterized by separation curves of the form (\ref{8.2}), are
constructed by the appropriate deformations of related Benenti systems. The
most fundamental and surprising result of the presented paper can be
formulated in the following way. If, for a given canonical coordinate system
$(q_{i},p_{i}),i=1,...,n,$ we have a pair of objects, i.e. contravariant
metric tensor $G^{(0)}$ and related special conformal Killing tensor $L,$
then we can construct systematically, in these coordinates, all Liouville
integrable and separable Hamiltonian systems (\ref{8.1}) with respective
separation curves of the form (\ref{8.2}) and explicit form of
transformation to separated coordinates. Observe, that accordind to what was
said in Section 4, the separation curve of geodesic motion for $G^{(0)\text{
}}$ is the following
\begin{equation}
E_{1}\xi ^{n-1}+E_{2}\xi ^{n-2}+...+E_{n}=\frac{1}{2}\mu ^{2}.  \tag{8.3}
\label{8.3}
\end{equation}%
So, the passage from the system (\ref{8.3}) to the systems (\ref{8.2}) is
constructive and determined completely by the pair $(G^{(0)},L).$

\subsection*{Acknowledgments}

The author would like to thanks prof. Franco Magri for many stimulating
discussions on the subject. This research was supported in part by KBN grant
No. 5P03B 004 20

\end{document}